\documentclass[12pt]{article}
\usepackage{amsmath,amssymb}
\usepackage{hyperref}

\usepackage{subfigure}
\usepackage{color}
\usepackage{tikz}
\usetikzlibrary{patterns}
\usepackage{wrapfig}

\usepackage{young}
\usepackage[vcentermath]{youngtab}

\definecolor{darkred}{rgb}{0.5,0,0.5}
\hypersetup{pdfborder={0 0 0},colorlinks=true,urlcolor=darkred,citecolor=blue,linkcolor=darkred}

\makeatletter

\newcommand{\ie}{i.e.,~}
\newcommand{\re}{\mathbb{R}}
\newcommand{\ce}{\mathbb{C}}

\newcommand{\ze}{\mathbb{Z}}

\@addtoreset{equation}{section}
\makeatother

\newcommand{\al}{\beta}
\newcommand{\bt}{\alpha}
\newcommand{\lb}{\lambda}

\newcommand{\g}{\mathfrak{g}}

\newcommand{\sgl}{\mathfrak{sl}}
\newcommand{\su}{\mathfrak{su}}
\newcommand{\so}{\mathfrak{so}}
\newcommand{\mfs}{\mathfrak{s}}
\newcommand{\mfu}{\mathfrak{u}}
\newcommand{\mfh}{\mathfrak{h}}
\newcommand{\mfk}{\mathfrak{k}}
\newcommand{\mfp}{\mathfrak{p}}
\newcommand{\mfa}{\mathfrak{a}}
\newcommand{\mft}{\mathfrak{t}}
\newcommand{\mfb}{\mathfrak{b}}
\newcommand{\mfn}{\mathfrak{n}}
\newcommand{\ssp}{\mathfrak{sp}}
\newcommand{\usp}{\mathfrak{usp}}

\begin{document}

{\flushright{AEI-2011-064\\ ULB-TH/11-23}\\}

\thispagestyle{empty}

\setcounter{page}{0}

\mbox{}
\vspace{8mm}

\begin{center} {\bf \Large  Real forms of extended Kac-Moody\\[3mm]  symmetries and higher spin gauge theories}

\vspace{1.6cm}

Marc Henneaux${}^{1,2,3,4}$, Axel Kleinschmidt${}^{3,4}$ and Hermann Nicolai${}^4$

\footnotesize
\vspace{.6 cm}

${}^1${\em Centro de Estudios Cient\'{\i}ficos (CECS), Casilla 1469, Valdivia, Chile}

\vspace{.1cm}

${}^2${\em Universit\'e Libre de Bruxelles, ULB-Campus Plaine CP231, B-1050 Brussels, Belgium} \\

\vspace{.1cm}

${}^3${\em International Solvay Institutes, ULB-Campus Plaine CP231, B-1050 Brussels, Belgium} \\

\vspace{.1cm}

${}^4${\em Max-Planck-Institut f\"ur Gravitationsphysik (Albert-Einstein-Institut),
M\"uhlenberg 1, D-14476 Potsdam, Germany}

\vspace {15mm}

\end{center}
\centerline{\bf Abstract}
\vspace{.6cm}
We consider the relation  between higher spin gauge fields and real Kac-Moody Lie 
algebras. These algebras are obtained by double and triple extensions of real forms $\g_0$ 
of the finite-dimensional simple algebras $\g$ arising in dimensional reductions of gravity and supergravity theories. Besides providing an exhaustive list of all such algebras, together with
their associated involutions and restricted root diagrams, we are able to prove general 
properties of their spectrum of generators with respect to a decomposition of the triple extension
of $\g_0$ under its gravity subalgebra ${\mathfrak{gl}}(D,\re)$. These results are then combined 
with known consistent models of higher spin gauge theory to prove that all but finitely many generators correspond to non-propagating fields and there are no higher spin fields contained 
in the Kac-Moody algebra. 
\vspace{.8cm}
\noindent

\newpage
\section{Introduction}

The existence of infinite-dimensional  Kac-Moody symmetries in models of 
matter-coupled gravity or supergravity theories extending the known symmetries of gravitational 
systems has been suggested repeatedly~\cite{Julia,Nicolai:1986jk}. In the context
of M-Theory (the putative non-perturbative formulation of superstring theory)
the most widely discussed recent proposals  involve  $E_{11}$~\cite{West} and 
$E_{10}$~\cite{Damour:2002cu}, both of which are  infinite-dimensional 
Kac-Moody extensions of the exceptional split Lie algebra $E_{8,8}$ arising in 
(ungauged) maximal supergravity in three dimensions. Being double and triple
extensions  of $E_{8,8}$, the split Lie algebras $E_{10}$ and $E_{11}$ are often denoted 
as $E_8^{++}$ and  $E_8^{+++}$, respectively. For any simple, complex and finite-dimensional Lie algebra $\g$ there exists a chain of embeddings of complex Lie algebras
 \begin{equation}
 \label{eq:extend}
 \g \subset \g^+ \subset \g^{++} \subset \g^{+++}
 \end{equation}
where $\g^+$ is the affine extension of $\g$,
while the further extensions $\g^{++}$ and $\g^{+++}$  are Kac-Moody algebras
of indefinite type (hence, the algebras $\g^+$, $\g^{++}$ and $\g^{+++}$ are all
infinite-dimensional). One of the objectives of this paper is to study the chain of embeddings of {\em real} Lie algebras
 \begin{equation}
 \label{eq:extendreal}
 \g_0 \subset \g^+_0 \subset \g^{++}_0 \subset \g^{+++}_0
 \end{equation}
associated to a real form $\g_0$ of the complex $\g$ in (\ref{eq:extend}). The three extensions are defined and analysed in detail below. For the case $\g_0=E_{8,8}$ one obtains the split real Lie algebras $E_{10}$ and $E_{11}$ of~\cite{West,Damour:2002cu}.

More generally, real forms $\g_0$ of a finite-dimensional Lie algebra $\g$ 
often arise as infinitesimal symmetries of  matter coupled gravity theories via dimensional reduction 
from $D\geq 4$ to $D=3$ \cite{BGM}. Well-known examples of the associated real Lie
groups $G_0$  include the Ehlers 
group $SL(2,\re)$ for pure gravity in $D=4$ dimensions (and more generally 
$SL(D-2,\re)$ for higher dimensional pure gravity), and $SU(2,1)$ for 
Maxwell-Einstein gravity in $D=4$. Affine symmetries $\g_0^+$ emerge upon
further reduction to $D=2$ (axisymmetric stationary or colliding plane wave
solutions), the best known example being
the Geroch group $A_1^{(1)} \equiv \widehat{SL(2,\re)}_{ce}$~\cite{Geroch:1970nt,Breitenlohner:1986um}. One can also study $D=3$ systems coupled to a maximal symmetric 
space of type $G_0/K(G_0)$ directly, where ${\rm Lie}(G_0)=\g_0$ and $K(G_0)$ is the maximal compact subgroup~\cite{CJP,Damour:2002fz,Keurentjes,Keurentjes:2002rc}. 

The Kac-Moody algebras $\g^{++}_0$ and $\g^{+++}_0$ occurring in the two proposals of~\cite{West,Damour:2002cu} are both infinite-dimensional, 
and not fully understood. In fact, only a finite number of  so-called `low level' 
generators have found an interpretation 
in terms of physical degrees of freedom to date. Here, the term {\em level} refers to the 
decomposition of either algebra into an infinite ordered `stack' of representations of
some {\em finite-dimensional} subalgebra. In this paper, this subalgebra is always
taken to be the $GL(D,\re)$ symmetry associated to the $D$-bein describing
gravity in $D$ space-time dimensions for $\g^{+++}_0$, or the spatial $(D-1)$-bein 
for $\g^{++}_0$. Therefore, the Tits-Satake diagrams of $\g^{++}_0$ and $\g^{+++}_0$ always
contain, respectively, an $A_{D-2}$ and an $A_{D-1}$ subdiagram. In the remainder
we will refer to this subdiagram as the {\em gravity line}.

Despite their evident technical similarities, the conjectures
of \cite{West} and \cite{Damour:2002cu}
are very different conceptually. The approach based on the `very extended' algebra
$\g^{+++} _0= E_{11}$ \cite{West} aims for a covariant description of the dynamics
underlying M-Theory. More precisely, the $\g^{+++}_0$ symmetries are supposed to 
give rise to space-time covariant descriptions of the underlying field equations via 
a non-linear realisation, such that $\g^{+++}_0$ acts as a generalized duality symmetry
on the field equations. Various gauge symmetries (such as space-time diffeomorphisms)
are conjectured to be contained in some extension of $\g^{+++}_0$ such as 
$\g^{+++}_0\ltimes l_1$ (where $l_1$ is the so-called `fundamental representation') or 
embedded in the yet further extended $\g^{++++}_0$ (see e.g. \cite{West:2003fc,West1}) or an even larger 
extension~\cite{Riccioni:2009hi}.

By contrast, the Hamiltonian approach \cite{Damour:2002cu} realizes only 
the over-extended algebras $\g^{++}_0$. Evidence for the
appearance of the indefinite (often hyperbolic) symmetries $\g^{++}_0$ mainly
comes from a BKL-type analysis of cosmological solutions of Einstein's equations near
a spacelike singularity; indeed, the most economical description of the chaotic
metric oscillations {\it \`a la} BKL is in terms of a `cosmological billiard' that takes place 
in the Weyl chamber of the algebra $\g^{++}_0$~\cite{Damour:2000hv,Damour:2001sa,DHN}.
The basic assumption of~\cite{Damour:2002cu} 
is then that the $\g^{++}_0$ symmetric formulation, a `geodesic' nonlinear
$\sigma$-model based on the coset $G^{++}_0/K(G^{++}_0)$ (where 
$K(G^{++}_0)$ is maximal compact subgroup of $G^{++}_0$) can be matched
with the field theory only after gauges have been fixed, as is the case
for the derivation of $E_7$ invariant supergravity from $D=11$ supergravity
by dimensional reduction \cite{CJ}. In this approach space-time symmetries
(as well as space-time itself) are supposed to be `emergent', and not necessarily 
part of the underlying theory. The full symmetry of M-Theory would
thus become apparent only {\em at} the singularity.

A key question in both approaches concerns 
the physical interpretation of the higher level generators in the 
decompositions of $\g^{++}_0$ and $\g^{+++}_0$ under their respective 
gravity subalgebras ${\mathfrak{gl}}(D,\re)$. From string theory, one would expect these degrees of freedom 
to be related to the higher excited string modes (perhaps after taking a
suitable zero slope limit). However, all indications so far point in a different 
direction. Namely, as has been noted in several places~\cite{Riccioni:2006az,Riccioni:2007au,Bergshoeff:2007qi,Riccioni:2007hm,deWit:2008ta,Henneaux:2008nr},
the higher level degrees of freedom either correspond to possible `deformations'
(such as gauged supergravities), or are `dual' to the low order degrees
of freedom (in a sense that remains to be specified).
For $\g^{++}_0$, a special role is played by the so-called gradient representations
\cite{Damour:2002cu} (further analyzed in \cite{KN,KNP}) which are conjectured
to correspond to the degrees of freedom arising in a gradient expansion
{\it \`a la} BKL. The latter can be related to an expansion in heights
of roots of the corresponding over-extended algebra. In the covariant $\g^{+++}_0$ approach 
of \cite{West,Riccioni:2006az} these representations  would  correspond to 
`dual fields' generalizing the relation between $p$-form fields
and $(D-p-2)$-form fields. However, covariant dualities so far exist
only for free fields, and several No-Go Theorems~\cite{Bekaert:2002uh,Bekaert:2004dz} indicate
that non-linear deformations may be altogether incompatible with space-time 
covariance. In fact, the only known example of an infinite-dimensional gravitational
duality symmetry that works at the non-linear level is the stationary axisymmetric
reduction of Einstein's equations, where the Geroch group relates the
basic physical fields to an infinite tower of `dual potentials'~\cite{Geroch:1970nt,Breitenlohner:1986um,Nicolai:1988jb}.
See also \cite{KN,KNP} for an analysis of the relation between the Geroch group and the affine subalgebra $\g^+_0\subset\g^{+++}_0$. 

The purpose of this paper is to address the issue of the higher level fields in full 
generality for triply extended $\g^{+++}_0$ algebras where $\g_0$ can be any real form 
of a finite-dimensional complex Lie algebra and the precise real form of $\g^{+++}_0$ 
is defined below. The generality of our analysis
is warranted by the fact that all these real forms actually do occur in gravitational models. While 
the split real forms are often relevant for (maximal) supergravities, there are many 
examples of non-split real forms that are present in other gravitational systems (as, for 
instance $\g_0=\mathfrak{su}(2,1)$ for Einstein-Maxwell gravity). Some of the magic supergravity theories~\cite{Gunaydin:1983rk} also possess global symmetries of non-split type.
We will obtain results on the structure of the higher level fields that group them 
into four different classes, depending on whether the fields arise from the 
$\g_0$, $\g^+_0$, $\g^{++}_0$ or $\g^{+++}_0$ part in the extension process (\ref{eq:extendreal}). Since many of the features can be obtained from properties of the complexified embedding (\ref{eq:extend}) we will often refer to the four classes as arising from $\g$, $\g^+$, $\g^{++}$ and $\g^{+++}$. The fields will 
be written in terms of representations of the gravity subalgebra ${\mathfrak{gl}}(D,\re)$ 
already mentioned above; the representation (Dynkin) labels are thus attached to the
gravity line. The four classes have very different but characteristic shapes in 
terms of Young tableaux of ${\mathfrak{gl}}(D,\re)$. Only the first class consisting 
of the fields arising from $\g$ are finite in number; all the other classes contain 
{\em infinitely many} fields. The class associated with $\g^+_0$ contains the 
`dual fields' or `gradient representations' mentioned above.

Having classified the fields arising in any $\g^{+++}_0$ we address the issue of the 
number of local degrees of freedom associated with these (generically mixed symmetry) fields. The assumption that we make and motivate further below is that they 
are described by the gauge symmetries and dynamics as they arise in the literature 
on higher spin fields~\cite{Curtright:1980yk,Labastida}. Under this assumption we show,
as one of our main results, that there are no local degrees of freedom associated with 
any of the fields arising from $\g^+$, $\g^{++}$ and $\g^{+++}$. In particular, and
contrary to what one might expect, the `dual fields' associated with the affine
algebra $\g^+$ are seen {\em not} to give covariant dual descriptions of the physical fields contained in $\g$ (which do carry degrees of freedom). This implies that there are 
no higher spin fields contained in $\g^{+++}$, not even at the linear level. All the  
fields required for making the $\g^{+++}$ symmetry manifest are therefore 
truly auxiliary and non-propagating.\footnote{The possible interpretation of the higher level fields has changed over time. Whereas early literature suggested that there are higher spin gauge fields contained in $\g^{+++}$~\cite{deBuyl:2004ps,West:2007mh}, more recent arguments point in the opposite direction~\cite{Riccioni:2009hi}. We will compare our results with those of~\cite{Riccioni:2009hi} in more detail below.}

Our article is structured as follows. First, we discuss the extension process for real 
forms $\g_0$ of $\g$ to real forms $\g^{+++}_0$ of $\g^{+++}$. This results in very special real forms of Kac-Moody algebras that have not been discussed fully in the literature before (see, however,~\cite{BackValente,BenMessaoud} for the general theory and~\cite{Riccioni:2008jz} for 
some special cases and low level field content). Then, in section~\ref{sec:levdec}, we 
proceed to prove general properties of the full field content for all $\g^{+++}_0$ and 
also properties of the $l_1$ representation (referred to as the $L(\Lambda_{D-1})$
representation in the mathematical literature ). In section~\ref{sec:lag}, we discuss the free 
field theory realisation of the higher level fields and their associated degrees of freedom. Several appendices contain background material on real forms of Lie algebras, and the explicit construction of all necessary involutions and restricted root system for all $\g^{+++}_0$.

\section{Extending real forms of finite-dimensional simple Lie algebras}
\label{AlmostSplit}

The real forms of the Kac-Moody algebras $\g^{+++}$, $\g^{++}$ and $\g^+$ considered 
here are of a particular type: they are obtained by extending in a direct manner the 
real forms of the underlying finite-dimensional algebras $\g$ which give rise to them.  
Complex Lie algebras $\g^+$, $\g^{++}$ and $\g^{+++}$ are obtained by first constructing 
the non-twisted affine extension of $\g$ and then extending twice from the affine node by 
a single line (see for instance~\cite{Gaberdiel:2002db}). Here, we will show how to do 
this for the real forms of interest and how to obtain the associated Tits-Satake diagrams.
Our discussion in this section is slightly more technical than in the other sections and relies on background material on real forms $\g_0$ of Lie algebras that is presented in appendix~\ref{app:real}. Our main statements in later sections can be understood without delving 
into the details here; so readers may skip them on a first perusal. We emphasize, however,
that the proofs do depend on the proper definitions explained here and in the appendices.

\subsection{Real forms of $\g^+$}

A real form $\g_0$ of a finite-dimensional complex Lie algebra $\g$ is defined by a conjugate-linear 
involution $\sigma'$, in such a way that the real form 
$\g_0$ is  the fixed point algebra of $\sigma'$.\footnote{The prime on $\sigma'$
  is to indicate that we are dealing with an {\em anti}-involution (i.e., conjugate linear). See appendix~\ref{app:real}.} 
Our aim now is to define appropriate extensions of $\sigma'$ when the finite-dimensional $\g$ is extended through the chain (\ref{eq:extend}).

The untwisted affine Kac-Moody algebra $\g^+$ associated with the complex
finite-dimensional simple Lie algebra $\g$ can be identified with 
$\g \otimes \ce[t,t^{-1}] \oplus \ce c \oplus \ce d$ where $\ce[t,t^{-1}]$ is the 
complex vector space of Laurent polynomials in the formal parameter $t$, 
$c$ is the central charge and $d$ the degree.  It  has non-trivial commutators given by
\begin{eqnarray}\label{affalg}
&& [x \otimes t^n, y \otimes t^m ] = [x,y] \otimes t^{n+m}  + g(x,y) n \delta_{n+m, 0} \,  c , \label{n,m}\nonumber\\
&& [ d, x \otimes t^n] = n \, x \otimes t^n
\end{eqnarray}
where $x \otimes t^n$ ($x \in \g$, $n \in \ze$) are the elements of the loop algebra $\g \otimes \ce[t,t^{-1}] $ of $\g$ and $g(x,y)$  is the Killing form of $\g$.

Now let $\sigma'$ be a conjugation of $\g$ defining the real form $\g_0$.  One 
extends $\sigma' $ to $\g^+$ as follows,
\begin{equation}
\sigma'(x \otimes t^n) = \sigma'(x) \otimes t^n, \; \; \; \; \sigma'(c) = c, \; \; \; \; \sigma'(d) = d
\end{equation}  
One easily verifies that $\sigma'$ is a conjugation of $\g^+$.  The corresponding 
real form of $\g^+$ is $\g^+_0 \equiv \g_0 \otimes \re[t,t^{-1}] \oplus \re c \oplus \re d$.  
There are other possible involutions on $\g^+$ (in particular also involving the 
formal parameter $t$), but this is the only type of real forms of $\g^+$ that we shall 
need to consider from now on.  

The simple roots of $\g^+$ are the simple roots of $\g$ together with the affine root $\bt_0$.  
In this section our labelling is such that the simple roots of $\g$ are called 
$\bt_i$ ($i=1,\ldots,{\rm rank}(\g)$) and the three extending nodes are 
$\bt_0$, $\bt_{-1}$ and $\bt_{-2}$. The root vector $e_{\bt_0}$ is $e_{-\mu} \otimes t$ 
where $\mu$ is the highest root of $\g$, and $e_{-\bt_0}  = e_{\mu} \otimes t^{-1}$. 
The standard Borel subalgebra of $\g^+$ is the linear span of $x \otimes t^n$ 
($n>0$),  $y \otimes t^0$ where $y$ belongs to the standard Borel subalgebra 
of $\g$, $c$, and $d$.

One has $\sigma'(e_{\bt_0}) = e_{-\sigma'(\mu)} \otimes t$. The complex linear Cartan involution $\theta$ associated with the real form  $\g^+_0$  is obtained by multiplying $\sigma'$ with the antilinear involution $\tau'$ associated with the standard compact form and defined in the appendix. It coincides on $\g$ with the Cartan involution $\theta$ associated with $\g_0$ and so it extends it (hence the same notation).  On $e_{\bt_0}$ it yields
\begin{equation}
\theta(e_{\bt_0}) = \theta(e_{- \mu}) \otimes t^{-1}.
\end{equation}
Therefore, since $\theta(\g_\al) = \g_{\theta (\al)}$ and $e_{-\bt_0}  = e_{\mu} \otimes t^{-1}$, one concludes that the affine root $\bt_0$ transforms as
\begin{equation} \theta (\bt_0)  = - \bt_0 - \mu - \theta(\mu) \label{transfaff}
\end{equation}
(use the obvious equality $- \theta(\mu)  = \mu - \mu - \theta(\mu)$ and observe that the first $\mu$ on the right hand side of this equality is precisely what is needed to make  $e_{-\bt_0}$ appear).  The action of the Cartan involution on the other simple roots, which are roots of $\g$, is just the one of the original Cartan involution.

To further analyse the properties of $\g^+_0$, one considers a maximally split Cartan subalgebra $\mfh_0 = \mft_0 \oplus \mfa_0$ of $\g$ adapted to the real form $\g_0$, where $\mft_0$ are the compact and $\mfa_0$ the non-compact generators in $\mfh$. Maximally split means that one chooses $\mfa_0$ as large as possible.
The affine root $\bt_0$ is never mapped on itself by the Cartan involution since $\bt_0$ and $\mu + \theta(\mu)$ are linearly independent.  Accordingly, $\bt_0$ never vanishes on the noncompact subalgebra $\mfa$ of the maximal split Cartan subalgebra\footnote{This is in fact rather obvious since the noncompact subalgebra $\mfa_{\g^+}$ is given by $\mfa_{\g^+}  = \mfa_{\g} \oplus \ce c \oplus \ce d$, and $\bt_0$ does not vanish on $d$, $[d, e_{\bt_0}] = e_{\bt_0}$.}.  It can vanish, however, on the compact part $\mft$.  This occurs if and only if $\theta(\mu) = - \mu$, that is, if and only if the highest root of $\g$ is `real' (in the terminology used in the study of real forms -- it is of course always real in the Kac-Moody sense since it is a root of the finite-dimensional 
algebra $\g$).
Note also that $\bt_0$ fulfills the `normality condition' \cite{Araki} that $\bt_0 + \theta(\bt_0)$ is not a root, since this condition is satisfied by the roots of the given real form of the finite-dimensional algebra $\g$, and in particular by the highest root $\mu$.

As is well known \cite{Kac} the root lattice $Q(\g^{++})$ of the over-extended algebra
$\g^{++}$ is related to the root lattice $R\equiv Q(\g)$ of the finite-dimensional 
subalgebra $\g$ via the canonical extension by a two-dimensional Lorentzian space
$V\equiv \mathbb{R}^{1,1}$ (taken to be orthogonal to $R$) with basis $\{u_1,u_2\}$ 
and Lorentzian bilinear form of signature $(-,+)$,
\begin{equation}
(u_1 \vert u_2) = 1,  \; \; \; \; (u_1 \vert u_1) = 0, \; \; \; \; (u_2 \vert u_2) = 0.
\end{equation}
In the canonical extension $R\oplus V$ there is a representation of the affine root $\bt_0$ 
as \cite{Kac}
\begin{equation}
\bt_0 = - u_2 - \mu
\end{equation}
From (\ref{transfaff}) the action of the Cartan involution on $u_2$ is then found to be
\begin{equation}\label{u2}
\theta(u_2) = - u_2.
\end{equation}

The real form $\g^+_0$ is almost split in the sense of~\cite{BackValente}.  Indeed, the image of the standard positive Borel subalgebra  of $\g$ by $\sigma'$ is $G$-conjugate to it through the adjoint action of some $\rho \in G$ since $\g$ is finite-dimensional.  The action of $\rho$ extended to $\g^+$ is such that $\rho(x \otimes t^n) = \rho(x) \otimes t^n$ and $\rho(c)= c$, $\rho(d) =d$.   Therefore, the image of the standard positive Borel subalgebra  of $\g^+$ by $\sigma'$ is conjugate to it through the adjoint action of the same (extended) $\rho$.

 \subsection{Real forms of $\g^{++}$ and $\g^{+++}$}
 
 The real forms of $\g^{++}$ relevant for our discussion are obtained by extending the above conjugation $\sigma'$ to the generators associated with the over-extended root $\bt_{-1}$ as
 \begin{equation}
 \sigma' (h_{-1}) = h_{-1}, \; \; \; \; \sigma' (e_{-1}) = e_{-1}, \; \; \; \; \sigma' (f_{-1}) = f_{-1}.
 \end{equation}
 This is permissible because this definition preserves all the Chevalley-Serre relations involving the generators $\{h_{-1},  e_{-1}, f_{-1}\}$.\footnote{This is a straightforward consequence of the facts that (i) the over-extended generators commute with $\g$; (ii) $h_{-1} = -d$; (iii) $\sigma'(e_0)= [e_{i_1}, [ e_{i_2}, \ldots, [e_{i_k}, e_0] \ldots ]]$ where the chain $e_{i_j}$ is the chain of raising operators of $\g$ necessary to go from the root vector $e_{-\mu}$ to $e_{- \sigma'(\mu)}$.}
 This definition implies 
 \begin{equation}
 \theta(\bt_{-1}) = - \bt_{-1}
 \end{equation}
 for the over-extended root.   In terms of the above orthogonal direct sum $V \oplus R$, $\bt_{-1}$ is given by $u_1 + u_2$ so that one has also 
 \begin{equation}
 \theta(u_1) = -u_1.
 \end{equation}
which complements (\ref{u2}), thereby extending the action of $\theta$ to all of $V$, and
thus to the full root space of $\g^{++}$

The extension of $\sigma'$ to the very extended generators is done in exactly the same way,
\begin{equation}
 \sigma' (h_{-2}) = h_{-2}, \; \; \; \; \sigma' (e_{-2}) = e_{-2}, \; \; \; \; \sigma' (f_{-2}) = f_{-2},
 \end{equation}
leading to
\begin{equation}
 \theta(\bt_{-2}) = - \bt_{-2}
 \end{equation}
for the very extended root $\bt_{-2}$.
 
 One sees in particular that (i) the roots $\bt_{-1}$  and $\bt_{-2}$ both vanish on the compact part of the Cartan subalgebra (they are `real roots' in the terminology used in the theory of real forms); accordingly, of the extended roots,  it is only the affine root that can be complex and on which the Cartan involution can have an action that differs from the simple one $\theta(\al) = - \al$; (ii) the real forms of $\g^{++}$ and $\g^{+++}$ so defined are almost split\footnote{The same $\rho$ as above, defined by conjugating with exponentials of elements of the finite-dimensional Lie algebra $\g \subset \g^{+++}$,  achieves the requested conjugation.}; and (iii) the normality condition that $\theta(\bt_{i}) + \bt_{i} $ is not a root is trivially satisfied by the over-extended and very extended roots since this sum is then zero.

\subsection{Tits-Satake diagrams}

We now aim to construct the Tits-Satake diagrams of $\g^{+++}$. For this we will also need to construct the restricted root systems of $\g^{+++}$.
The restricted roots of the real forms of the extensions of $\g$ can be easily described in terms of the restricted roots of the real forms of $\g$ itself.  Recalling the maximal split of the Cartan subalgebra $\mfh_0$ of $\g_0$ according to $\mfh_0=\mft_0\oplus\mfa_0$, one has that
elements in $\mfa_0$  have real eigenvalues when acting via the adjoint action on the real form of $\g$; the eigenvalues are the roots and form the restricted root system.  The restricted root system can be obtained easily by knowing the Cartan involution $\theta$ and the associated projection $\pi=\frac12(1-\theta)$: One projects all roots of $\g$ (or $\g^{+++}$) using $\pi$ and thus arrives at the restricted root system. The Tits-Satake diagram is constructed from the knowledge of $\theta$ as we will describe in more detail below.

The restricted root system of a real form of a finite-dimensional simple Lie algebra $\g$ is known to be one of the standard root systems of $A$, $B$, $C$, $D$, $G$, $F$ or $E$ type, or to be of $BC$ type when it is `nonreduced'.  The $BC_n$ root system is obtained by combining together the $B_n$ and $C_n$ root systems in such a way that the long roots of $B_n$ are the short roots of $C_n$.  It is the only root system for which non trivial multiples of roots can also be roots (hence the terminology `nonreduced').  If $\al$ is a short root of $B_n$, $2 \al$ is a long root of $C_n$.  The highest root of the nonreduced root system $BC_n$ is twice the highest short root of $B_n$.  The multiciplity of the restricted roots can be non trivial since two distinct roots can project on the same restricted root.

Let $\mfh = \mft \oplus \mfa$ be a maximally split Cartan algebra of $\g$, $\Gamma = \{\bt_i\}$ ($i = 1, \ldots , n$) a basis of simple roots and $\Gamma'= \{\lambda_m\}$ ($m = 1, \ldots, r$) the corresponding basis of restricted roots\footnote{The set $\{\pi(\bt_i)\}$ is in general not a basis of the restricted root system because the vectors $\pi(\bt_i)$ are in general not linearly independent. Some of the $\bt_i$'s project to zero, and two distinct $\bt_i$'s might project on the same restricted root. The set $\{\lambda_m\}$ is a linearly independent subset of the set $\{\pi(\bt_i)\}$.}. It is convenient to split the basis of simple roots of $\g$ into two subsets, $\{\bt_i\} = \{\gamma_k, \delta_p\}$.  The subset $B_0 = \{\gamma_k\}$ ($k = 1, \ldots, l$) contains the simple roots that vanish on $\mfa$ and hence project to zero, $\pi(\gamma_k) = 0$.  The subset $\Gamma \backslash B_0 = \{\delta_p\}$ ($p = 1, \ldots, s \geq r$) contains the remaining simple roots, which do not vanish on $\mfa$.

It can be shown that the Cartan involution acts as follows on the roots $\delta_p$,
\begin{equation}\label{Cartan}
\theta(\delta_p) = - \delta_{P(p)} + \sum_{k = 1}^l a_p^k \gamma_k
\end{equation}
where $P$ is a permutation of the indices $\{1, \ldots, s\}$ that squares to one and 
so is a product of commuting $1$-cycles or $2$-cycles.  If  $P(p) = p$ ($p$ belongs to a $1$-cycle), the only simple root that projects on $\pi(\delta_p)$ is $\delta_p$ (but note that non-simple roots may also project on $\pi(p)$).  If $P(p) = q$ and $P(q) = p$ ($p$ and $q$ belong to 
the same $2$-cycle),   the roots $\delta_p$ and $\delta_{P(p)}$ project on the same 
restricted root, $\pi(\delta_p) = \pi(\delta_q) $ and no other simple root projects 
on $\pi(\delta_p)$ (but non simple roots again may).  That (\ref{Cartan}) defines an 
involution (that is, $\theta^2 =1$) is a non-trivial constraint on the linear combinations
of roots that may arise in (\ref{Cartan}).

The Tits-Satake diagram of the real form $\g_0$ of $\g$ is obtained by adding further information  to the Dynkin diagram of $\g$ as follows.  If the simple root $\bt_i$ belongs to $B_0$, it is painted in black.  If it belongs to $\Gamma \backslash B_0$, it remains white. Furthermore, the orbits of the permutation $P$ are indicated through double arrows pointing to the roots belonging to the same $2$-cycle.  This extra information completely characterizes the real form $\g_0$.  

The Tits-Satake diagrams of the real forms of the extensions of $\g$ are drawn by following the same rules.
\vspace{.2cm}

\noindent
{\bf Affine extension $\g^+$:} The affine root does not vanish on $\mfa$ and so it belongs to  $\Gamma \backslash B_0$.  It is therefore a white root. Thus, when going from $\g_0$ to the corresponding real form of $\g^+$, one enlarges the set of white roots $\Gamma \backslash B_0$ by adding to it the affine root, while the set $B_0$ of black roots is unchanged. Furthermore, $\theta(\bt_0) = - \bt_0 - \mu - \theta(\mu)$, where $\mu + \theta(\mu)$ project to zero and is accordingly a combination of roots in $B_0$.  This implies that $\bt_0$ belongs to a $1$-cycle of $P$ and is the only simple root that projects on $\pi(\bt_0)$.   There is thus no $P$-arrow connecting  the affine root to any other root. We denote $\pi(\bt_0)$ by $\lambda_0$. 

One has
\begin{equation}
\lambda_0 = - u_2 - \pi(\mu)
\end{equation} 
since $\pi(u_2) = u_2$.  One easily verifies that the highest root of $\g$ projects on the highest root of the restricted root system.  Therefore, the restricted root system of the real form of the affine extension is the affine extension of the restricted root system of $\g_0$.

There is a slight subtlety when the restricted root system is of $BC_n$ type since in that case, the basis of $BC_n$ is taken to be a basis of roots of $B_n$.  The highest root of $BC_n$ is however not the highest root of $B_n$ and so it connects differently to the Dynkin diagram of $B_n$, leading to a `twisting'.  This phenomenon was described in \cite{Henneaux:2003kk}, where it was pointed out that it occurs for $N=2$, $N=3$ and $N=5$ supergravities in $D=4$ dimensions.  Except in this case, the Dynkin diagram of the restricted root system of the real form of the affine extension is the affine extension of the restricted root system of the real form $\g_0$.

The multiplicity of $\lambda_0$ is equal to the multiplicity of the highest root.  It may be non trivial as non simple roots might project on $\lambda_0$ besides the simple root $\bt_0$.

{} Finally, one observes that the norm of $\lambda_0$  is equal to the norm of $\bt_0$ if and only if the norm of $\pi(\mu)$ is equal to the norm of $\mu$.  This occurs if and only if $\pi(\mu) = \mu$, i.e., $\theta(\mu) = - \mu$. This then yields $\theta(\bt_0) = - \bt_0$ and $\pi(\bt_0) = \bt_0$.  When this property does not hold, the norm of $\lambda_0$ is strictly smaller than the norm of $\bt_0$.

\vspace{.2cm}

\noindent
{\bf Double extension $\g^{++}$:}
The over-extended root does not vanish on $\mfa$ and so it belongs to  $\Gamma \backslash B_0$.  It is therefore also a white root. The set $B_0$ of black roots is again unchanged. Furthermore, $\theta(\bt_{-1}) = - \bt_{-1}$.  This implies that $\bt_{-1}$ belongs to a $1$-cycle of $P$ and is the only simple root that projects on $\pi(\bt_{-1}) \equiv \bt_{-1}$.   There is thus no arrow connecting  the over-extended root to any other root. We denote $\pi(\bt_{-1}) \equiv \bt_{-1}$ by $\lambda_{-1}$.   

Because the projection $\lambda_{-1}$ of $\bt_{-1}$ coincides with $\bt_{-1}$, these two vectors have the same norm.  The restricted root $\lambda_{-1}$ connects only to the restricted root $\lambda_0$.  The link is simple, as between $\bt_{-1}$ and $\bt_0$, if and only if $\pi(\bt_0) = \bt_0$ since otherwise $\lambda_0$ is shorter.  Adding the over-extended root yields the double extension of the restricted root system of $\g_0$ when this condition is realized, provided that in addition the previous affine step has yielded the affine extension of the restricted root system.

{}Finally, the multiplicity of $\lambda_{-1}$ is equal to one.  Indeed, if the root $\al = k \bt_{-1} + k' \bt_0 + \phi$, $\phi \in \Delta^\g$, projects  on $\lambda_{-1}$ then $k=1$, $k' = 0$ and $\pi(\phi)= 0$.  But there is no root of the form $\bt_{-1} + \phi$ with $\phi \in \Delta^\g$ since $\bt_{-1}$ has no link with the Dynkin diagram of the finite-dimensional algebra $\g$. The vector $\phi$ must therefore vanish and $\al = \bt_{-1}$.

\vspace{.2cm}

\noindent
{\bf Triple extension $\g^{+++}$:}
The very extended root does not vanish on $\mfa$ and so it belongs to  $\Gamma \backslash B_0$.  It is therefore also a white root. The set $B_0$ of black roots is again unchanged. Furthermore, $\theta(\bt_{-2}) = - \bt_{-2}$.  This implies that $\bt_{-2}$ belongs to a $1$-cycle of $P$ and is the only simple root that projects on $\pi(\bt_{-2}) \equiv \bt_{-2}$.   There is thus no arrow connecting  the very extended root to any other root. We denote $\pi(\bt_{-2}) \equiv \bt_{-2}$ by $\lambda_{-2}$.   

Because the projection $\lambda_{-2}$ of $\bt_{-2}$ coincides with $\bt_{-2}$, these two vectors have the same norm.  The restricted root $\lambda_{-2}$ connects only to the restricted root $\lambda_{-1}$, with a simple  link  (as between $\bt_{-2}$ and $\bt_{-1}$).  

{}Finally, the multiplicity of $\lambda_{-2}$ is also equal to one.

\vspace{.2cm}

The exact form of all the possible Cartan involutions and all reduced root systems are presented in appendix~\ref{app:TS}.

\section{Gravity and Iwasawa decomposition}

Consider gravity coupled to the scalar quotient model $G_0/K_0$ in three dimensions, where $G_0$ is the real Lie group associated with the real form $\g_0$ of $\g$ and $K_0$ its maximal compact subgroup.  It is well known that many  interesting gravitational theories reduce to such  models upon toroidal dimensional reduction to three dimensions.  
The theory admits a description in terms of billiards and billiard walls.  The analysis in three dimensions was performed in  \cite{Damour:2002fz} for the split case but proceeds similarly in the general case because the Iwasawa decomposition holds for general real forms, as we have recalled.  

Employing exactly the same methods as in \cite{Damour:2002fz},  as well as the Iwasawa decomposition for the finite-dimensional scalar model $G_0/K_0$, one finds  billiard walls of three types:
\begin{enumerate} 
\item A `dominant gravitational (symmetry) wall'  coming from the Einstein-Hilbert 
Lagrangian: this wall is non degenerate and corresponds to the very extended root.
\item `Dominant electric walls' coming from the scalar Lagrangian; these are in bijective correspondence (including multiplicities) with the simple restricted roots of the real form $\g_0$.
\item A `dominant magnetic wall' coming from the scalar Lagrangian; this wall corresponds 
to the affine root of the restricted root system  of the real form $\g_0$.  In particular, if the 
system is of $BC_n$ type, it is the highest root of the $BC_n$ system that appears, leading 
to the above-mentioned twisting \cite{Henneaux:2003kk}.  The dominant magnetic wall is degenerate as many times as the affine root.
\end{enumerate}

Thus, the billiard region is exactly determined by the simple restricted roots of the
real form  $\g^{++}_0$ constructed above.   The multiplicities also match.  This is
 the key fact which motivates consideration of these real forms.

The gravity line in three dimensions consists only of the over-extended root. If the restricted affine root is non degenerate and of the same length as the over-extended root (trivial projection), one can oxidize the theory to 4 dimensions.  Otherwise, there is an obstruction \cite{deBuyl:2003ub}. The affine root appears in four dimensions as a gravitational wall, and these are non degenerate.  We note that oxidation and obstructions to oxidation of three-dimensional cosets based on different real forms have been studied in detail in~\cite{CJP,Keurentjes,Keurentjes:2002rc}.

It is mainly the emergence of the restricted roots of $\g^{++}_0$ in the BKL limit~\cite{Damour:2000hv,Damour:2001sa}, together with the Iwasawa decomposition valid for infinite-dimen\-sional almost split Kac-Moody algebras that 
suggests the real forms $\g^{++}_0$ as hidden symmetries 
of the theory.

\section{Structure of the adjoint and fundamental representations of $\g^{+++}_0$}
\label{sec:levdec}

In this section, we present general arguments on the level decomposition~\cite{Damour:2002cu,Nicolai:2003fw,Kleinschmidt:2003mf} of real forms $\g^{+++}_0$ of triple extensions of finite-dimensional simple Lie algebras $\g$ as constructed in the previous section.

\begin{table}[t]
\centering
\begin{tabular}{c|c|c}
Longest Young tableau column & Number of tableaux & Algebra needed\\
\hline\hline\\[-4mm]
$D$ boxes & Infinite & $\g^{+++}$\\
$D-1$ boxes & Infinite & $\g^{++\phantom{+}}$\\
$D-2$ boxes & $\mathbb{Z}\times$finite & $\g^{+\phantom{++}}$\\
$<(D-2)$ boxes & finite & $\g^{\phantom{+++}}$
\end{tabular}
\caption{\label{tab:spectrum}\sl The generators contained in $\g^{+++}_0$ decomposed under a gravity ${\mathfrak{gl}}(D,\re)$ subalgebra.}
\end{table}

For the reader's convenience let us begin with an {\it aper\c{c}u} of the main results of 
this section, which are summarised in table~\ref{tab:spectrum}. It shows all the generators contained in any $\g^{+++}_0$ seen from a so-called `gravity subalgebra' ${\mathfrak{gl}}(D,\re)\subset \g^{+++}_0$ grouped into four different classes according to the extension process (\ref{eq:extend}). This extends results for split forms of~\cite{Riccioni:2006az}. The gravity subalgebra is such that any generator can be represented as a tensor of $GL(D,\re)$ in terms of a Young tableau. A characteristic feature of any tableau is the length of its longest column as this will have important consequences for its physical interpretation as we will show later. We will prove below that asking for the longest Young tableau column to be of a given length places the associated generator into one of the four classes displayed in table~\ref{tab:spectrum}. One notes that there is only a finite number of generators with longest column shorter than $D-2$ boxes.

\subsection{Fields in the adjoint of $\g^{+++}_0$}

\begin{figure}[t]
\centering
\begin{picture}(265,100)
\multiput(10,20)(50,0){3}{\circle{10}}
\multiput(15,20)(50,0){2}{\line(1,0){40}}
\multiput(115,20)(10,0){9}{\line(1,0){5}}
\multiput(205,20)(50,0){2}{\circle{10}}
\put(210,20){\line(1,0){40}}
\put(-7,5){$D-1$}
\put(42,5){$D-2$}
\put(92,5){$D-3$}
\put(202,5){$2$}
\put(252,5){$1$}
\multiput(105,80)(0,10){2}{\line(1,0){10}}
\multiput(105,80)(10,0){2}{\line(0,1){10}}
\multiput(150,80)(0,10){2}{\line(1,0){10}}
\multiput(150,80)(10,0){2}{\line(0,1){10}}
\multiput(200,80)(0,10){2}{\line(1,0){10}}
\multiput(200,80)(10,0){2}{\line(0,1){10}}
\multiput(250,40)(0,10){2}{\line(1,0){10}}
\multiput(250,40)(10,0){2}{\line(0,1){10}}
\put(110,80){\line(0,-1){55}}
\multiput(115,82)(0,6){2}{\line(1,0){35}}
\put(130,95){\line(1,-1){10}}
\put(130,75){\line(1,1){10}}
\put(155,80){\line(-2,-3){40}}
\multiput(155,80)(0,-10){6}{\line(0,-1){5}}
\multiput(160,85)(10,0){4}{\line(1,0){5}}
\multiput(155,80)(12,-16){4}{\line(3,-4){8}}
\multiput(252,40)(6,0){2}{\line(0,-1){16}}
\end{picture}
\caption{\label{fig:resrt}\sl The generic form of the diagram of the restricted root system of a real Lie algebra $\g^{+++}_0$. The nodes $D-1$, $D-2$,$\ldots$, $1$ represent a `gravity line' of type $A_{D-1}\cong {\mathfrak{sl}}(D,\re)$ with the (round) nodes $D-1$, $D-2$ and $D-3$ corresponding to the three extending nodes of the triple extension $\g^{+++}_0$. The square nodes represent the `level nodes'. The diagram is described in more detail in the text.}
\end{figure}
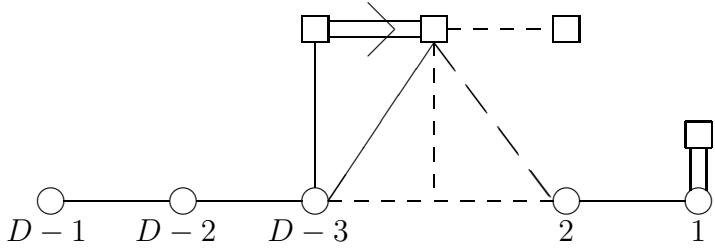

The diagram of the {\em restricted} root system is always of the form depicted in figure~\ref{fig:resrt}. The important property of {\em all} the restricted root systems that appear is that they contain gravity lines that originate at the triply extended end. A gravity line is an $A_{D-1}$ subalgebra of $\g^{+++}_0$. This means that the generalized Cartan matrix $A_{IJ}$ describing the restricted root system can be arranged in such a way that there is a sub-block of size $(D-1)\times(D-1)$ identical with the standard Cartan matrix of $A_{D-1}$. Since all the nodes of the restricted root system correspond to ad-diagonalisable generators (over $\re$)  this means that one has chosen an ${\mathfrak{sl}}(D,\re)$ subalgebra of $\g^{+++}_0$. By including an appropriate additional Cartan generator from the remainder of the real Lie algebra, this subalgebra gets enlarged to ${\mathfrak{gl}}(D,\re)$. `Level decomposition' means that one writes all generators of $\g^{+++}_0$ in terms of representations of this ${\mathfrak{gl}}(D,\re)$; all our subsequent results will only depend on the representation theory of ${\mathfrak{gl}}(D,\re)$.

Let us explain figure~\ref{fig:resrt} in some more detail. The round nodes depicted in the horizontal line correspond to the gravity line just discussed. The square nodes are the remaining nodes and can be attached in an (almost) arbitrary way to the gravity line and among themselves.\footnote{We even allow for the possibility that the restricted root system only admits a generalized Cartan matrix associated with a Borcherds algebra. As shown in the appendix this happens when starting from the compact real form of $\g$.} They are drawn as squares rather than circles only to indicate that they are not part of the chosen gravity line, otherwise there is no difference between the various nodes. In the way the diagram is drawn, the triple extension of $\g$ to $\g^{+++}$ is on the left. The three extending nodes are now called $D-3$, $D-2$ and $D-1$.\footnote{Note that we are using a different notation and labelling convention here compared to section~\ref{AlmostSplit}, as different features of $\g^{+++}$ are in the focus. To minimise confusion, the restricted roots will be called $\al$ here instead of $\lambda$ there.} By virtue of the extension process one has as only constraint that there are {\em no} square nodes connected to the triply extended node. The node $D-2$ also always belongs to the gravity line and, if $D>3$, there are no square nodes attached to it either. If $D=3$ there is single square node attached to the second node $D-2$ and this node arises from the projection of the affine extension. Otherwise there are no constraints on the number and types of connections between the various nodes.

The choice of a gravity line $A_{D-1}$ gives a natural split of the Cartan matrix $A_{IJ}$ of the restricted root system. We divide the indices $I$ into $I=(i,a)$, where $i=1,\ldots,D-1$ labels the (round) nodes of the gravity line and $a$ the (square) nodes outside the gravity line. We will also call these nodes `level nodes'. The components $A_{ij}$ of $A_{IJ}$ form the standard Cartan matrix of $A_{D-1}$. We are now interested in decomposing the adjoint representation of $\g^{+++}_0$ under the gravity subalgebra ${\mathfrak{gl}}(D,\re)$.

The generators of $\g^{+++}_0$ lie in root spaces of the restricted root system. Let
\begin{align}
\label{genalpha}
\al = \sum_i m^i \al_i + \sum_a \ell^a \al_a
\end{align}
be the expansion of any restricted root $\al$ on the basis of simple roots $\al_I$ of the restricted root system. 
Clearly, we have the usual root space decomposition of the adjoint of $\g^{+++}_0$
\begin{align}
\g^{+++}_0 = \bigoplus_{\al} \g^{+++}_{\al}=\bigoplus_{(m^i,\ell^a)} \g^{+++}_{(m^i,\ell^a)},
\end{align}
where we have included the Cartan generators as $\al=0$. The root spaces $\g^{+++}_\al$ can be of very high dimension, both (i) because there are multiplicities associated with restricted root systems and (ii) because $\g^{+++}_0$ is a (real) Kac--Moody algebra. For the level decomposition, we are interested in a slightly `coarser' description
\begin{align}
\g^{+++}_0 = \bigoplus_{(\ell^a)} \g^{+++}_{(\ell^a)},
\end{align} 
where $(\ell^a)$ now only ranges over the level components of $\al$ and hence we have carried out the sum over the $m^i$. For each fixed level $(\ell^a)$, there is only a finite number of generators of $\g^{+++}_0$ and they can be grouped into representations of ${\mathfrak{gl}}(D,\re)$. 

Which representations arise is described most easily using the Dynkin labels that can be translated directly into Young tableaux of $A_{D-1}$. The $A_{D-1}$ Dynkin labels are obtained by converting $\al$ of (\ref{genalpha}) with the Cartan matrix. For this purpose we re-write the root $\al$ in (\ref{genalpha}) in the basis of fundamental weights $\Lambda^I$ dual to the simple restricted roots via $(\Lambda^I | \alpha_J)=\delta^I_J$. In the resulting expression 
for $\al$ as a weight
\begin{equation}
\al = \sum_I p_I \Lambda^I = \sum_j p_j \Lambda^j + \cdots
\end{equation}
we focus on the weights $\Lambda^j$ associated with the gravity algebra $A_{D-1}$, 
and then project out the Dynkin labels $p_i$ by multiplication with the relevant simple 
root of $A_{D-1}$.\footnote{Note that there is an extra overall minus sign in this relation. This is customary for such level decompositions as one is effectively describing lowest weight representations of $A_{D-1}$ rather than highest weight representations.} 
\begin{align}
\label{dynklab}
p_i = -\sum_{j} A_{ij}m^j - \sum_a A_{ia} \ell^a.
\end{align}
The representation associated with a set of $p_i\geq 0$ is a tableau with $p_{D-1}$ columns consisting of $D-1$ boxes, $p_{D-2}$ columns with $D-2$ boxes and so on. As we are dealing with ${\mathfrak{gl}}(D,\re)$ rather than ${\mathfrak{sl}}(D,\re)$ we also have to keep 
track of the overall weight of a representation. This weight is here represented by the
number of columns with $D$ boxes.\footnote{Such columns do not occur 
for Young tableaux of the {\em special} linear group $SL(D,\re)$ which can have 
only columns of at most $D-1$ boxes.}

We can also count the number of boxes from knowing which level $(\ell^a)$ we are at. Indeed, if a square level node $a$ is attached to node $i$ of the gravity line, it will give rise to a multiple of $i$ boxes. The multiple is given by the way it is connected (i.e. the value of $-A_{ia}$) and $\ell^a$ itself. In all, we arrive at two different ways of counting the number of boxes of a given ${\mathfrak{gl}}(D,\re)$ representation:
\begin{align}
qD + \sum_i i p_i = -\sum_{i,a} i A_{ia}\ell^a .
\end{align}
Here, $q$ denotes the number of columns with $D$ boxes. Multiplying (\ref{dynklab}) by $i$ and then summing on $i$ we deduce therefore
\begin{align}
q D = \sum_{i,j} i A_{ij}m^j = D m^{D-1},
\end{align}
where we have used the explicit form of the $A_{D-1}$ Cartan matrix to simplify the telescoping sum. Therefore we see that no columns of $D$ boxes occur if and only if $m^{D-1}=0$, i.e., the root belongs to $\g^{++}_0\subset \g^{+++}_0$.

Let us continue this to columns of $D-1$ boxes. From (\ref{dynklab}) one deduces immediately
\begin{align}
p_{D-1} = -2m^{D-1}+m^{D-2}
\end{align}
as there are no level nodes attached to the node $D-1$. This equation implies that there are no columns of $D$ boxes and no columns of $D-1$ boxes if and only if $m^{D-1}=m^{D-2}=0$.

In the next step one has to distinguish between $D=3$ and $D>3$. For $D>3$ one obtains
\begin{align}
\label{affbox}
p_{D-2} = m^{D-1}-2m^{D-2}+m^{D-3}.
\end{align}
The number of columns of $D-2$ boxes is therefore equal to $m^{D-3}$ if one demands that there are no columns with $D$ nor $D-1$ boxes. But $m^{D-3}$ is the number of times the (projected) affine root appears in an element of $\g^+_0$. However, the structure of $\g^{+}_0$ is well-known; it consists only of repetitions of the generators of $\g_0$ as shown in (\ref{affalg}). Therefore, any tableau for which the longest column has $D-2$ boxes will contain a tableau that is contained solely in $\g_0$. More precisely, for an element $x\otimes t^n\in\g^+_0$ the number of columns with $D-2$ boxes is equal to the affine level $n$  and then the remaining part of the diagram is given by that of $x\in\g_0$. Demanding also the absence of columns with $D-2$ boxes one is forced to consider only elements of the finite-dimensional $\g_0$. As there are only finitely many such elements, we conclude that almost all tableaux have at least columns with $D-2$ boxes. This will be important below when we analyse the number of degrees of freedom associated with the various ${\mathfrak{gl}}(D,\re)$ representations contained in $\g^{+++}_0$.

The case $D=3$ appears only if the affine root $\al_{D-3}$ is projected onto a root of length different from those of the other extending roots $\al_{D-1}$ and $\al_{D-2}$. (This happens for example in the case ${\mathfrak{su}}^*(n+1)^{+++}$.) Since the projected affine root is always shorter than $\al_{D-1}$ and $\al_{D-2}$, we have $A_{D-2,D-3}=-1$ and equation (\ref{affbox}) gets replaced by
\begin{align}
p_{D-2}=m^{D-1}-2m^{D-2}+\ell^{D-3},
\end{align}
where we have now written $\ell^{D-3}$ since the node $D-3$ is now a level node. But we see that the same argument as above still applies. Therefore the structure of $\g^{+++}_0$ is the same in both cases and we summarize the result in table~\ref{tab:spectrum}.

\subsection{Structure of the fundamental $\g^{+++}_0$ representation}

We now consider a specific lowest weight representation of $\g^{+++}_0$. This is the one with lowest weight equal to (minus) the fundamental weight corresponding to node $D-1$ of diagram~\ref{fig:resrt}. The representation has occurred in~\cite{West:2003fc,West1} and has been named $l_1$ representation there due to different labelling conventions. We will call it the fundamental or $L(\Lambda_{D-1})$ representation.  

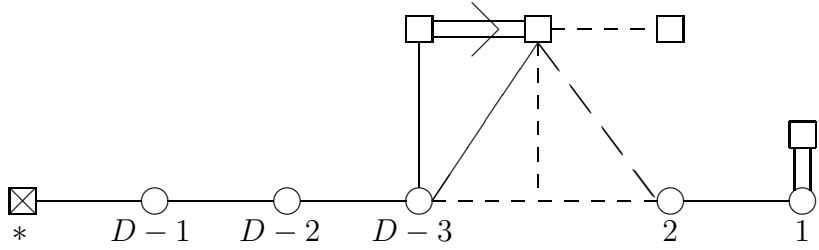
\begin{figure}
\centering
\begin{picture}(265,100)
\put(-35,15){\line(-1,1){10}}
\put(-45,15){\line(1,1){10}}
\multiput(-45,15)(0,10){2}{\line(1,0){10}}
\multiput(-45,15)(10,0){2}{\line(0,1){10}}
\put(-35,20){\line(1,0){40}}
\put(-44,5){$*$}
\multiput(10,20)(50,0){3}{\circle{10}}
\multiput(15,20)(50,0){2}{\line(1,0){40}}
\multiput(115,20)(10,0){9}{\line(1,0){5}}
\multiput(205,20)(50,0){2}{\circle{10}}
\put(210,20){\line(1,0){40}}
\put(-7,5){$D-1$}
\put(42,5){$D-2$}
\put(92,5){$D-3$}
\put(202,5){$2$}
\put(252,5){$1$}
\multiput(105,80)(0,10){2}{\line(1,0){10}}
\multiput(105,80)(10,0){2}{\line(0,1){10}}
\multiput(150,80)(0,10){2}{\line(1,0){10}}
\multiput(150,80)(10,0){2}{\line(0,1){10}}
\multiput(200,80)(0,10){2}{\line(1,0){10}}
\multiput(200,80)(10,0){2}{\line(0,1){10}}
\multiput(250,40)(0,10){2}{\line(1,0){10}}
\multiput(250,40)(10,0){2}{\line(0,1){10}}
\put(110,80){\line(0,-1){55}}
\multiput(115,82)(0,6){2}{\line(1,0){35}}
\put(130,95){\line(1,-1){10}}
\put(130,75){\line(1,1){10}}
\put(155,80){\line(-2,-3){40}}
\multiput(155,80)(0,-10){6}{\line(0,-1){5}}
\multiput(160,85)(10,0){4}{\line(1,0){5}}
\multiput(155,80)(12,-16){4}{\line(3,-4){8}}
\multiput(252,40)(6,0){2}{\line(0,-1){16}}
\end{picture}
\caption{\label{fig:l1}\sl The diagram relevant for analysing the $L(\Lambda_{D-1})$ representation of $\g^{+++}_0$. Compared to figure~\ref{fig:resrt}, a new node $*$ has been added on the left, here drawn as a crossed square.}
\end{figure}

In order to analyse it, we employ the diagram technique of~\cite{KMW,Kleinschmidt:2003pt,West1} by drawing an extended diagram whose associated algebra is then analysed. More precisely, we consider the diagram by adjoining yet another node, called $*$,  to figure~\ref{fig:resrt} on the left. This is shown in figure~\ref{fig:l1}. The thus quadruply extended algebra will be denoted $\g^{++++}_0$ and its (restricted) roots decompose as
\begin{align}
\hat{\al} = \al + \ell^* \al_* = \sum_i m^i\al_i +\sum_a \ell^a \al_a +\ell^*\al_*,
\end{align}
where $\al$ is a (restricted) root of $\g^{+++}_0$. If $\ell^*=0$, one is describing the adjoint representation of $\g^{+++}_0$, if $\ell^*=1$, one recovers the $L(\Lambda_{D-1})$ representation. (Other values of $\ell^*$ do not matter here.)

Consider now a representation of ${\mathfrak{gl}}(D,\re)$ occurring in the adjoint of $\g^{+++}_0$, \ie$\ell^*=0$. We can represent it by Dynkin labels $p_i$ of $A_{D-1}$. But in fact now the gravity line can also be thought of as one node longer by including the node $*$ and we can wonder whether there is an associated representation in $A_D$. Let the $p_i$ belong to some root $\al$ of $\g^{+++}_0$, indicating the lowest element in the representation. This we can also we view trivially as a root $\hat{\al}$ of $\g^{++++}_0$ as $\ell^*=0$. The Dynkin labels under $A_D$ are then
\begin{align}
[p_*; p_i]_{A_D},\quad\text{where $p_* = m^{D-1}$.}
\end{align}
Since $m^{D-1}\geq 0$ for positive roots, we also obtain a representation of $A_{D}$ from any such representation of $A_{D-1}$. We have indicated $A_D$ on the Dynkin labels in order to avoid confusion.

Now assume that there is a column with $i<D$ boxes in the above representation of $A_D$, i.e., $p_i>0$. Then $A_D$ representation theory tells us that we can act on the lowest element by the corresponding $i$-th raising generator. This will have the following effect on the root $\hat{\al}$ at $\ell^*=0$ and the Dynkin labels:
\begin{align}\label{orgrep}
&\hat{\al}&&\Leftrightarrow&& [p_*; p_{D-1}, \ldots, p_{i+1}, p_i , p_{i-1}, \ldots ,p_1]_{A_D}&\\
\text{(raise in $i$)}&&&\downarrow&&& \nonumber\\
&\hat{\al}+ \al_i &&\Leftrightarrow&& [p_*;p_{D-1},\ldots, p_{i+1}+1, p_i-2, p_{i-1}+1,\ldots,p_1]_{A_D}&
\end{align}
In fact, we can now continue by applying the raising operators $i+1$, $i+2$ and so forth and obtain
\begin{align}
\hat{\al}+\sum_{k=i}^{D-1}\al_k&&\Leftrightarrow&& [p_*+1; p_{D-1}-1, \ldots, p_{i+1}, p_i-1 , p_{i-1}+1, \ldots ,p_1]_{A_D}&.
\end{align}
Now, we can finally also apply the $*$ raising operator and get
\begin{align}
\hat{\al}+\sum_{k=i}^{D-1}\al_k + \al_*&&\Leftrightarrow&& [p_*-1; p_{D-1}, \ldots, p_{i+1}, p_i-1 , p_{i-1}+1, \ldots ,p_1]_{A_D}&.
\end{align}
Since we added $\al_*$ we are now no longer in the adjoint of $\g^{+++}_0$ but in fact in the $L(\Lambda_{D-1})$ representation. Restricting the above to the old gravity line $A_{D-1}$ in terms of which we are decomposing $L(\Lambda_{D-1})$ we find the Dynkin labels
\begin{align}
[p_{D-1}, \ldots, p_{i+1}, p_i-1 , p_{i-1}+1, \ldots ,p_1]_{A_{D-1}}.
\end{align}
This representation is very similar to the original (\ref{orgrep}) except for that the number of columns with $i$ boxes has been decreased by one and the number of columns with $i-1$ boxes has been increased by one.

In other words, given a ${\mathfrak{gl}}(D,\re)$ representation in the adjoint of $\g^{+++}_0$, there is an associated representation in $L(\Lambda_{D-1})$ where any chosen column has been made shorter by one box. Therefore, the $L(\Lambda_{D-1})$ representation contains at least all tableaux of the adjoint with a single box removed in all possible ways. We say ``at least'' here since the (outer) multiplicity of the thence obtained tableaux is always greater or equal to that of the adjoint by the above argument. That there can be additional tableaux (or enhanced multiplicities) can be seen by studying the tables of~\cite{West1}.

All possible ways of removing a single box from a given representation of ${\mathfrak{gl}}(D,\re)$ is exactly the way gauge parameters of general mixed symmetry potentials are constructed in the standard formulation. This fact, together with the results of table~\ref{tab:spectrum} will be important now when analysing the degrees of freedom.

\section{Covariant free field Lagrangians for $\g^{+++}$}
\label{sec:lag}

If a space-time covariant Lagrangian exists which involves all the fields of the 
adjoint representation of $\g^{+++}$ it is natural to require that in the free limit, 
this Lagrangian will reduce to a sum of free Lagrangians describing these fields 
separately.  All these fields are massless. The corresponding Lagrangians for 
massless tensor fields with arbitrary Young symmetry in flat space were already written explicitly 
long ago by Curtright in \cite{Curtright:1980yk} for the case of two columns, 
who also gave indications on the general case, treated in detail later by 
Labastida \cite{Labastida} and Bekaert/Boulanger~\cite{Bekaert:2002dt,Bekaert:2003az,Bekaert:2006ix}. Such theories might arise in the zero slope limit of
string theory where one expects all degrees of freedom to be described by
massless fields, and this has been one of the main motivations behind recent studies
of higher spin gauge theories~\cite{Fradkin:1987ks,Francia:2002aa,Campoleoni:2008jq,Bekaert:2010hw,Boulanger:2011dd}. In this section we examine
the prospects for realizing this idea within the framework of $\g^{+++}$ duality
symmetry.

The crucial requirement guiding the construction of the free, flat space Lagrangians is that 
there should be just enough gauge freedom to go to the light cone gauge, where 
the physical degrees of freedom are transverse and  described by $SO(D-2)$ representations 
{\em characterized by the same Young tableau} as the corresponding $GL(D,\re)$
tableau in the covariant formulation.  Namely, the physical degrees of freedom then 
are purely transverse and hence described by a tensor in $D-2$ dimensions.  This 
tensor has  the same symmetry properties on the indices as its covariant ancestor 
field but in addition, all its traces are zero (as appropriate for an $SO(n)$ Young
tableau). As suggested in \cite{Curtright:1980yk,Labastida} and proved in~\cite{Bekaert:2006ix}
this procedure ensures unitarity and absence of ghosts, and can be achieved 
by introducing, for each field represented by a given $GL(D,\re)$ Young  tableau, a 
collection of gauge parameters corresponding to all those tableaux obtained 
by removing one box from the original tableau in all possible ways. 

The construction is most easily understood in a pictorial way and  in terms of a simple 
example (see \cite{Labastida} for the technical details). For instance, if the 
gauge field is given by  the $GL(D,\re)$ Young tableau in the figure below
\begin{equation}\label{GFYoung}
\yng(4,2,1)
\end{equation}
the associated gauge parameters are represented by the tableaux
\begin{equation}
\yng(3,2,1) \quad 
\quad
\yng(4,1,1) \quad 
\quad
\yng(4,2)
\end{equation}
The variation of the gauge field is then obtained by acting with a derivative operator 
$\partial$ on any of these Young tableaux and then applying the Young projector
corresponding to (\ref{GFYoung}) to obtain the tableau representing the gauge 
field itself. A new feature in  comparison with ordinary gauge theories is that 
{\em the gauge invariant field strength 
is of higher order in the derivatives, as it must involve as many derivatives  as there 
are columns in the Young tableau}. The necessity of higher derivatives
is easily seen as follows: when varying the gauge field strength corresponding 
to a given Young tableau with respect to any of the associated gauge parameters 
there must occur (at least) two derivatives in one of the columns so the variation of
the field strength vanishes by antisymmetry. The best known example of this is, of course, gravity: the graviton being a symmetric tensor we have two columns (of one box each), 
hence the gauge invariant field strength (Riemann tensor) involves two derivatives.  For the 
above example the gauge invariant field strength would thus be of fourth order in the 
derivatives, with associated tableau
\begin{equation}
\partial^4 \; \;\yng(4,2,1) \quad \sim \quad \young(\hfil\hfil\hfil\hfil,\hfil\hfil\partial\partial,\hfil\partial,\partial)
\end{equation}
where a derivative operator $\partial$ is associated with each extra box,
and the action of the corresponding Young projector required for the right
hand side is understood. 

In a space-time covariant formulation we will thus have to allow for derivatives of
arbitrarily high order in the free part of the Lagrangian if the resulting theory
is to be unitary and free of ghosts. The equations of motion are then 
expressed in terms of the gauge invariant field strength~\cite{Labastida,Bekaert:2002dt}.
Let us point out here that the Cartan form which usually serves as the basis for 
the non-linear realization of the $\g^{+++}$ symmetry involves only first order derivatives,  
so it is unclear how one would implement this in terms of  a single $\g^{+++}$ covariant Lagrangian or $\g^{+++}$ covariant field equations even at the free 
field level. To be sure, higher derivatives can possibly be avoided by 
imposing trace conditions on the gauge fields and/or gauge 
parameters \cite{Fronsdal:1978rb}. However, 
within the framework of a $\g^{+++}$ covariant  theory this avenue is unavailable 
because the background metric $\eta^{\mu\nu}$ is not an invariant tensor of $\g^{+++}$, 
hence there is no way of imposing trace conditions without breaking this symmetry.
Besides, the representations corresponding to the traces that one would like
to eliminate do not even occur in $\g^{+++}$. There is a way to avoid trace conditions
\cite{Francia:2002aa,Campoleoni:2008jq} but the corresponding Lagrangians are non-local. For these reasons,
we will not pursue this option here.

A main point now is that the so-called $L(\Lambda_{D-1})$ (fundamental) representation
 of $\g^{+++}$ (whose relevance in this context was already emphasized in~\cite{West:2003fc,Riccioni:2009hi} and called `$l_1$-representation' there) combines all
the gauge transformations. Namely, as we showed in the preceding section, for any field occurring 
in the adjoint representation of $\g^{+++}$ the $L(\Lambda_{D-1})$ representation of $\g^{+++}$ 
contains all the requisite gauge parameter representations, and with the correct 
multiplicities, as required by the formalism of \cite{Curtright:1980yk,Labastida} 
(in addition $L(\Lambda_{D-1})$ contains many other  representations). Consequently, 
if there is a covariant field theoretic realization of $\g^{+++}$,  the gauge transformations 
will automatically include the ones known from higher spin gauge theories, as 
outlined above. In particular, at the free field level, the analysis for each individual
Young tableau reduces to that of  \cite{Curtright:1980yk,Labastida}. Assuming that
there is such a covariant formulation, we now show that the higher level fields in this 
formalism carry no propagating physical degrees of freedom. In other words, at least 
at the free field level, the formalism of  \cite{Curtright:1980yk,Labastida} implies that, 
out of the infinite tower of fields occurring in the  adjoint representation of $\g^{+++}$  
only the (finitely many)  fields associated with the finite dimensional subalgebra $\g$ 
can possibly correspond to true physical degrees of freedom. This includes the so-called `dual graviton'.
   
In the foregoing section we analyzed the field content of the adjoint of $\g^{+++}$. This analysis showed that one can  distinguish between two types of 
higher level representations, corresponding to the `gradient' and 'non-gradient' representations of~\cite{Damour:2002cu,KN}, depending on whether the longest Young tableau column has $\leq D-2$ boxes or more. For each representation contained in $\g_0$ there is an infinite tower of tableaux obtained 
by attaching any number of columns with  $D-2$ boxes. In addition to these infinite 
towers, there will be a (vastly larger) set  of representations with any number 
of columns  of $D-1$ and $D$ boxes attached to some $GL(D,\re)$ Young tableau.

The `gradient' representations in the $\g^{++}$ formulation of~\cite{Damour:2002cu,KN,KNP} correspond
to the towers of tableaux with any number of columns of $D-2$ boxes attached to a 
basic $\g$ tableau. These representations are associated to the affine subroots 
of $\g^+\subset \g^{+++}$, in such a way that the number of columns with $D-2$
boxes is equal to the affine level of the $\g^+$ algebra.
Because $D-2$ form fields in $D$ dimensions are dual 
to scalar fields, one might expect these higher level fields to be `dual' to the original 
field in the $\g$ representation in the same sense that a (massless) scalar in 
$D$ dimensions can be equivalently described by a $D-2$ form. However, this turns 
out not to be the case.  Namely, as representations of $SO(D-2)$ such tableaux
make no sense (while they do as $GL(D,\re)$ tableaux) because such columns can be
replaced by Levi-Civita $\varepsilon$  symbols and can thus be `peeled off'. However, 
for $SO(D-2)$ there is now the extra condition of vanishing trace, and this in fact 
eliminates the representation altogether, implying that the corresponding fields must 
be set to zero. Consequently, these fields carry no propagating degrees of freedom. 

Similarly, it is clear that fields with a column of $D$ boxes simply vanish in $D-2$ dimensions, 
so the corresponding $\g^{+++}$ fields are altogether absent in the light cone gauge 
and thus carry no local degrees of freedom (nevertheless, such fields are thought to 
play a role in connection with `space filling branes', to which they couple). The same 
holds true for diagrams with columns of $D-1$ boxes. Recall that in the covariant formulation 
a $(D-1)$-form field is worth a constant, and may thus correspond to a `deformation' of 
the original theory by a constant parameter (such as a gauge coupling constant,
a cosmological constant, or a Romans mass parameter). This is, however, no longer 
the case when the column of $D-1$ boxes comes attached with a non-trivial Young tableau:
in this case a constant vacuum expectation value would break Lorentz invariance. 
Hence the higher level fields with columns of $D-1$ boxes do not even carry constant 
degrees of freedom.

Let us emphasize that the on-shell vanishing of the higher level dual fields 
with columns of $D-2$ boxes is not at  all what one would expect for a 
chain of dual fields. Instead one would expect 
the polarizations of the higher level dual fields to be expressible as (non-vanishing)
functions of the lowest level field. 
This is well known 
from the duality between electric and magnetic vector fields in Maxwell theory. An 
example involving the affine subgroup $G^+$ is provided by the infinite hierarchy of dual gravitational potentials emerging under the action of the Geroch group in stationary 
axisymmetric  solutions of Einstein's equations: the dual potentials are non-linear and 
non-local functionals of the level 0 fields, but do not vanish.\footnote{In fact,
   the {\em non-linearity} of the affine duality transformations is an essential
   part of the Geroch group action, and possibly also an essential feature that
   is lost in any linearized analysis. This possible caveat of our analysis
   should always be kept in mind.}  On the other 
hand, if the theory is to be invariant under $G^{+++}$ gauge transformations, its
affine subgroup $G^+$ is surely realized as a subsymmetry that is preserved under
dimensional reduction to $D=2$. We arrive at the conclusion that the (putative)
field theoretic realizations of the dual degrees of freedom in the framework of an
$G^{+++}$ symmetry is necessarily different from the one expected from ordinary
electromagnetic duality, or the known realization of affine symmetries in 
axisymmetric reductions of gravity and supergravity.

We thus conclude that -- even independently of the existence and consistency of
self-interactions -- the only fields carrying local dynamics in the putative 
space-time covariant field theoretic realization of the $G^{+++}$ symmetry are those 
associated with the roots of the original finite dimensional duality symmetry $\g$, which are  
finite in number (see also~\cite{Riccioni:2009hi} where similar conclusions have been reached\footnote{A crucial difference between our results and~\cite{Riccioni:2009hi} is, however, that there the dual graviton was argued not to be needed and that one could eliminate it completely by means of the so-called inverse Higgs effect. In our analysis, the dual graviton does 
carry degrees of freedom and is hence a dynamical field.}).  In terms of table~\ref{tab:spectrum} of section~\ref{sec:levdec} our results can be summarised by saying that only the finitely many fields associated with the last line of the table correspond to propagating degrees of freedom. All the infinitely many fields of the first three lines do not.

Furthermore, the $\g^{+++}$ theory {\em contains no fields of higher spin $s>2$} 
because there are no higher spin diagrams among those of the finite 
dimensional subsymmetry $\g$. This indicates that the $\g^{+++}$ theory is
neither a higher spin theory in the sense of Vasiliev nor a naive zero slope
limit of string field theory.

There are various options to evade these conclusions, although none of them
appears particularly compelling to us. One would be to look for a gauge covariant formulation 
with less gauge invariance (for instance, by dropping some of the gauge parameter
Young tableaux), in which case there might survive propagating degrees of freedom
in the light cone gauge for the higher level fields. However, we cannot see any systematic 
and consistent procedure for eliminating a subset of the gauge parameters, and we are 
also not aware of a single working example of this type in higher spin gauge theory.
Another option is to enlarge the adjoint representation of $\g^{+++}$ by more fields 
(such as the extra towers of fields labelled `${\rm Og\,} n$' in~\cite{Riccioni:2009hi}). However, in this case, the 
$\g^{+++}$ symmetry would only be a tiny subsymmetry of a vastly larger structure 
(assuming a manageable algebraically closed structure containing $\g^{+++}$ exists). Another possibility might be to weaken the dynamical equations but, again, we are not aware of any consistent and interesting system constructed in this way.
We finally note that, irrespective of their role as physical degrees of freedom, the mixed 
symmetry tensors occurring in $E_{10}$ and $E_{11}$ have been useful for classifying the potentials appearing in the p-form hierarchies in various dimensions
\cite{Riccioni:2007au,Bergshoeff:2007qi,deWit:2008ta} and supersymmetric solutions
\cite{Englert:2003py,Englert:2007qb,West:2004st,Cook:2004er,Bergshoeff:2011zk,Bergshoeff:2011ee,Kleinschmidt:2011vu,Bergshoeff:2011se,Cook:2011ir}.

\section{Conclusions}
In summary, it appears that none of the `obvious' possibilities for dealing with the higher
level states in a way that preserves space-time covariance is viable. Of course, there are assumptions in our reasoning that can be weakened. For example,  allowing for different 
gauge transformations of the mixed symmetry fields one can turn them into propagating 
fields. As an example we mention the possibility of decomposing a mixed tableau into 
traces and tracefree parts and eliminating the tracefree parts by appropriate shift 
symmetries. The resulting traces then will be propagating. However, this construction 
is not well-motivated by the $\g^{+++}$ symmetries where one does not have a trace operator.

We are thus left with the seeming paradox that none of the huge extra structures 
introduced by enlarging the finite dimensional duality symmetry $\g$ to its very 
extended version $\g^{+++}$ leads to any `added value' in terms of extra physical 
degrees of freedom, at least not in the framework of conventional higher spin gauge
theories. Furthermore, this structure does not appear to link up with 
some currently popular ideas on M-theoretic extensions of string theory, such as the 
higher spin gauge theories of \cite{Fradkin:1987ks} or string field theory in 
the zero tension limit. We see here an interesting and remarkable analogy with 
supermembrane theory, or equivalently,  M(atrix) theory (see \cite{HN}).
There as well, only the lowest states (representations) admit a  particle-like 
(field theoretic) interpretation as propagating one-particle degrees of freedom: 
only the massless supermultiplet corresponds to a discrete eigenvalue in the 
spectrum, whereas the excitations in the continuous spectrum cannot be 
associated to states of a definite particle number. Similarly, in the present set-up, 
the higher rank representations do not appear to correspond to propagating 
degrees of  freedom in a field theoretic realization, but require a more
sophisticated interpretation.

\vspace{.2cm}

\noindent {\bf Acknowledgements.} 
We would like to thank A.~Campoleoni and D.~Francia for useful discussions on higher spin theory. We also thank N.~Boulanger and the referee for useful comments on the first version of this paper.  M. H. gratefully acknowledges support from the Alexander von Humboldt Foundation through a Humboldt Research Award and support from the ERC through the ``SyDuGraM" Advanced Grant. The work of M. H. is also partially supported by IISN
- Belgium (conventions 4.4511.06 and 4.4514.08), by the Belgian Federal Science Policy Office through the Interuniversity Attraction Pole P6/11 and by the ``Communaut\'e Fran\c{c}aise de Belgique" through the ARC program.

\appendix

\section{Background on real forms}
\label{app:real}

The purpose of this appendix is to recall the general facts about real forms of (symmetrizable) Kac-Moody algebras that are necessary to grasp the results given in the paper.  No proofs are reproduced here.  For more information, we refer the reader to \cite{Cartan,Araki,Helgason,Knapp,Henneaux:2007ej} for the finite-dimensional case and to \cite{KacWang,BackValente,BenMessaoud} for the Kac-Moody general case.

\subsection{Definitions}

Let $\g$ be a complex Lie algebra. If $\g$ (viewed as a real vector space of double dimension) can be written as
\begin{equation}
\g = \g_0 \oplus i \g_0
\end{equation}
where $\g_0$ is a real Lie algebra, one says that $\g_0$ is a {\em real form} of $\g$.  
Conversely, we re-obtain $\g$ by complexifying $\g_0$
\begin{equation}
\g = \g_0 \otimes_\re \ce. 
\end{equation}
Every real form $\g_0$ of $\g$ determines an associated conjugation $\sigma'$ through
\begin{equation}
\sigma' (x+i y) := x - i y, \; \; \; x,y \in \g_0.
\end{equation}
$\sigma'$ is easily verified to be antilinear (i.e., conjugate-linear), to preserve the bracket structure and to square to the identity.  [Antilinear transformations will systematically denoted with a prime.] The real form $\g_0$, viewed as a subalgebra of $\g$, is the subalgebra of fixed points of $\sigma'$,
\begin{equation}
\sigma'(x) = x \; \; \; \Leftrightarrow \; \; \; x \in \g_0.
\end{equation}
Therefore, the problem of determining all real forms $\g_0$ of a given complex Lie algebra 
$\g$ is equivalent to determining all its conjugations.

\subsection{Conjugations and involutions of $\g$}

There are two standard real forms of $\g$ with associated standard conjugations, namely 
the {\em split real form}  and the {\em compact real form}. For instance, for $\sgl(n,\ce)$ 
these are $\sgl(n,\re)$ and $\su(n)$, respectively. To define the standard conjugations 
more generally we employ the 
Chevalley-Serre presentation.  Accordingly, let $\g$ be a complex Lie algebra with 
symmetrizable Cartan matrix $A_{ij}$ and Chevalley-Serre generators $\{h_i,e_i,f_i\}$. 
At this point $\g$ could be finite-dimensional or infinite-dimensional Kac-Moody; in  the latter 
case there are some additional notions that will be explained in section~\ref{app:KM}.
The {\em Chevalley-Cartan involution} $\omega$ of $\g$ is the linear involutive automorphism 
defined by
\begin{equation}
\omega(h_i) = -h_i\;,\quad \omega(e_i) = -f_i\; , \quad \omega(f_i) = -e_i. 
\end{equation}
and extends to the whole algebra $\g$ by the invariance property
$\omega([x,y]) = [\omega(x), \omega (y)]$.
The real form $\mfs_0$ of $\g$ obtained by taking only real combinations of the 
Chevalley-Serre generators and their (multi-)commutators 
is called the `standard split' (or `standard maximally 
non-compact') form of $\g$. The corresponding conjugation $\sigma'$ is just 
complex conjugation (in that basis) and is denoted by $\sigma'\equiv\sigma'_s$.  
The real form $\mfu_0$ of $\g$ defined by the conjugation $\tau'= \sigma'_s \omega = \omega \sigma'_s$ is called the {\em standard compact form}. It is generated over the reals by $i h_i$, $i(e_i + f_i)$ and $e_i- f_i$. Note that the Killing form on $\mfu_0$ is negative definite only in the 
finite-dimensional case.  Nevertheless, the terminology `compact'  is used.
One has clearly $\g = \mfs_0 \otimes \ce = \mfu_0 \otimes \ce$.

Note that the standard split form and the standard compact form are `aligned', in the 
sense that the corresponding conjugations commute, $\sigma'_s \tau' = \tau' \sigma'_s$. 
It is in fact a general result that by conjugation, one can align {\em any} real form  with 
the standard compact form $\mfu_0$, and this will be assumed in the sequel. That is, 
for any real form $\g_0$, we shall assume
\begin{equation}
\label{eq:cartinv}
\sigma' \tau' = \tau' \sigma'
\end{equation}
for the associated conjugation $\sigma'$.  This relation implies
\begin{eqnarray}
\sigma'(\mfu_0) & \subset& \mfu_0, \\
\tau'(\g_0) &\subset & \g_0.
\end{eqnarray}
The {\em Cartan involution} associated with a given real form $\g_0$ of $\g$ is then 
the complex linear involutive automorphism
\begin{equation}
\theta := \sigma' \tau' = \tau' \sigma'
\end{equation}
defined by (\ref{eq:cartinv}). So for the standard split form $\mfs_0$ we have 
$\theta = \omega$, while for the standard compact form (for which $\sigma'=\tau'$)
$\theta$ is the identity. While $\omega$ and $\tau'$ are always the same for a given $\g$,
we thus see that $\theta$ and $\sigma'$ are different for different real forms.

Let us illustrate these abstract definitions with the simple example of $\g=\sgl(n,\ce)$,
that is, the algebra of complex traceless $n$-by-$n$ matrices $M$. In this case,
the various (anti-)involutions are given by
\begin{equation}
\omega(M) = - M^T\; , \quad \sigma'_s(M) = \bar{M} \;, \quad
\tau'(M) = \sigma'_s\omega(M) = - M^\dagger
\end{equation}
so, in particular, $\sigma'_s$ acts by standard complex element-wise matrix conjugation.
As the Cartan subalgebra one may take the diagonal traceless $n$-by-$n$ matrices;
the Chevalley generators are given by $e_i \equiv E_{i \, i+1}$,  $f_i \equiv E_{i +1\, i}$ 
($i = 1, \ldots, n-1$) with $(E_{i \, j})_{kl} = \delta_{ik} \delta_{jl}$, that is,  the $n$ by $n$ matrix with a $1$ in position $(i,j)$ and $0$'s elsewhere, the standard split form is $\sgl(n, \re)$ and the standard compact form is $\su(n)$.  

The Cartan involution $\theta$ leaves $\g_0$ invariant and leads to the Cartan decomposition
\begin{equation}
\g_0 = \mfk_0 \oplus \mfp_0
\end{equation}
where $\mfk_0$ and $\mfp_0$ are the $\theta$-eigenspaces of eigenvalues $+1$ and $-1$, respectively, corresponding to the compact and non-compact generators.
It is easy to check that 
\begin{equation}
[\mfk_0, \mfk_0 ] \subset \mfk_0, \; \; \; \; 
[\mfk_0, \mfp_0 ] \subset \mfp_0, \; \; \; \; 
[\mfp_0, \mfp_0 ] \subset \mfk_0.
\end{equation}
{}For $x \in \mfk_0$, one has $\tau'(x) = x$, i.e. $x \in \mfu_0$, and similarly $\tau'(y) = - y$ for $y \in \mfp_0$, so that $iy \in \mfu_0$.  This implies that one can write for the standard compact form $\mfu_0$,
\begin{equation}
\mfu_0 = \mfk_0 \oplus i \mfp_0, \; \; \; \mfk_0 = \g_0 \cap \mfu_0, \; \; \; \mfp_0 = \g_0 \cap i \mfu_0.
\end{equation}
Since $\theta$ is linear, the complexifications $\mfk = \mfk_0 \otimes \ce$ and $\mfp = \mfp_0 \otimes \ce$ of $\mfk_0$ and $\mfp_0$ are also $\theta$-eigenspaces for the respective eigenvalues $1$ and $-1$.  To summarize, for $x\in \mfk_0$, $y \in \mfp_0$ and 
$\bt \in \ce$, one has
\begin{align}
 \theta (\bt x) &= \bt x,&\theta (\bt y)& = - \bt y,& \nonumber\\
 \sigma' (\bt x) &= \bar{\bt} x,& \sigma' (\bt y) &=  \bar{\bt} y, &\nonumber\\
 \tau' (\bt x) &= \bar{\bt} x,& \tau' (\bt y) &= - \bar{\bt} y.&
\end{align}

A Cartan subalgebra $\mfh_0$ of the real form $\g_0$ is a subalgebra of $\g_0$ whose complexification $\mfh_0 \otimes \ce$ is a Cartan subalgebra of $\g$.  Without loss of generality, one may assume that the Cartan subalgebra is $\theta$-stable, i.e., $\theta(\mfh_0) \subset \mfh_0$ and this will always be done in the sequel.   
Given a $\theta$-stable Cartan subalgebra $\mfh_0$ of $\g_0$, one can decompose it into compact and noncompact parts,
\begin{equation}
\mfh_0 = \mft_0 \oplus  \mfa_0\; , \; \; \; \mft_0 = \mfh_0 \cap \mfk_0\; , \; \; \; \mfa_0 = \mfh_0 \cap  \mfp_0.
\end{equation}
We define the complexifications $\mft = \mft_0 \otimes \ce$ and $\mfa = \mfa_0 \otimes \ce$.  
In $\g_0$, the elements of $\mfa_0$ are ad-diagonalizable over the real numbers ($\mfa_0$ 
is a `split toral subalgebra') while those of $\mft_0$ are not (they have imaginary eigenvalues).

Although Cartan subalgebras are conjugate over the complex numbers, this is not 
the case over the real numbers.  For instance, both $h$ and $e-f$ are Cartan subalgebras 
of $\sgl(2,\re)$ but, although they are conjugate in $\sgl(2, \ce)$, they are not conjugate 
in $\sgl(2,\re)$ since ad$_h$ (being non-compact) can be diagonalized over the 
reals, but not ad$_{e-f}$ (being compact). This is not the only case for which one 
can take the Cartan subalgebra either completely in $\mfk_0$ or completely in $\mfp_0$. 
A less simple example is  the split form $E_{8,8}$ of $E_8$, whose Cartan subalgebra can obviously be taken entirely in $\mfp_0$ since this is the split form entirely in $\mfk_0$ since the maximal compact subalgebra of $E_{8,8}$ is $\so(16)$, which has rank $8$.

One has
\begin{equation}
\mbox{rank}(\g)= \mbox{dim}(\mft_0) + \mbox{dim}(\mfa_0).
\end{equation}
Below we shall only consider `maximally split' (or `maximally noncompact') 
Cartan subalgebras of $\g_0$, i.e. Cartan subalgebras $\mfh_0$ whose non-compact 
part $\mfa_0$ has a dimension as big as possible (and so the dimension of $\mft_0$ 
is as small as possible). For instance, for $\sgl(2,\re)$ this means choosing $h$ and not 
$e-f$ as the Cartan generator. One can show that maximally split Cartan subalgebras 
are conjugate. The real rank of the real form $\g_0$ is the dimension of such 
maximal $\mfa_0$'s.  It is for instance $n-1$ for $\sgl(n,\re)$ and $0$ for $\su(n)$. 

By taking a $G$-conjugate of  $\sigma'$ if necessary, one may assume without loss of generality that  the complexification of $\mfh_0$ is  the standard Cartan subalgebra $\mfh$ generated by the $h_i$'s.  This will be assumed in the sequel.

\subsection{Cartan involution and roots}
Let $\al$ be a root of $\g$.  The number $\al(h) $, $h \in \mfh_0$, is real if $h \in \mfa_0$ and imaginary if $h \in \mft_0$.  Accordingly, $\al(h) $ is real if $\al$ vanishes on $\mft_0$ (and by linearity, on $\mft$),
\begin{equation}
\al(h) \in \re \; \; \; \; \forall h \in \mfh_0 \; \; \; \; \Leftrightarrow \; \; \; \; \al(\mft) = 0
\end{equation}
 and that it is imaginary if it vanishes on $\mfa_0$ (and $\mfa$),
 \begin{equation}
i\al(h) \in \re \; \; \; \; \forall h \in \mfh_0 \; \; \; \; \Leftrightarrow \; \; \; \; \al(\mfa) = 0.
\end{equation} In the case of finite-dimensional algebras,  roots of the first type are called `real' while roots of the the second type are called `imaginary'.  Roots which neither vanish on $\mft$ or $\mfa$ are called `complex'.  This terminology will {\it not} be used here as it may lead to confusion with the Kac-Moody notion of `real' or `imaginary'.

Since $\mfh$ is a $\theta$-stable Cartan subalgebra, one can extend the action of $\theta$ from $\mfh$ to its dual $\mfh^*$ through,
\begin{equation}
\theta(\al)(h) = \al(\theta(h)).
\end{equation} 
For roots that vanish on $\mfa$, $\theta(\al) = \al$ while for real roots that vanish on $\mft$, $\theta(\al) = - \al$.  One has also $\theta(\g_\al) = \g_{\theta(\al)}$.  It follows that  $\g_\al$ is $\theta$-stable for  roots that vanish on $\mfa$, i.e., $\theta(\g_\al) = \g_\al$. 

\subsection{Restricted roots}
Let $\mfh_0$ be a maximally split Cartan subalgebra  of $\g_0$ and let $\mft_0$ and $\mfa_0$ be the corresponding compact and noncompact subalgebras.  We denote by $\Delta$ the root system of $\g$. For a root $\al \in \Delta$ that does not vanish on $\mfa$, one defines the restricted root $\al'$ as the restriction of $\al$ to the noncompact (split toral) subalgebra $\mfa_0$.  The restricted root system $\Delta'$ is the set of restricted roots. 

Since $\mfa$ is a subspace of the eigenspace of $\theta$ for the eigenvalue $-1$, one has
$ \al (h) = - \theta(\al)(h) \; \; \; \forall h \in \mfa $ and thus 
\begin{equation}
\al(h) = \frac{1}{2} \left(\al - \theta(\al)\right)(h) \; \; \; \forall h \in \mfa.
\end{equation}
In a similar way, 
\begin{equation}
 \frac{1}{2} \left(\al - \theta(\al)\right)(h) = 0\; \; \; \forall h \in \mfh
\end{equation}
since in that case $\theta(h) = h$.  

The linear operator
\begin{equation}
\pi(\al) = \frac{1}{2} \left(\al - \theta(\al)\right)
\end{equation}
is a projection operator, $\pi^2= \pi$. Its kernel is given by the subspace of $\mfh^*$ that vanishes on $\mfa$, its image is the subspace that vanishes on $\mft$. Two roots $\al$ and $\bt$ yield the same restricted root, i.e., coincide on $\mfa$ if and only if they have the same projection,
\begin{equation}
\al' = \bt' \; \; \; \;  \Leftrightarrow \; \; \; \; \pi( \al) = \pi (\bt).
\end{equation}
For this reason, one may identify the restricted root $\al'$ with $\pi(\al)$.

The set of restricted roots $\Delta'$ of an almost split Kac-Moody algebra is studied in detail in \cite{BackValente} where it is shown to be associated with a generalized Cartan matrix, which  might be of Borcherds type (one might have $A'_{ii} \leq 0$ for some diagonal matrix elements).  For the particular Kac-Moody algebras occuring in gravity, further information will be given below.

\subsection{Iwasawa decomposition}

Define
\begin{equation}
\mfn'_{(\pm)} = \oplus_{\al' \in \Delta'_{\pm}} \g_{\al'} \label{Defn}
\end{equation}
where $\Delta'_{\pm}$ are the sets of positive and negative restricted roots.  

In (\ref{Defn}), $\g_{\al'} \subset \g_0$ is the root space associated with the restricted root $\al'$.  Even if $\al$ is non degenerate, the space $\g_{\al'}$ might be multidimensional as two distinct roots might project on the same restricted root $\al'$. 

One has the Iwasawa decomposition
\begin{equation}
\g_0 = \mfk_0 \oplus \mfa_0 \oplus \mfn_{(+)}
\end{equation}
A similar decomposition holds with $\mfn_{(+)}$ replaced by $\mfn_{(-)}$.

The Iwasawa decomposition can be shown to exponentiate to the group.

\subsection{(Almost split) real forms of Kac-Moody algebras}
\label{app:KM}

In the finite-dimensional case all Borel subalgebras are conjugate under the action of $G$ but this is no longer true in the infinite-dimensional Kac-Moody case. If we denote by $\mfb_{(\pm)}$ the standard positive and negative Borel subalgebras, $\mfb_{(\pm)} = \mfh \oplus (\oplus_{\bt \in \Delta_\pm} \g_\bt)$, then it is no longer possible to map all of the negative Borel subalgebra $\mfb_-$ onto the positive Borel subalgebra $\mfb_+$ since there is no $G$ elements that moves all the infinitely many negative roots to positive roots. One can show that there are exactly two $G$-conjugate classes of Borel subalgebras, namely those of $\mfb_{(+)}$ and that of $\mfb_{(-)}$~\cite{KacWang}.  This leads to two different classes of real forms.

A real form is called  `almost split' if $\sigma'(\mfb_{(+)})$ is conjugate to  $\mfb_{(+)}$ under the adjoint action of $G$ (`$G$-conjugate'), where $G$ is the group associated with $\g$ constructed in \cite{Peterson}. It is `almost compact' if $\sigma'(\mfb_{(+)})$ is $G$-conjugate to $\mfb_{(-)}$.  
We consider here only almost split real forms for physical reasons explained in subsection \ref{AlmostSplit}.

\section{Tits-Satake diagrams of $\g^+$, $\g^{++}$ and $\g^{+++}$}
\label{app:TS}
In this appendix, we list the Tits-Satake diagrams for the extensions of all the real forms of all the finite-dimensional simple complex Lie algebras $\g$.  Given what has been stated above, the affine, over-extended and very extended roots are always white roots with no arrow connecting them to any other root. The Tits-Satake diagrams with these properties are a subclass of the Tits-Satake diagrams describing all the split real forms of the complex affine extensions $\g^+$ and double extensions $\g^{++}$, given in  \cite{BenMessaoud,Tripathy}.

We draw explicitly the diagrams for $\g^{+++}$.  To get the diagrams for $\g^{++}$, $\g^+$ or $\g$, it suffices to remove the very extended root, or the very extended and over-extended roots, or the very extended, over-extended and affine roots, respectively.

\vspace{.3cm}

\noindent
{\bf Notations and conventions}

\vspace{.2cm}

In this appendix, the simple roots are denoted by $\bt_i$. As already done in section~\ref{AlmostSplit}, we denote in particular the affine root by $\bt_0$, the over-extended root by $\bt_{-1}$ and the very extended root by $\bt_{-2}$.  

The highest root of the complex Lie algebra $\g$ is called $\mu$.  It is given by the following expression (see Dynkin diagrams below for numbering of the roots)
\begin{eqnarray}
A_n &:& \mu =  \bt_1 + \bt_2 + \bt_3 + \ldots + \bt_n \\
B_n &:& \mu =  \bt_1 + 2 \bt_2 + 2 \bt_3 + \ldots + 2 \bt_n\\
C_n &:& \mu =  2 \bt_1 + 2 \bt_2 + \ldots + 2 \bt_{n-1} + \bt_n \\
D_n &:& \mu =  \bt_1 + 2 \bt_2 + 2 \bt_3 + \ldots + 2 \bt_{n-2} + \bt_{n-1} + \bt_n\\
G_2 &:& \mu =  2\bt_1+3\bt_2\\
F_4 &:& \mu =  2 \bt_1 + 3 \bt_2 + 4 \bt_3 + 2 \bt_4\\
E_6 &:& \mu =  \bt_1 + 2 \bt_2 + 3 \bt_3 + 2 \bt_4 + \bt_5 + 2 \bt_6\\
E_7 &:& \mu =  2 \bt_1 + 3 \bt_2 + 4 \bt_3 + 3 \bt_4 + 2 \bt_5 + \bt_6 + 2 \bt_7\\
E_8 &:& \mu =  2 \bt_1 + 3 \bt_2 + 4 \bt_3 + 5 \bt_4 + 6 \bt_5 + 4 \bt_6 + 2 \bt_7 + 3 \bt_8 \\
BC_n &:& \mu =  2\bt_1 + 2 \bt_2 + 2 \bt_3 + \ldots + 2 \bt_n
\end{eqnarray}

The restricted roots, i.e., the roots of the restricted root system, are denoted by $\lambda_i$.

\subsection{The split real form and the compact real forms}
As we have seen, the Cartan involution acts trivially on the over-extended and very extended roots as
\begin{equation}
\theta(\bt_{-1}) = - \bt_{-1}, \; \; \; \; \theta(\bt_{-2}) = -\bt_{-2}
\end{equation}
The only non trivial question is how the Cartan involution acts on the affine root.

The action on $\bt_0$ is on general grounds
\begin{equation}
\theta (\bt_0) = - \bt_0 - \mu - \theta(\mu)
\end{equation}
In order to completely describe the Cartan involution, the problem is therefore to compute  $\theta(\mu)$ for all the real forms $\g$ of the complex, simple, finite-dimensional Lie algebras $\g_\ce$.  

There are two cases that can be treated straightforwardly.

\subsubsection{The split real form $\mfs_0$}
In that case, all the nodes of the Tits-Satake diagram are white: all the roots of $\g$ are real ($\theta(\bt) = - \bt$ for all of them) so that $\theta(\mu) = - \mu$.  This implies $\theta(\bt_0) = - \bt_0$, which is a root, and the normality condition is obviously satisfied.

 {}Furthermore, the restricted root system coincides with the root system of $\g$ since the projection $\pi$ reduces to the identity, $\pi (\bt) = \bt$ for all roots.

\subsubsection{The compact real form $\mfu_0$}
In that case, all the nodes of the Tits-Satake diagram of the finite-dimensional algebra $\g$ are black: all the roots of $\g$ are compact imaginary ($\theta(\al) =  \al$ for all of them) so that $\theta(\mu) =  \mu$.  The only white nodes of the extensions are the affine, over-extended and very extended roots. 

One has $\theta(\bt_0) = - \bt_0 - 2 \mu$.  In terms of the current algebra description, one easily sees that $\bt_0 + 2 \mu$ is a root whose root vector is $e^1_\mu$ (and so is also its negative).  Here the index $1$ is the level.  We recall that the root vector of $\bt_0$ is $e^1_{-\mu}$.  The normality condition is also obviously satisfied since $-2 \mu$ is not a root.

Because all the roots of $\g$ project to zero, the restricted affine root $\lambda_0= \bt_0 + \mu$ has norm zero, $(\lambda_0\vert \lambda_0) = 0$.    Because of this,  the Cartan matrix of the restricted root system is of Borcherds type in the compact case,
\begin{equation}
A =   \begin{pmatrix} 0&-1 & 0 \\ -1 & 2 & -1 \\ 0 & -1 & 2 \end{pmatrix}  .
\end{equation}
yielding as restricted root diagram
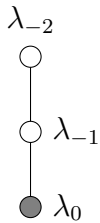
\begin{center}
\begin{tikzpicture} 
[place/.style={circle,draw=black, fill=white, inner sep=0pt,minimum size=8}] 
\draw (0,0) -- (0,1);\draw (0,1) -- (0,2);
\node at (0,0) [circle,draw=black, fill=gray, inner sep=0pt,minimum size=8,label=right:$\lambda_0$] {};
\node at (0,1) [place,label=right:$\lambda_{-1}$] {};
\node at (0,2) [place,label=above:$\lambda_{-2}$] {};
\end{tikzpicture}
\end{center}
The zero-norm restricted root $\lambda_0$ has been colored in grey.  Its degeneracy is equal to the dimension of $\g$.

\vspace{.3cm}
We now list explicitly the Tits-Satake diagrams for the extensions of all the real forms of all the finite-dimensional simple complex Lie algebras $\g$.  In each case except the split and compact cases which have just been covered, we provide: (i) the Tits-Satake diagram of the corresponding real form of $\g^{+++}$; (ii) the action of the Cartan involution $\theta$ on the highest root $\mu$ of  $\g$; (ii) the action of $\theta$ on the affine root $\bt_0$; (iv) the restricted root system. In the split and compact cases, we only provide (i). The Tits-Satake diagrams of the real forms of the finite-dimensional simple complex Lie algebras $\g$ themselves are taken from \cite{Araki,Helgason}.

\subsection{The $A_n$ case}
\subsubsection{$A_nI \equiv \sgl(n+1,\re)$ (split real form)}

\noindent
{\bf Tits-Satake diagram of $\sgl(n+1,\re)^{+++}$}
\begin{center}
\begin{tikzpicture} 
[place/.style={circle,draw=black, fill=white, inner sep=0pt,minimum size=8}] 
\draw (-1,0)--(0,0); \draw (0,0) -- (1,0); \draw[dashed](1,0)--(3,0); \draw (3,0) -- (4,0); \draw (-1,0) -- (1.5,1);\draw (4,0) -- (1.5,1);\draw (1.5,1)--(1.5,2); \draw (1.5,2) -- (1.5,3);
\node at (-1,0) [place,label=below:$\bt_1$] {};
\node at (0,0) [place,label=below:$\bt_2$] {};
\node at (1,0) [place,label=below:$\bt_3$] {};
\node at (3,0) [place,label=below:$\bt_{n-2}$] {};
\node at (4,0) [place,label=below:$\bt_n$] {};
\node at (1.5,1) [place,label=below:$\bt_0$] {};
\node at (1.5,2) [place,label=right:$\bt_{-1}$] {};
\node at (1.5,3) [place,label=above:$\bt_{-2}$] {};
\end{tikzpicture}
\end{center}

\subsubsection{$A_nII \equiv \su^*(n+1)$ ($n$ odd)}

\noindent
{\bf Tits-Satake diagram of $\su^*(n+1))^{+++}$}
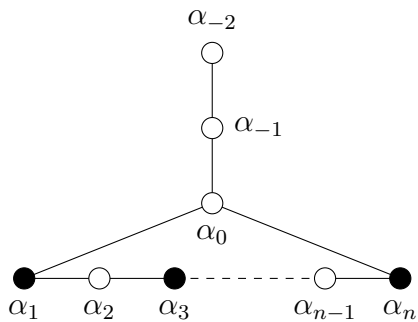
\begin{center}
\begin{tikzpicture} 
[place/.style={circle,draw=black, fill=white, inner sep=0pt,minimum size=8}] 
\draw (-1,0)--(0,0); \draw (0,0) -- (1,0); \draw[dashed](1,0)--(3,0); \draw (3,0) -- (4,0); \draw (-1,0) -- (1.5,1);\draw (4,0) -- (1.5,1);\draw (1.5,1)--(1.5,2); \draw (1.5,2) -- (1.5,3);
\node at (-1,0) [circle, draw=black, fill = black, inner sep=0pt, minimum size=8,label=below:$\bt_1$] {};
\node at (0,0) [place,label=below:$\bt_2$] {};
\node at (1,0) [circle, draw=black, fill = black, inner sep=0pt, minimum size=8,label=below:$\bt_3$] {};
\node at (3,0) [place,label=below:$\bt_{n-1}$] {};
\node at (4,0) [circle, draw=black, fill = black, inner sep=0pt, minimum size=8,label=below:$\bt_n$] {};
\node at (1.5,1) [place,label=below:$\bt_0$] {};
\node at (1.5,2) [place,label=right:$\bt_{-1}$] {};
\node at (1.5,3) [place,label=above:$\bt_{-2}$] {};
\end{tikzpicture}
\end{center}

\noindent
{\bf Action on $\mu$ and $\bt_0$}

 \begin{equation}\theta(\mu) = - \sum_{i=2}^{n-1}\bt_i , \; \; \; \; \theta(\bt_0) =- \bt_0 - \bt_1 - \bt_n \; \; \; \;  (\su^*(n+1)^{+++})\end{equation}  
 
\vspace{.3cm}

\noindent
{\bf Restricted root system of $\su^*(n+1)^{+++}$}

\noindent
$n>3$:
\begin{center}
\begin{tikzpicture} 
[place/.style={circle,draw=black, fill=white, inner sep=0pt,minimum size=8}] 
\draw (0,0) -- (1,0); \draw[dashed](1,0)--(3,0); \draw (0,0) -- (1.5,1);\draw (3,0) -- (1.5,1);\draw (1.4,1)--(1.4,2); \draw (1.6,1)--(1.6,2); \draw (1.5,2) -- (1.5,3);
\draw (1.3,1.6)--(1.5,1.4); \draw(1.7,1.6)--(1.5,1.4);
\node at (0,0) [place,label=below:$\lambda_2$] {};
\node at (1,0) [circle, draw=black, fill = white, inner sep=0pt, minimum size=8,label=below:$\lambda_4$] {};
\node at (3,0) [place,label=below:$\lambda_{n-1}$] {};
\node at (1.5,1) [place,label=below:$\lambda_0$] {};
\node at (1.5,2) [place,label=right:$\lambda_{-1}$] {};
\node at (1.5,3) [place,label=above:$\lambda_{-2}$] {};
\end{tikzpicture}
\end{center}
$n=3$:
\begin{center}
\begin{tikzpicture} 
[place/.style={circle,draw=black, fill=white, inner sep=0pt,minimum size=8}] 
\draw (1.4,0) -- (1.4,1);\draw (1.6,0) -- (1.6,1);\draw (1.4,1)--(1.4,2); \draw (1.6,1)--(1.6,2); \draw (1.5,2) -- (1.5,3);
\draw (1.3,1.6)--(1.5,1.4); \draw(1.7,1.6)--(1.5,1.4);
\node at (1.5,0) [place,label=below:$\lambda_2$] {};
\node at (1.5,1) [place,label=right:$\lambda_0$] {};
\node at (1.5,2) [place,label=right:$\lambda_{-1}$] {};
\node at (1.5,3) [place,label=above:$\lambda_{-2}$] {};
\draw (1.3,0.4)--(1.5,0.2); \draw(1.7,0.4)--(1.5,0.2);
\draw (1.3,0.6)--(1.5,0.8); \draw(1.7,0.6)--(1.5,0.8);
\end{tikzpicture}
\end{center}

The roots $\lambda_0$ through $\lambda_{n-1}$ have multipicity $4$, while the roots $\lambda_{-1}$ and $\lambda_{-2}$ are non degenerate. For no $i$ is $2 \lambda_i$ a root. The restricted root $\lambda_0$ has norm squared equal to $1$ and is shorter than the restricted root $\lambda_{-1}$, which has norm squared equal to $2$. The restricted Dynkin diagram of the over-extended extension $\g^{++}$ is hyperbolic for $n=3,5,7$ (the twisted affine algebra obtained by removing the node $\lambda_4$ for $n=7$ is called $A_5^{(2)}$).  The gravity line contains only $\lambda_{-2}$ and $\lambda_{-1}$ and the system cannot be oxidized above $3$ dimensions.

\subsubsection{$A_nIII \equiv \su(p, q)$, $p,q>1$, $p+q = n+1$}

\noindent
{\bf Tits-Satake diagram of $\su(p,q)^{+++}$}

(i) $p \not=q = n+1-p$ (we then assume for definiteness $p<q $)
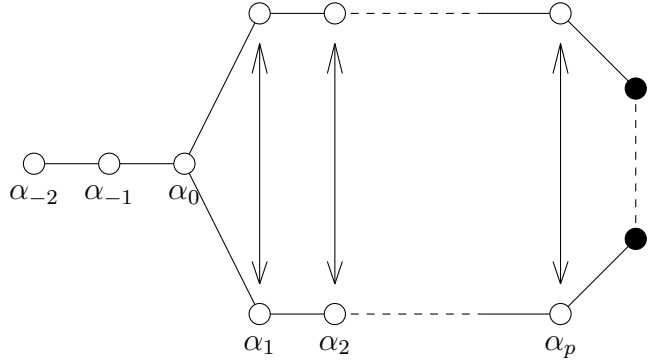
\begin{center}
\begin{tikzpicture} 
[place/.style={circle,draw=black, fill=white, inner sep=0pt,minimum size=8}] 
\draw (-1,0)--(0,0); \draw[dashed](0,0)--(2,0); \draw (2,0) -- (3,0); \draw (3,0) -- (4,1);  \draw[dashed](4,1)--(4,3);\draw (4,3)--(3,4); \draw (2,4) -- (3,4);
\draw(-1,0)--(-2,2); \draw(-1,4)--(-2,2);\draw(-2,2)--(-3,2); \draw(-3,2)--(-4,2);
\draw (-1,4)--(0,4); \draw[dashed](0,4)--(2,4); 
\node at (-1,0) [circle, draw=black, fill=white, inner sep=0pt, minimum size=8,label=below:$\bt_1$] {};
\node at (0,0) [circle, draw=black, fill = white, inner sep=0pt, minimum size=8,label=below:$\bt_2$] {};
\node at (3,0) [circle, draw=black, fill = white, inner sep=0pt, minimum size=8,label=below:$\bt_{p}$] {};
\node at (4,1) [circle, draw=black, fill = black, inner sep=0pt, minimum size=8] {};
\node at (4,3) [place,fill=black] {};
\node at (3,4) [place] {};
\node at (0,4) [place] {};
\node at (-1,4) [place] {};
\draw (-1,3.6)--(-1,.4); 
\draw (-.9,.7)--(-1,.4); \draw(-1.1,.7)--(-1,.4);
\draw (-.9,3.2)--(-1,3.6); 
\draw(-1.1,3.2)--(-1,3.6);
\draw (0,3.6)--(0,.4); 
\draw (0.1,.7)--(0,.4); \draw(-.1,.7)--(0,.4);
\draw (0.1,3.2)--(0,3.6); 
\draw(-0.1,3.2)--(0,3.6);
\draw (3,3.6)--(3,.4); 
\draw (3.1,.7)--(3,.4); \draw(2.9,.7)--(3,.4);
\draw (3.1,3.2)--(3,3.6); 
\draw(2.9,3.2)--(3,3.6);
\node at (-2,2)[place,label=below:$\bt_0$] {};
\node at (-3,2)[place,label=below:$\bt_{-1}$] {};
\node at (-4,2)[place,label=below:$\bt_{-2}$] {};
\end{tikzpicture}
\end{center}
There are $n-2p \geq 0$ black roots (no black root for $q = p+1$ since then $n = 2p$).

\vspace{.3cm}

\noindent
(ii) $p=q$ ($su(p,p)^{+++}$):
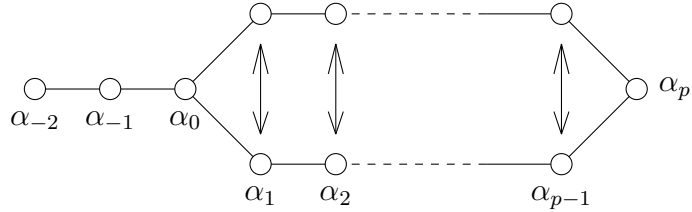
\begin{center}
\begin{tikzpicture} 
[place/.style={circle,draw=black, fill=white, inner sep=0pt,minimum size=8}] 
\draw (-1,0)--(0,0); \draw[dashed](0,0)--(2,0); \draw (2,0) -- (3,0); \draw (3,0) -- (4,1);  \draw (4,1)--(3,2); \draw (2,2) -- (3,2);
\draw (-1,2)--(0,2); \draw[dashed](0,2)--(2,2); 
\draw(-1,0)--(-2,1); \draw(-1,2)--(-2,1);\draw(-2,1)--(-3,1); \draw(-3,1)--(-4,1);
\node at (-1,0) [circle, draw=black, fill=white, inner sep=0pt, minimum size=8,label=below:$\bt_1$] {};
\node at (0,0) [circle, draw=black, fill = white, inner sep=0pt, minimum size=8,label=below:$\bt_2$] {};
\node at (3,0) [circle, draw=black, fill = white, inner sep=0pt, minimum size=8,label=below:$\bt_{p-1}$] {};
\node at (4,1) [circle, draw=black, fill = white, inner sep=0pt, minimum size=8,label=right:$\bt_p$] {};
\node at (3,2) [place] {};
\node at (0,2) [place] {};
\node at (-1,2) [place] {};
\draw (-1,1.6)--(-1,.4); 
\draw (-.9,.7)--(-1,.4); \draw(-1.1,.7)--(-1,.4);
\draw (-.9,1.2)--(-1,1.6); 
\draw(-1.1,1.2)--(-1,1.6);
\draw (0,1.6)--(0,.4); 
\draw (0.1,.7)--(0,.4); \draw(-.1,.7)--(0,.4);
\draw (0.1,1.2)--(0,1.6); 
\draw(-0.1,1.2)--(0,1.6);
\draw (3,1.6)--(3,.4); 
\draw (3.1,.7)--(3,.4); \draw(2.9,.7)--(3,.4);
\draw (3.1,1.2)--(3,1.6); 
\draw(2.9,1.2)--(3,1.6);
\node at (-2,1)[place,label=below:$\bt_0$] {};
\node at (-3,1)[place,label=below:$\bt_{-1}$] {};
\node at (-4,1)[place,label=below:$\bt_{-2}$] {};
\end{tikzpicture}
\end{center}

\vspace{.3cm}

\noindent
{\bf Action on $\mu$ and $\bt_0$}

 \begin{equation}\theta(\mu) = - \mu , \; \; \; \; \theta(\bt_0) =- \bt_0  \; \; \; \;  (\su(p,q)^{+++})\end{equation}  
 
\vspace{.3cm}

\noindent
{\bf Restricted root system of $\su(p,q)^{+++}$}

(i) $p \not=q = n+1-p$
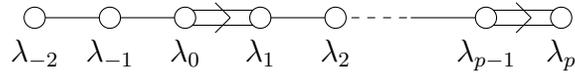
\begin{center}
\begin{tikzpicture} [place/.style={circle,draw=black,fill=white, inner sep=0pt,minimum size=8}] 
\draw(-1,.1)--(0,.1);  \draw(-1,-.1)--(0,-.1);
\draw (0,0) -- (1,0); \draw[dashed] (1,0) -- (2,0); \draw (2,0) -- (3,0);  
\draw (3,.1) -- (4,.1); \draw (3,-.1) -- (4,-.1); 
\draw (-0.6,0.2) -- (-0.4,0); \draw (-0.6,-0.2) -- (-0.4,0); 
\draw (3.4,0.2) -- (3.6,0); \draw (3.4,-0.2) -- (3.6,0); 
\draw(-1,0)--(-2,0);\draw(-2,0)--(-3,0);
\node at (0,0) [place,label=below:$\lambda_1$] {}; 
\node at (1,0) [place,label=below:$\lambda_{2}$] {};     
\node at (3,0) [place,label=below:$\lambda_{p-1}$] {}; 
\node at (4,0) [place,label=below:$\lambda_{p}$] {}; 
\node at (-1,0) [place,label=below:$\lambda_0$] {}; 
\node at (-2,0) [place,label=below:$\lambda_{-1}$] {}; 
\node at (-3,0) [place,label=below:$\lambda_{-2}$] {}; 
\end{tikzpicture}
\end{center}
The degeneracy of the roots $\lambda_i$ ($1 \leq i \leq p-1$) is equal to $2$; the degeneracy of the root $\lambda_p$ is equal to $2(n -2p +1)$.  The restricted root system of $\g_0$ is of $BC$-type, with $2 \lambda_p$ a root of multiplicity $1$.  This leads to a restricted Dynkin diagram of twisted $A_{2p}^{(2)++}$-type \cite{Henneaux:2003kk}. The restricted Dynkin diagram $A_{2p}^{(2)+}$ of the over-extended extension $\su(p,q)^{++}$ is hyperbolic for $p\leq 4$.  The affine root has the same length squared as the over-extended root and is non degenerate.  It is part of the gravity line.  The system can be oxidized to $4$ dimensions.

\vspace{.3cm}

(ii) $p=q$ ($su(p,p)^{+++}$)
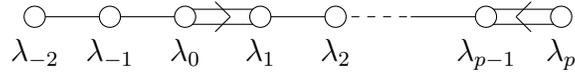
\begin{center}
\begin{tikzpicture} [place/.style={circle,draw=black,fill=white, inner sep=0pt,minimum size=8}] 
\draw(-1,.1)--(0,.1);  \draw(-1,-.1)--(0,-.1);
\draw (0,0) -- (1,0); \draw[dashed] (1,0) -- (2,0); \draw (2,0) -- (3,0);  
\draw (3,.1) -- (4,.1); \draw (3,-.1) -- (4,-.1); 
\draw (-0.6,0.2) -- (-0.4,0); \draw (-0.6,-0.2) -- (-0.4,0); 
\draw (3.6,0.2) -- (3.4,0); \draw (3.6,-0.2) -- (3.4,0); 
\draw(-1,0)--(-2,0);\draw(-2,0)--(-3,0);
\node at (0,0) [place,label=below:$\lambda_1$] {}; 
\node at (1,0) [place,label=below:$\lambda_{2}$] {};   
\node at (3,0) [place,label=below:$\lambda_{p-1}$] {}; 
\node at (4,0) [place,label=below:$\lambda_{p}$] {}; 
\node at (-1,0) [place,label=below:$\lambda_0$] {}; 
\node at (-2,0) [place,label=below:$\lambda_{-1}$] {}; 
\node at (-3,0) [place,label=below:$\lambda_{-2}$] {}; 
\end{tikzpicture}
\end{center}

The restricted root system is of $C_p^{+++}$ type.  The degeneracy of the roots $\lambda_i$ ($1 \leq i \leq p-1$) is equal to $2$; the degeneracy of the root $\lambda_p$  is equal to $1$.  For no root $i$ is $2 \lambda_i$ a root. The affine root is part of the gravity line (same length as $\lambda_{-1}$ and non degenerate) so the system can be oxidized to $4$ dimensions.

\subsubsection{$A_nIV \equiv \su(n,1)$}

\noindent
{\bf Tits-Satake diagram of $\su(n,1)^{+++}$}

\begin{center}
\begin{tikzpicture} 
[place/.style={circle,draw=black, fill=white, inner sep=0pt,minimum size=8}] 
\draw (-1,0)--(0,0); \draw[dashed](0,0)--(2,0); \draw (2,0) -- (3,0);
\draw(-1,0)--(1,-1);\draw(3,0)--(1,-1); \draw(1,-1)--(1,-2); \draw(1,-2)--(1,-3);
\node at (-1,0) [circle, draw=black, fill = white, inner sep=0pt, minimum size=8,label=below:$\bt_1$] {};
\node at (0,0) [circle, draw=black, fill = black, inner sep=0pt, minimum size=8,label=above:$\bt_2$] {};
\node at (2,0) [circle, draw=black, fill = black, inner sep=0pt, minimum size=8,label=above:$\bt_{n-1}$] {};
\node at (3,0) [circle, draw=black, fill = white, inner sep=0pt, minimum size=8,label=below:$\bt_n$] {};
\draw (-1,1)--(-1,.4); \draw (-1,1)--(3,1);
\draw (-.9,.7)--(-1,.4); \draw(-1.1,.7)--(-1,.4);
\draw (3,1)--(3,.4); 
\draw (3.1,.7)--(3,.4); \draw(2.9,.7)--(3,.4);
\node at (1,-1)[place,label=right:$\bt_0$]{};
\node at (1,-2)[place,label=right:$\bt_{-1}$]{};
\node at (1,-3)[place,label=right:$\bt_{-2}$]{};
\end{tikzpicture}
\end{center}

\vspace{.3cm}

\noindent
{\bf Action on $\mu$ and $\bt_0$}

 \begin{equation}\theta(\mu) = - \mu , \; \; \; \; \theta(\bt_0) =- \bt_0  \; \; \; \;  (\su(n,1)^{+++})\end{equation}  
 
\vspace{.3cm}

\noindent
{\bf Restricted root system of $\su(n,1)^{+++}$}

\begin{center}
\begin{tikzpicture} 
[place/.style={circle,draw=black, fill=white, inner sep=0pt,minimum size=8}] 
\draw(1.5,0)--(1.5,1); \draw(1.5,1)--(1.5,2); 
\draw(1.35,2)--(1.35,3); \draw(1.45,2)--(1.45,3); \draw(1.55,2)--(1.55,3); \draw(1.65,2)--(1.65,3);
\draw (1.3,2.4)--(1.5,2.6); \draw(1.7,2.4)--(1.5,2.6);
\node at (1.5,0) [place,label=below:$\lambda_{-2}$] {};
\node at (1.5,1) [place,label=right:$\lambda_{-1}$] {};
\node at (1.5,2) [place,label=right:$\lambda_{0}$] {};
\node at (1.5,3) [place,label=above:$\lambda_{1}$] {};
\end{tikzpicture}
\end{center}
The restricted root $\lambda_1$ has multiplicity $2(n-1)$ and $2 \lambda_1$ is a root with multiplicity $1$.  The root $\lambda_0$ is non degenerate and part of the gravity line so the system can be oxidized to $4$ dimensions. The restricted root system is of twisted $A_{2}^{(2)++}$-type \cite{Henneaux:2003kk}.

\subsubsection{Compact real form $\su(n+1)^{+++}$}

\begin{center}
\begin{tikzpicture} 
[place/.style={circle,draw=black, fill=white, inner sep=0pt,minimum size=8}] 
\draw (-1,0)--(0,0); \draw (0,0) -- (1,0); \draw[dashed](1,0)--(3,0); \draw (3,0) -- (4,0); \draw (-1,0) -- (1.5,1);\draw (4,0) -- (1.5,1);\draw (1.5,1)--(1.5,2); \draw (1.5,2) -- (1.5,3);
\node at (-1,0) [circle, draw=black, fill = black, inner sep=0pt, minimum size=8,label=below:$\alpha_1$] {};
\node at (0,0) [circle, draw=black, fill = black, inner sep=0pt, minimum size=8,label=below:$\alpha_2$] {};
\node at (1,0) [circle, draw=black, fill = black, inner sep=0pt, minimum size=8,label=below:$\alpha_3$] {};
\node at (3,0) [circle, draw=black, fill = black, inner sep=0pt, minimum size=8,label=below:$\alpha_{n-2}$] {};
\node at (4,0) [circle, draw=black, fill = black, inner sep=0pt, minimum size=8,label=below:$\alpha_n$] {};
\node at (1.5,1) [place,label=below:$\alpha_0$] {};
\node at (1.5,2) [place,label=right:$\alpha_{-1}$] {};
\node at (1.5,3) [place,label=above:$\alpha_{-2}$] {};
\end{tikzpicture}
\end{center}

\subsection{The $B_n$ case}
\subsubsection{$B_nI \equiv \so(p,q)$, $p,q>1$, $p+q = 2n+1$}
We assume for definiteness $p<q$. 

\vspace{.3cm}

\noindent
{\bf Tits-Satake diagram of $\so(p,q)^{+++}$, $p,q>1$, $p+q = 2n+1$, $p<q$}
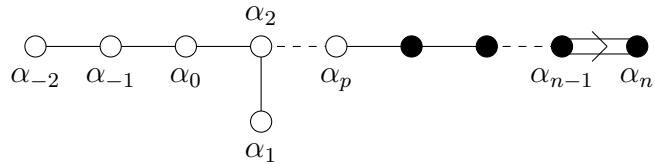
\begin{center}
\begin{tikzpicture} [place/.style={circle,draw=black,fill=white, inner sep=0pt,minimum size=8}] 
\draw(-1,0)--(0,0);\draw[dashed] (0,0) -- (1,0); \draw (1,0) -- (2,0); \draw (2,0) -- (3,0); \draw[dashed] (3,0) -- (4,0); \draw(0,0)--(0,-1);
\draw (4,.1) -- (5,.1); \draw (4,-.1) -- (5,-.1); 
\draw (4.4,0.2) -- (4.6,0); \draw (4.4,-0.2) -- (4.6,0); 
\draw(-1,0)--(-2,0); \draw(-2,0)--(-3,0);
\node at (0,-1) [place,label=below:$\bt_1$] {}; 
\node at (0,0) [place,label=above:$\bt_2$] {}; 
\node at (1,0) [place,label=below:$\bt_{p}$] {}; 
\node at (2,0) [circle,draw=black,fill=black, inner sep=0pt,minimum size=8] {};     
\node at (3,0) [circle,draw=black,fill=black, inner sep=0pt,minimum size=8] {}; 
\node at (4,0) [circle,draw=black,fill=black, inner sep=0pt,minimum size=8,label=below:$\bt_{n-1}$] {}; 
\node at (5,0) [circle,draw=black,fill=black, inner sep=0pt,minimum size=8,label=below:$\bt_n$] {}; 
\node at (-1,0) [place,label=below:$\bt_0$] {}; 
\node at (-2,0) [place,label=below:$\bt_{-1}$] {}; 
\node at (-3,0) [place,label=below:$\bt_{-2}$] {}; 
\end{tikzpicture}
\end{center}

Because $p \geq 2$, $\bt_2$ is a white node. Note also that the split real form $\so(n, n+1)$ has all its nodes unpainted and corresponds to $p=n$.

\vspace{.3cm}

\noindent
{\bf Action on $\mu$ and $\bt_0$}

 \begin{equation}\theta(\mu) = - \mu , \; \; \; \; \theta(\bt_0) =- \bt_0  \; \; \; \;  (\so(p,q)^{+++}, p,q>1)\end{equation}  
 
\vspace{.3cm}

\noindent
{\bf Restricted root system of $\so(p,q)^{+++}$ ($p,q >1$)}

(i) $p >2$
\begin{center}
\begin{tikzpicture} [place/.style={circle,draw=black,fill=white, inner sep=0pt,minimum size=8}] 
\draw(-1,0)--(0,0);\draw[dashed] (0,0) -- (1,0); 
\draw(0,0)--(0,-1);
\draw (1,.1) -- (2,.1); \draw (1,-.1) -- (2,-.1); 
\draw (1.4,0.2) -- (1.6,0); \draw (1.4,-0.2) -- (1.6,0); 
\draw(-1,0)--(-2,0); \draw(-2,0)--(-3,0);
\node at (0,-1) [place,label=below:$\lambda_1$] {}; 
\node at (0,0) [place,label=above:$\lambda_2$] {}; 
\node at (1,0) [place,label=below:$\lambda_{p-1}$] {}; 
\node at (2,0) [place,label=below:$\lambda_{p}$] {};     
\node at (-1,0) [place,label=below:$\lambda_0$] {}; 
\node at (-2,0) [place,label=below:$\lambda_{-1}$] {}; 
\node at (-3,0) [place,label=below:$\lambda_{-2}$] {}; 
\end{tikzpicture}
\end{center}

(ii) $p =2$
\begin{center}
\begin{tikzpicture} [place/.style={circle,draw=black,fill=white, inner sep=0pt,minimum size=8}]  
\draw(-.1,0)--(-.1,-1); \draw(.1,0)--(.1,-1);
\draw (-0.2,-0.6) -- (0,-0.4); \draw (0.2,-0.6) -- (0,-0.4);
\draw (-1,.1) -- (0,.1); \draw (-1,-.1) -- (0,-.1); 
\draw (-0.6,0.2) -- (-0.4,0); \draw (-0.6,-0.2) -- (-0.4,0); 
\draw(-1,0)--(-2,0); \draw(-2,0)--(-3,0);
\node at (0,-1) [place,label=below:$\lambda_1$] {}; 
\node at (0,0) [place,label=above:$\lambda_2$] {};  
\node at (-1,0) [place,label=below:$\lambda_0$] {}; 
\node at (-2,0) [place,label=below:$\lambda_{-1}$] {}; 
\node at (-3,0) [place,label=below:$\lambda_{-2}$] {}; 
\end{tikzpicture}
\end{center}

The restricted root system is of $B_p^{+++}$-type ($C_2^{+++}$-type for $p=2$).
The roots are non degenerate, except $\lambda_p$ which has multiplicity $2(n-p)+1$.  The affine root is part of the gravity line.

\subsubsection{$B_nII \equiv \so(1,2n)$}
\noindent
{\bf Tits-Satake diagram of $\so(1,2n)^{+++}$}
\begin{center}
\begin{tikzpicture} [place/.style={circle,draw=black,fill=white, inner sep=0pt,minimum size=8}] 
\draw(-1,0)--(0,0);\draw (0,0) -- (1,0); \draw[dashed] (1,0) -- (2,0); 
\draw (2,.1) -- (3,.1); \draw (2,-.1) -- (3,-.1); 
\draw (2.4,0.2) -- (2.6,0); \draw (2.4,-0.2) -- (2.6,0); 
\draw(0,0)--(0,-1);
\draw(-1,0)--(-2,0); \draw(-2,0)--(-3,0);
\node at (0,-1) [place,label=below:$\bt_1$] {}; 
\node at (0,0) [circle,draw=black,fill=black, inner sep=0pt,minimum size=8,label=above:$\bt_2$] {};    
\node at (1,0) [circle,draw=black,fill=black, inner sep=0pt,minimum size=8] {}; 
\node at (2,0) [circle,draw=black,fill=black, inner sep=0pt,minimum size=8,label=below:$\bt_{n-1}$] {}; 
\node at (3,0) [circle,draw=black,fill=black, inner sep=0pt,minimum size=8,label=below:$\bt_n$] {}; 
\node at (-1,0) [place,label=below:$\bt_0$] {}; 
\node at (-2,0) [place,label=below:$\bt_{-1}$] {}; 
\node at (-3,0) [place,label=below:$\bt_{-2}$] {}; 
\end{tikzpicture}
\end{center}

\vspace{.3cm}

\noindent
{\bf Action on $\mu$ and $\bt_0$}

 \begin{equation}\theta(\mu) = - \bt_1 , \; \; \; \; \theta(\bt_0) =- \bt_0 - 2 \bt_2 - 2 \bt_3 - \ldots - 2 \bt_n \; \; \; \;  (\so(1,2n)^{+++})\end{equation} 
 One recognizes $-\theta(\bt_0)$ as the highest root of the $B_n$ subalgebra associated with the roots $\{\bt_0, \bt_2, \bt_3, \ldots, \bt_n \}$. 
 
\vspace{.3cm}

\noindent
{\bf Restricted root system of $\so(1,2n)^{+++}$} 

The restricted root system is the same as for $\su^*(4)^{+++}$, 
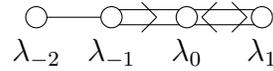
\begin{center}
\begin{tikzpicture} [place/.style={circle,draw=black,fill=white, inner sep=0pt,minimum size=8}]  
\draw (-1,.1) -- (0,.1); \draw (-1,-.1) -- (0,-.1); 
\draw (-1.6,0.2) -- (-1.4,0); \draw (-1.6,-0.2) -- (-1.4,0); 
\draw(-1,.1)--(-2,.1); \draw(-1,-.1)--(-2,-.1);
\draw (-0.4,0.2) -- (-0.2,0); \draw (-0.4,-0.2) -- (-0.2,0); 
\draw (-0.6,0.2) -- (-0.8,0); \draw (-0.6,-0.2) -- (-0.8,0); 
\draw(-2,0)--(-3,0);
\node at (0,0) [place,label=below:$\lambda_1$] {};  
\node at (-1,0) [place,label=below:$\lambda_0$] {}; 
\node at (-2,0) [place,label=below:$\lambda_{-1}$] {}; 
\node at (-3,0) [place,label=below:$\lambda_{-2}$] {}; 
\end{tikzpicture}
\end{center}
The roots $\lambda_1$ and $\lambda_0$ are $2n-1$ times degenerate.

\subsubsection{Compact real form $\so(2n+1)^{+++}$}

\begin{center}
\begin{tikzpicture} [place/.style={circle,draw=black,fill=white, inner sep=0pt,minimum size=8}] 
\draw(-1,0)--(0,0);\draw (0,0) -- (1,0); \draw[dashed] (1,0) -- (2,0); 
\draw (2,.1) -- (3,.1); \draw (2,-.1) -- (3,-.1); 
\draw (2.4,0.2) -- (2.6,0); \draw (2.4,-0.2) -- (2.6,0); 
\draw(0,0)--(0,-1);
\draw(-1,0)--(-2,0); \draw(-2,0)--(-3,0);
\node at (0,-1) [circle,draw=black,fill=black, inner sep=0pt,minimum size=8,label=below:$\bt_1$] {}; 
\node at (0,0) [circle,draw=black,fill=black, inner sep=0pt,minimum size=8,label=above:$\bt_2$] {};    
\node at (1,0) [circle,draw=black,fill=black, inner sep=0pt,minimum size=8] {}; 
\node at (2,0) [circle,draw=black,fill=black, inner sep=0pt,minimum size=8,label=below:$\bt_{n-1}$] {}; 
\node at (3,0) [circle,draw=black,fill=black, inner sep=0pt,minimum size=8,label=below:$\bt_n$] {}; 
\node at (-1,0) [place,label=below:$\bt_0$] {}; 
\node at (-2,0) [place,label=below:$\bt_{-1}$] {}; 
\node at (-3,0) [place,label=below:$\bt_{-2}$] {}; 
\end{tikzpicture}
\end{center}

\subsection{The $C_n$ case}
\subsubsection{$C_nI \equiv \ssp(n,\re)$ (split form)}

\noindent
{\bf Tits-Satake diagram of $\ssp(n,\re)^{+++}$}

\begin{center}
\begin{tikzpicture} [place/.style={circle,draw=black,fill=white, inner sep=0pt,minimum size=8}] 
\draw(-1,.1)--(0,.1);  \draw(-1,-.1)--(0,-.1);
\draw (0,0) -- (1,0); \draw[dashed] (1,0) -- (2,0); \draw (2,0) -- (3,0);  
\draw (3,.1) -- (4,.1); \draw (3,-.1) -- (4,-.1); 
\draw (-0.6,0.2) -- (-0.4,0); \draw (-0.6,-0.2) -- (-0.4,0); 
\draw (3.6,0.2) -- (3.4,0); \draw (3.6,-0.2) -- (3.4,0); 
\draw(-1,0)--(-2,0);\draw(-2,0)--(-3,0);
\node at (0,0) [place,label=below:$\bt_1$] {}; 
\node at (1,0) [place,label=below:$\bt_{2}$] {}; 
\node at (2,0) [place] {};     
\node at (3,0) [place,label=below:$\bt_{n-1}$] {}; 
\node at (4,0) [place,label=below:$\bt_{n}$] {}; 
\node at (-1,0) [place,label=below:$\bt_0$] {}; 
\node at (-2,0) [place,label=below:$\bt_{-1}$] {}; 
\node at (-3,0) [place,label=below:$\bt_{-2}$] {}; 
\end{tikzpicture}
\end{center}

\subsubsection{$C_nII \equiv \ssp(p,n-p)$}
\noindent
{\bf Tits-Satake diagram of $\ssp(p,n-p)^{+++}$}

(i) $2 \leq 2p \leq n-1$ ($n\geq 3 $): $\ssp(p, q)^{+++}$ with $p<q$
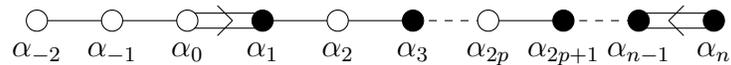
\begin{center}
\begin{tikzpicture} [place/.style={circle,draw=black,fill=white, inner sep=0pt,minimum size=8}] 
\draw(-1,.1)--(0,.1);  \draw(-1,-.1)--(0,-.1);
\draw (0,0) -- (1,0); \draw(1,0) -- (2,0); \draw[dashed]  (2,0) -- (3,0); \draw(3,0) -- (4,0);  \draw[dashed]  (4,0) -- (5,0);
\draw (5,.1) -- (6,.1); \draw (5,-.1) -- (6,-.1); 
\draw (-0.6,0.2) -- (-0.4,0); \draw (-0.6,-0.2) -- (-0.4,0);
\draw (5.6,0.2) -- (5.4,0); \draw (5.6,-0.2) -- (5.4,0); 
\draw(-1,0)--(-2,0); \draw(-2,0)--(-3,0);
\node at (0,0) [circle,draw=black,fill=black, inner sep=0pt,minimum size=8,label=below:$\bt_1$] {}; 
\node at (1,0) [place,label=below:$\bt_{2}$] {}; 
\node at (2,0) [circle,draw=black,fill=black, inner sep=0pt,minimum size=8,label=below:$\bt_3$] {}; 
\node at (3,0) [place,label=below:$\bt_{2p}$] {};  
\node at (4,0) [circle,draw=black,fill=black, inner sep=0pt,minimum size=8,label=below:$\bt_{2p+1}$] {};   
\node at (5,0) [circle,draw=black,fill=black, inner sep=0pt,minimum size=8,label=below:$\bt_{n-1}$] {}; 
\node at (6,0) [circle,draw=black,fill=black, inner sep=0pt,minimum size=8,label=below:$\bt_{n}$] {}; 
\node at (-1,0) [place,label=below:$\bt_0$] {}; 
\node at (-2,0) [place,label=below:$\bt_{-1}$] {}; 
\node at (-3,0) [place,label=below:$\bt_{-2}$] {}; 
\end{tikzpicture}
\end{center}
The first roots starting from $\bt_1$ are alternatively black and white up to $2p$.  Then, they are all black ($p$ white roots from $\g_0$ plus $3$ white roots from the extension).

\vspace{.3cm}

(ii) $2 \leq 2p = n-2$ ($n \geq 4$): $\ssp(p+1,p+1)^{+++}$
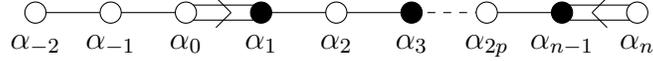
\begin{center}
\begin{tikzpicture} [place/.style={circle,draw=black,fill=white, inner sep=0pt,minimum size=8}] 
\draw(-1,.1)--(0,.1);  \draw(-1,-.1)--(0,-.1);
\draw (0,0) -- (1,0); \draw(1,0) -- (2,0); \draw[dashed]  (2,0) -- (3,0); \draw(3,0) -- (4,0); 
\draw (4,.1) -- (5,.1); \draw (4,-.1) -- (5,-.1); 
\draw (-0.6,0.2) -- (-0.4,0); \draw (-0.6,-0.2) -- (-0.4,0);
\draw (4.6,0.2) -- (4.4,0); \draw (4.6,-0.2) -- (4.4,0); 
\draw(-1,0)--(-2,0); \draw(-2,0)--(-3,0);
\node at (0,0) [circle,draw=black,fill=black, inner sep=0pt,minimum size=8,label=below:$\bt_1$] {}; 
\node at (1,0) [place,label=below:$\bt_{2}$] {}; 
\node at (2,0) [circle,draw=black,fill=black, inner sep=0pt,minimum size=8,label=below:$\bt_3$] {}; 
\node at (3,0) [place,label=below:$\bt_{2p}$] {};     
\node at (4,0) [circle,draw=black,fill=black, inner sep=0pt,minimum size=8,label=below:$\bt_{n-1}$] {}; 
\node at (5,0) [circle,draw=black,fill=white, inner sep=0pt,minimum size=8,label=below:$\bt_{n}$] {}; 
\node at (-1,0) [place,label=below:$\bt_0$] {}; 
\node at (-2,0) [place,label=below:$\bt_{-1}$] {}; 
\node at (-3,0) [place,label=below:$\bt_{-2}$] {}; 
\end{tikzpicture}
\end{center}
The roots starting from $\bt_1$ are alternatively black and white ($p$ white roots from $\g_0$ plus $3$ white roots from the extension).

\vspace{.3cm}

\noindent
{\bf Action on $\mu$ and $\bt_0$}

 \begin{equation}\theta(\mu) = - 2\bt_2 - 2 \bt_3 - \ldots - 2 \bt_{n-1} - \bt_n , \; \; \; \; \theta(\bt_0) =- \bt_0 - 2 \bt_1 \; \; \; \;  (\ssp(p,n-p)^{+++})\end{equation} 
 One recognizes $-\theta(\mu)$ as the highest root of the $C_{n-1}$ subalgebra associated with the roots $\{\bt_2, \bt_3,  \ldots, \bt_n \}$, and $- \theta(\bt_0)$  as the highest root of the $C_{2}$ subalgebra associated with the roots $\{\bt_0, \bt_1\}$. 
 
\vspace{.3cm}

\noindent
{\bf Restricted root system of $\ssp(p,q)^{+++}$ ($p,q >1$)}

(i) $2 \leq 2p \leq n-1$ ($n\geq 3 $): $\ssp(p, q)^{+++}$ with $p<q$

The restricted root system of $\ssp(p, q)$  is of $BC_p$-type, leading to the twisted extension $A_{2p}^{(2)++}$.
\begin{center}
\begin{tikzpicture} [place/.style={circle,draw=black,fill=white, inner sep=0pt,minimum size=8}] 
\draw(-1,.1)--(0,.1);  \draw(-1,-.1)--(0,-.1);
\draw (0,0) -- (1,0); \draw[dashed] (1,0) -- (2,0); \draw (2,0) -- (3,0);  
\draw (3,.1) -- (4,.1); \draw (3,-.1) -- (4,-.1); 
\draw (-0.6,0.2) -- (-0.4,0); \draw (-0.6,-0.2) -- (-0.4,0); 
\draw (3.4,0.2) -- (3.6,0); \draw (3.4,-0.2) -- (3.6,0); 
\draw(-1,0)--(-2,0);\draw(-2,0)--(-3,0);
\node at (0,0) [place,label=below:$\lambda_1$] {}; 
\node at (1,0) [place,label=below:$\lambda_{2}$] {};     
\node at (3,0) [place,label=below:$\lambda_{p-1}$] {}; 
\node at (4,0) [place,label=below:$\lambda_{p}$] {}; 
\node at (-1,0) [place,label=below:$\lambda_0$] {}; 
\node at (-2,0) [place,label=below:$\lambda_{-1}$] {}; 
\node at (-3,0) [place,label=below:$\lambda_{-2}$] {}; 
\end{tikzpicture}
\end{center}
The roots $\lambda_i$ are degenerate $4$ times for $1 \leq i \leq p-1$.  The root $\lambda_p$ is degenerate $4 (n-2p)$ times.  $2 \lambda_p$ is a root with multiplicity $3$.  The affine root has also multiplicity $3$ and so the gravity line contains only $\lambda_{-1}$ and $\lambda_{-2}$.

\vspace{.3cm}

(ii) $2 \leq 2p = n-2$ ($n \geq 4$): $\ssp(p+1,p+1)^{+++}$

The restricted root system is of $C_{p+1}^{+++}$-type,
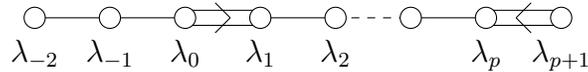
\begin{center}
\begin{tikzpicture} [place/.style={circle,draw=black,fill=white, inner sep=0pt,minimum size=8}] 
\draw(-1,.1)--(0,.1);  \draw(-1,-.1)--(0,-.1);
\draw (0,0) -- (1,0); \draw[dashed] (1,0) -- (2,0); \draw (2,0) -- (3,0);  
\draw (3,.1) -- (4,.1); \draw (3,-.1) -- (4,-.1); 
\draw (-0.6,0.2) -- (-0.4,0); \draw (-0.6,-0.2) -- (-0.4,0); 
\draw (3.6,0.2) -- (3.4,0); \draw (3.6,-0.2) -- (3.4,0); 
\draw(-1,0)--(-2,0);\draw(-2,0)--(-3,0);
\node at (0,0) [place,label=below:$\lambda_1$] {}; 
\node at (1,0) [place,label=below:$\lambda_{2}$] {}; 
\node at (2,0) [place] {};     
\node at (3,0) [place,label=below:$\lambda_{p}$] {}; 
\node at (4,0) [place,label=below:$\lambda_{p+1}$] {}; 
\node at (-1,0) [place,label=below:$\lambda_0$] {}; 
\node at (-2,0) [place,label=below:$\lambda_{-1}$] {}; 
\node at (-3,0) [place,label=below:$\lambda_{-2}$] {}; 
\end{tikzpicture}
\end{center}
The roots $\lambda_i$ are degenerate $4$ times for $1 \leq i \leq p$.  The root $\lambda_{p+1}$ is degenerate $3$ times.  The affine root has also multiplicity $3$ and so the gravity line contains only $\lambda_{-1}$ and $\lambda_{-2}$.

\subsubsection{Compact real form $\usp(n)^{+++}$}

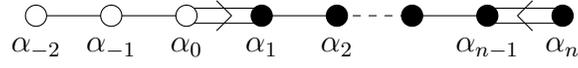
\begin{center}
\begin{tikzpicture} [place/.style={circle,draw=black,fill=white, inner sep=0pt,minimum size=8}] 
\draw(-1,.1)--(0,.1);  \draw(-1,-.1)--(0,-.1);
\draw (0,0) -- (1,0); \draw[dashed] (1,0) -- (2,0); \draw (2,0) -- (3,0);  
\draw (3,.1) -- (4,.1); \draw (3,-.1) -- (4,-.1); 
\draw (-0.6,0.2) -- (-0.4,0); \draw (-0.6,-0.2) -- (-0.4,0); 
\draw (3.6,0.2) -- (3.4,0); \draw (3.6,-0.2) -- (3.4,0); 
\draw(-1,0)--(-2,0);\draw(-2,0)--(-3,0);
\node at (0,0) [circle,draw=black,fill=black, inner sep=0pt,minimum size=8,label=below:$\bt_1$] {}; 
\node at (1,0) [circle,draw=black,fill=black, inner sep=0pt,minimum size=8,label=below:$\bt_{2}$] {}; 
\node at (2,0) [circle,draw=black,fill=black, inner sep=0pt,minimum size=8] {};     
\node at (3,0) [circle,draw=black,fill=black, inner sep=0pt,minimum size=8,label=below:$\bt_{n-1}$] {}; 
\node at (4,0) [circle,draw=black,fill=black, inner sep=0pt,minimum size=8,label=below:$\bt_{n}$] {}; 
\node at (-1,0) [place,label=below:$\bt_0$] {}; 
\node at (-2,0) [place,label=below:$\bt_{-1}$] {}; 
\node at (-3,0) [place,label=below:$\bt_{-2}$] {}; 
\end{tikzpicture}
\end{center}

\subsection{The $D_n$ case ($n \geq 4$)}

\subsubsection{$D_nI \equiv \so(p,q)$, $p,q>1$, $p+q = 2n$} 

\vspace{.3cm}

\noindent
{\bf Tits-Satake diagram of $\so(p,2n-p)^{+++}$

(i) $p\geq 2$, $p\leq n-2$}
\begin{center}
\begin{tikzpicture} [place/.style={circle,draw=black,fill=white, inner sep=0pt,minimum size=8}] 
\draw(-1,0)--(0,0);\draw[dashed] (0,0) -- (1,0); \draw (1,0) -- (2,0); \draw (2,0) -- (3,0); \draw[dashed] (3,0) -- (4,0); 
\draw(0,0)--(0,-1);
\draw (4,0) -- (5,1); \draw (4,0) -- (5,-1); 
\draw(-1,0)--(-2,0); \draw(-2,0)--(-3,0);
\node at (0,-1) [place,label=below:$\bt_1$] {}; 
\node at (0,0) [place,label=above:$\bt_2$] {}; 
\node at (1,0) [place,label=below:$\bt_{p}$] {}; 
\node at (2,0) [circle,draw=black,fill=black, inner sep=0pt,minimum size=8] {};     
\node at (3,0) [circle,draw=black,fill=black, inner sep=0pt,minimum size=8] {}; 
\node at (4,0) [circle,draw=black,fill=black, inner sep=0pt,minimum size=8,label=below:$\bt_{n-2}$] {}; 
\node at (5,1) [circle,draw=black,fill=black, inner sep=0pt,minimum size=8,label=right:$\bt_{n-1}$] {}; 
\node at (5,-1) [circle,draw=black,fill=black, inner sep=0pt,minimum size=8,label=right:$\bt_n$] {}; 
\node at (-1,0) [place,label=below:$\bt_0$] {}; 
\node at (-2,0) [place,label=below:$\bt_{-1}$] {}; 
\node at (-3,0) [place,label=below:$\bt_{-2}$] {}; 
\end{tikzpicture}
\end{center}

\vspace{.3cm}

(ii) $p = n-1$ ($\so(n-1,n+1)^{+++}$)
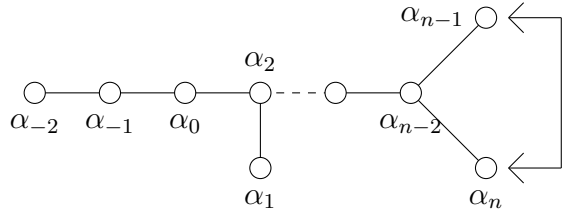
\begin{center}
\begin{tikzpicture} [place/.style={circle,draw=black,fill=white, inner sep=0pt,minimum size=8}] 
\draw(-1,0)--(0,0);\draw[dashed] (0,0) -- (1,0); \draw (1,0) -- (2,0); 
\draw (3.5,1.2) -- (3.3,1); \draw (3.5,.8) -- (3.3,1);
\draw (3.5,-.8) -- (3.3,-1); \draw (3.5,-1.2) -- (3.3,-1);
\draw(0,0)--(0,-1);
\draw (2,0) -- (3,1); \draw (2,0) -- (3,-1); 
\draw(-1,0)--(-2,0); \draw(-2,0)--(-3,0);
\draw(3.3,1)--(4,1);\draw(3.3,-1)--(4,-1);
\draw(4,-1)--(4,1);
\node at (0,-1) [place,label=below:$\bt_1$] {}; 
\node at (0,0) [place,label=above:$\bt_2$] {}; 
\node at (1,0) [place] {}; 
\node at (2,0) [circle,draw=black,fill=white, inner sep=0pt,minimum size=8,label=below:$\bt_{n-2}$] {}; 
\node at (3,1) [circle,draw=black,fill=white, inner sep=0pt,minimum size=8,label=left:$\bt_{n-1}$] {}; 
\node at (3,-1) [circle,draw=black,fill=white, inner sep=0pt,minimum size=8,label=below:$\bt_n$] {}; 
\node at (-1,0) [place,label=below:$\bt_0$] {}; 
\node at (-2,0) [place,label=below:$\bt_{-1}$] {}; 
\node at (-3,0) [place,label=below:$\bt_{-2}$] {}; 
\end{tikzpicture}
\end{center}

\vspace{.3cm}

(iii) $p = n$ ($\so(n,n)^{+++}$, split real form)
\begin{center}
\begin{tikzpicture} [place/.style={circle,draw=black,fill=white, inner sep=0pt,minimum size=8}] 
\draw(-1,0)--(0,0);\draw[dashed] (0,0) -- (1,0); \draw (1,0) -- (2,0); 
\draw(0,0)--(0,-1);
\draw (2,0) -- (3,1); \draw (2,0) -- (3,-1); 
\draw(-1,0)--(-2,0); \draw(-2,0)--(-3,0);
\node at (0,-1) [place,label=below:$\bt_1$] {}; 
\node at (0,0) [place,label=above:$\bt_2$] {}; 
\node at (1,0) [place] {}; 
\node at (2,0) [circle,draw=black,fill=white, inner sep=0pt,minimum size=8,label=below:$\bt_{n-2}$] {}; 
\node at (3,1) [circle,draw=black,fill=white, inner sep=0pt,minimum size=8,label=right:$\bt_{n-1}$] {}; 
\node at (3,-1) [circle,draw=black,fill=white, inner sep=0pt,minimum size=8,label=right:$\bt_n$] {}; 
\node at (-1,0) [place,label=below:$\bt_0$] {}; 
\node at (-2,0) [place,label=below:$\bt_{-1}$] {}; 
\node at (-3,0) [place,label=below:$\bt_{-2}$] {}; 
\end{tikzpicture}
\end{center}

\vspace{.3cm}

\noindent
{\bf Action on $\mu$ and $\bt_0$}

 \begin{equation}
\theta({\mu}) = - \mu, \; \; \; \; \theta({\bt_0}) = - \bt_0
\end{equation}
 
\vspace{.3cm}

\noindent
{\bf Restricted root system of $\so(p,2n-p)^{+++}$ ($p>1$, $n\geq p$)}

The restricted root system of $\so(p,2n-p)^{+++}$ is of $B_p^{+++}$-type, 
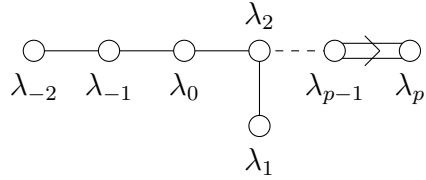
\begin{center}
\begin{tikzpicture} [place/.style={circle,draw=black,fill=white, inner sep=0pt,minimum size=8}] 
\draw(-1,0)--(0,0);\draw[dashed] (0,0) -- (1,0); 
\draw(0,0)--(0,-1);
\draw (1,.1) -- (2,.1); \draw (1,-.1) -- (2,-.1); 
\draw (1.4,0.2) -- (1.6,0); \draw (1.4,-0.2) -- (1.6,0); 
\draw(-1,0)--(-2,0); \draw(-2,0)--(-3,0);
\node at (0,-1) [place,label=below:$\lambda_1$] {}; 
\node at (0,0) [place,label=above:$\lambda_2$] {}; 
\node at (1,0) [place,label=below:$\lambda_{p-1}$] {}; 
\node at (2,0) [place,label=below:$\lambda_{p}$] {};     
\node at (-1,0) [place,label=below:$\lambda_0$] {}; 
\node at (-2,0) [place,label=below:$\lambda_{-1}$] {}; 
\node at (-3,0) [place,label=below:$\lambda_{-2}$] {}; 
\end{tikzpicture}
\end{center}
except for the split form $\so(n,n)^{+++}$ where it coincides with the Dynkin diagram of $\so(n,n)^{+++}$, 
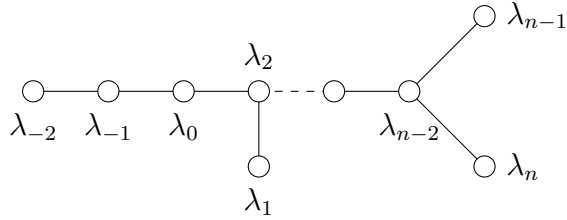
\begin{center}
\begin{tikzpicture} [place/.style={circle,draw=black,fill=white, inner sep=0pt,minimum size=8}] 
\draw(-1,0)--(0,0);\draw[dashed] (0,0) -- (1,0); \draw (1,0) -- (2,0); 
\draw(0,0)--(0,-1);
\draw (2,0) -- (3,1); \draw (2,0) -- (3,-1); 
\draw(-1,0)--(-2,0); \draw(-2,0)--(-3,0);
\node at (0,-1) [place,label=below:$\lambda_1$] {}; 
\node at (0,0) [place,label=above:$\lambda_2$] {}; 
\node at (1,0) [place] {}; 
\node at (2,0) [circle,draw=black,fill=white, inner sep=0pt,minimum size=8,label=below:$\lambda_{n-2}$] {}; 
\node at (3,1) [circle,draw=black,fill=white, inner sep=0pt,minimum size=8,label=right:$\lambda_{n-1}$] {}; 
\node at (3,-1) [circle,draw=black,fill=white, inner sep=0pt,minimum size=8,label=right:$\lambda_n$] {}; 
\node at (-1,0) [place,label=below:$\lambda_0$] {}; 
\node at (-2,0) [place,label=below:$\lambda_{-1}$] {}; 
\node at (-3,0) [place,label=below:$\lambda_{-2}$] {}; 
\end{tikzpicture}
\end{center}
In the $B_p$ case, the roots have multiplicity $1$ except $\lambda_p$ which is degenerate $2(n-p)$ times.

\subsubsection{$D_nII \equiv \so(1,2n-1)$} 

\vspace{.3cm}

\noindent
{\bf Tits-Satake diagram of $\so(1,2n-1)^{+++}$}

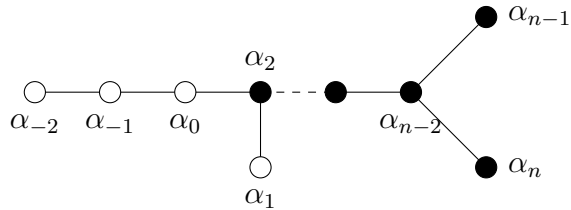
\begin{center}
\begin{tikzpicture} [place/.style={circle,draw=black,fill=white, inner sep=0pt,minimum size=8}] 
\draw(-1,0)--(0,0);\draw[dashed] (0,0) -- (1,0); \draw (1,0) -- (2,0); 
\draw(0,0)--(0,-1);
\draw (2,0) -- (3,1); \draw (2,0) -- (3,-1); 
\draw(-1,0)--(-2,0); \draw(-2,0)--(-3,0);
\node at (0,-1) [place,label=below:$\bt_1$] {}; 
\node at (0,0) [circle,draw=black,fill=black, inner sep=0pt,minimum size=8,label=above:$\bt_2$] {}; 
\node at (1,0) [circle,draw=black,fill=black, inner sep=0pt,minimum size=8] {}; 
\node at (2,0) [circle,draw=black,fill=black, inner sep=0pt,minimum size=8,label=below:$\bt_{n-2}$] {}; 
\node at (3,1) [circle,draw=black,fill=black, inner sep=0pt,minimum size=8,label=right:$\bt_{n-1}$] {}; 
\node at (3,-1) [circle,draw=black,fill=black, inner sep=0pt,minimum size=8,label=right:$\bt_n$] {}; 
\node at (-1,0) [place,label=below:$\bt_0$] {}; 
\node at (-2,0) [place,label=below:$\bt_{-1}$] {}; 
\node at (-3,0) [place,label=below:$\bt_{-2}$] {}; 
\end{tikzpicture}
\end{center}

\vspace{.3cm}

\noindent
{\bf Action on $\mu$ and $\bt_0$}

 \begin{equation}
\theta({\mu}) = - \bt_1, \; \; \; \; \theta({\bt_0}) = - \bt_0 - 2 \bt_2 - 2 \bt_3 - \ldots - 2 \bt_{n-2} - \bt_{n-1} - \bt_n
\end{equation}
One recognizes $- \theta(\bt_0)$ as the highest root of the $D_n$ subalgebra associated with the simple roots $\{\bt_0, \bt_2, \bt_3, \ldots, \bt_n\}$.
 
\vspace{.3cm}

\noindent
{\bf Restricted root system of $\so(1,2n-1)^{+++}$}

The restricted root system is the same as for $\su^*(4)^{+++}$, 
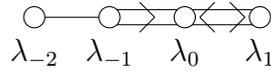
\begin{center}
\begin{tikzpicture} [place/.style={circle,draw=black,fill=white, inner sep=0pt,minimum size=8}]  
\draw (-1,.1) -- (0,.1); \draw (-1,-.1) -- (0,-.1); 
\draw (-1.6,0.2) -- (-1.4,0); \draw (-1.6,-0.2) -- (-1.4,0); 
\draw(-1,.1)--(-2,.1); \draw(-1,-.1)--(-2,-.1);
\draw (-0.4,0.2) -- (-0.2,0); \draw (-0.4,-0.2) -- (-0.2,0); 
\draw (-0.6,0.2) -- (-0.8,0); \draw (-0.6,-0.2) -- (-0.8,0); 
\draw(-2,0)--(-3,0);
\node at (0,0) [place,label=below:$\lambda_1$] {};  
\node at (-1,0) [place,label=below:$\lambda_0$] {}; 
\node at (-2,0) [place,label=below:$\lambda_{-1}$] {}; 
\node at (-3,0) [place,label=below:$\lambda_{-2}$] {}; 
\end{tikzpicture}
\end{center}
The roots $\lambda_1$ and $\lambda_0$ are degenerate $2(n-1)$ times.

\subsubsection{$D_nIII \equiv \so^*(2n)$}
\vspace{.3cm}

\noindent
{\bf Tits-Satake diagram of $\so^*(2n)^{+++}$}

(i) $n = 2p$

\begin{center}
\begin{tikzpicture} [place/.style={circle,draw=black,fill=white, inner sep=0pt,minimum size=8}] 
\draw(-1,0)--(0,0);\draw[dashed] (0,0) -- (1,0); \draw (1,0) -- (2,0); 
\draw(0,0)--(0,-1);
\draw (2,0) -- (3,1); \draw (2,0) -- (3,-1); 
\draw(-1,0)--(-2,0); \draw(-2,0)--(-3,0);
\node at (0,-1) [circle,draw=black,fill=black, inner sep=0pt,minimum size=8,label=below:$\bt_1$] {}; 
\node at (0,0) [circle,draw=black,fill=white, inner sep=0pt,minimum size=8,label=above:$\bt_2$] {}; 
\node at (1,0) [circle,draw=black,fill=black, inner sep=0pt,minimum size=8] {}; 
\node at (2,0) [circle,draw=black,fill=white, inner sep=0pt,minimum size=8,label=below:$\bt_{2(p-1)}$] {}; 
\node at (3,1) [circle,draw=black,fill=black, inner sep=0pt,minimum size=8,label=right:$\bt_{n-1}$] {}; 
\node at (3,-1) [circle,draw=black,fill=white, inner sep=0pt,minimum size=8,label=right:$\bt_n$] {}; 
\node at (-1,0) [place,label=below:$\bt_0$] {}; 
\node at (-2,0) [place,label=below:$\bt_{-1}$] {}; 
\node at (-3,0) [place,label=below:$\bt_{-2}$] {}; 
\end{tikzpicture}
\end{center}
The roots starting from $\bt_1$ are alternatively black and white up to $\bt_{2(p-1)} \equiv \bt_{n-2}$.  This branching point is white. The pair of roots $(\bt_{n-1}, \bt_n)$ contains one black root and one white root.

\vspace{.3cm}

(ii) $n = 2p+1$

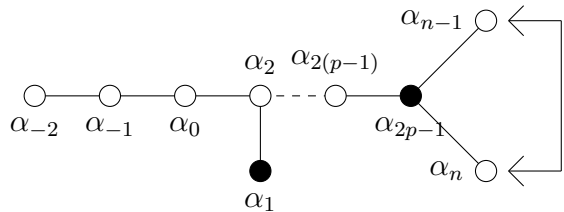
\begin{center}
\begin{tikzpicture} [place/.style={circle,draw=black,fill=white, inner sep=0pt,minimum size=8}] 
\draw(-1,0)--(0,0);\draw[dashed] (0,0) -- (1,0); \draw (1,0) -- (2,0); 
\draw(0,0)--(0,-1);
\draw (2,0) -- (3,1); \draw (2,0) -- (3,-1); 
\draw(-1,0)--(-2,0); \draw(-2,0)--(-3,0);
\draw(3.3,1)--(4,1);\draw(3.3,-1)--(4,-1);
\draw(4,-1)--(4,1);
\draw (3.5,1.2) -- (3.3,1); \draw (3.5,.8) -- (3.3,1);
\draw (3.5,-.8) -- (3.3,-1); \draw (3.5,-1.2) -- (3.3,-1);
\node at (0,-1) [circle,draw=black,fill=black, inner sep=0pt,minimum size=8,label=below:$\bt_1$] {}; 
\node at (0,0) [circle,draw=black,fill=white, inner sep=0pt,minimum size=8,label=above:$\bt_2$] {}; 
\node at (1,0) [circle,draw=black,fill=white, inner sep=0pt,minimum size=8,label=above:$\bt_{2(p-1)}$] {}; 
\node at (2,0) [circle,draw=black,fill=black, inner sep=0pt,minimum size=8,label=below:$\bt_{2p-1}$] {}; 
\node at (3,1) [circle,draw=black,fill=white, inner sep=0pt,minimum size=8,label=left:$\bt_{n-1}$] {}; 
\node at (3,-1) [circle,draw=black,fill=white, inner sep=0pt,minimum size=8,label=left:$\bt_n$] {}; 
\node at (-1,0) [place,label=below:$\bt_0$] {}; 
\node at (-2,0) [place,label=below:$\bt_{-1}$] {}; 
\node at (-3,0) [place,label=below:$\bt_{-2}$] {}; 
\end{tikzpicture}
\end{center}
The roots starting from $\bt_1$ are alternatively black and white up to $\bt_{2p-1} \equiv \bt_{n-2}$.  This branching point is black. 

\vspace{.3cm}

\noindent
{\bf Action on $\mu$ and $\bt_0$}

 \begin{equation}
\theta({\mu}) = - \mu, \; \; \; \; \theta({\bt_0}) = - \bt_0  
\end{equation}

\vspace{.3cm}

\noindent
{\bf Restricted root system of $\so^*(2n)^{+++}$}

(i) $n = 2p$

The restricted root system is of $C_p^{+++}$-type,
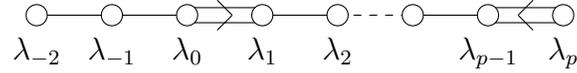
\begin{center}
\begin{tikzpicture} [place/.style={circle,draw=black,fill=white, inner sep=0pt,minimum size=8}] 
\draw(-1,.1)--(0,.1);  \draw(-1,-.1)--(0,-.1);
\draw (0,0) -- (1,0); \draw[dashed] (1,0) -- (2,0); \draw (2,0) -- (3,0);  
\draw (3,.1) -- (4,.1); \draw (3,-.1) -- (4,-.1); 
\draw (-0.6,0.2) -- (-0.4,0); \draw (-0.6,-0.2) -- (-0.4,0); 
\draw (3.6,0.2) -- (3.4,0); \draw (3.6,-0.2) -- (3.4,0); 
\draw(-1,0)--(-2,0);\draw(-2,0)--(-3,0);
\node at (0,0) [place,label=below:$\lambda_1$] {}; 
\node at (1,0) [place,label=below:$\lambda_{2}$] {}; 
\node at (2,0) [place] {};     
\node at (3,0) [place,label=below:$\lambda_{p-1}$] {}; 
\node at (4,0) [place,label=below:$\lambda_{p}$] {}; 
\node at (-1,0) [place,label=below:$\lambda_0$] {}; 
\node at (-2,0) [place,label=below:$\lambda_{-1}$] {}; 
\node at (-3,0) [place,label=below:$\lambda_{-2}$] {}; 
\end{tikzpicture}
\end{center}
The roots $\lambda_i$ are degenerate $4$ times for $1 \leq i \leq p-1$.  The root $\lambda_{p}$ is non-degenerate, as is the affine root. The gravity line contains $\lambda_{-2} $,$\lambda_{-1}$ and $\lambda_{0}$.

(ii) $n=2p +1$

The restricted root system of $\so^*(4p + 2)$  is of $BC_p$-type, leading to the twisted extension $A_{2p}^{(2)++}$,
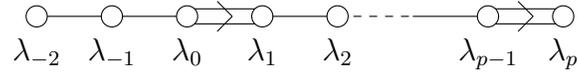
\begin{center}
\begin{tikzpicture} [place/.style={circle,draw=black,fill=white, inner sep=0pt,minimum size=8}] 
\draw(-1,.1)--(0,.1);  \draw(-1,-.1)--(0,-.1);
\draw (0,0) -- (1,0); \draw[dashed] (1,0) -- (2,0); \draw (2,0) -- (3,0);  
\draw (3,.1) -- (4,.1); \draw (3,-.1) -- (4,-.1); 
\draw (-0.6,0.2) -- (-0.4,0); \draw (-0.6,-0.2) -- (-0.4,0); 
\draw (3.4,0.2) -- (3.6,0); \draw (3.4,-0.2) -- (3.6,0); 
\draw(-1,0)--(-2,0);\draw(-2,0)--(-3,0);
\node at (0,0) [place,label=below:$\lambda_1$] {}; 
\node at (1,0) [place,label=below:$\lambda_{2}$] {};     
\node at (3,0) [place,label=below:$\lambda_{p-1}$] {}; 
\node at (4,0) [place,label=below:$\lambda_{p}$] {}; 
\node at (-1,0) [place,label=below:$\lambda_0$] {}; 
\node at (-2,0) [place,label=below:$\lambda_{-1}$] {}; 
\node at (-3,0) [place,label=below:$\lambda_{-2}$] {}; 
\end{tikzpicture}
\end{center}
The roots $\lambda_i$ are degenerate $4$ times for $1 \leq i \leq p$.  $2 \lambda_p$ is a root with multiplicity $1$.  The affine root has also multiplicity $1$ and so the gravity line contains $\lambda_{-2} $,$\lambda_{-1}$ and $\lambda_{0}$.

\subsubsection{Compact real form $\so(2n)^{+++}$}

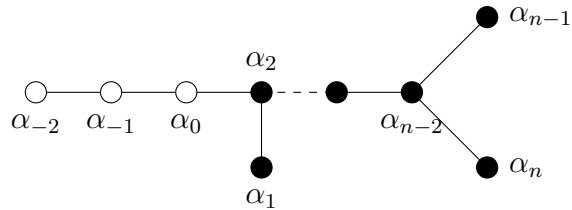
\begin{center}
\begin{tikzpicture} [place/.style={circle,draw=black,fill=white, inner sep=0pt,minimum size=8}] 
\draw(-1,0)--(0,0);\draw[dashed] (0,0) -- (1,0); \draw (1,0) -- (2,0); 
\draw(0,0)--(0,-1);
\draw (2,0) -- (3,1); \draw (2,0) -- (3,-1); 
\draw(-1,0)--(-2,0); \draw(-2,0)--(-3,0);
\node at (0,-1) [circle,draw=black,fill=black, inner sep=0pt,minimum size=8,label=below:$\bt_1$] {}; 
\node at (0,0) [circle,draw=black,fill=black, inner sep=0pt,minimum size=8,label=above:$\bt_2$] {}; 
\node at (1,0) [circle,draw=black,fill=black, inner sep=0pt,minimum size=8] {}; 
\node at (2,0) [circle,draw=black,fill=black, inner sep=0pt,minimum size=8,label=below:$\bt_{n-2}$] {}; 
\node at (3,1) [circle,draw=black,fill=black, inner sep=0pt,minimum size=8,label=right:$\bt_{n-1}$] {}; 
\node at (3,-1) [circle,draw=black,fill=black, inner sep=0pt,minimum size=8,label=right:$\bt_n$] {}; 
\node at (-1,0) [place,label=below:$\bt_0$] {}; 
\node at (-2,0) [place,label=below:$\bt_{-1}$] {}; 
\node at (-3,0) [place,label=below:$\bt_{-2}$] {}; 
\end{tikzpicture}
\end{center}

\subsection{The $G_2$ case}
\subsubsection{Split form $G_{2,2}$}
\vspace{.3cm}

\noindent
{\bf Tits-Satake diagram of $G_{2,2}^{+++}$}
\begin{center}
\begin{tikzpicture} [place/.style={circle,draw=black,fill=white, inner sep=0pt,minimum size=8}] 
\draw(-1,0)--(0,0);
\draw (0,.1) -- (1,.1); \draw (0,-.1) -- (1,-.1);  \draw (0,0) -- (1,0);
\draw (0.4,0.2) -- (0.6,0); \draw (0.4,-0.2) -- (0.6,0); 
\draw(-1,0)--(-2,0); \draw(-2,0)--(-3,0);
\node at (0,0) [place,label=below:$\bt_1$] {}; 
\node at (1,0) [place,label=below:$\bt_{2}$] {}; 
\node at (-1,0) [place,label=below:$\bt_0$] {}; 
\node at (-2,0) [place,label=below:$\bt_{-1}$] {}; 
\node at (-3,0) [place,label=below:$\bt_{-2}$] {}; 
\end{tikzpicture}
\end{center}
\begin{equation}
\theta({\bt_0}) = - \bt_0
\end{equation}

\subsubsection{Compact real form $G_{2,-14}$}
\noindent
{\bf Tits-Satake diagram of $G_{2,-14}^{+++}$}
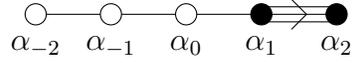
\begin{center}
\begin{tikzpicture} [place/.style={circle,draw=black,fill=white, inner sep=0pt,minimum size=8}] 
\draw(-1,0)--(0,0);
\draw (0,.1) -- (1,.1); \draw (0,-.1) -- (1,-.1);  \draw (0,0) -- (1,0);
\draw (0.4,0.2) -- (0.6,0); \draw (0.4,-0.2) -- (0.6,0); 
\draw(-1,0)--(-2,0); \draw(-2,0)--(-3,0);
\node at (0,0) [circle,draw=black,fill=black, inner sep=0pt,minimum size=8,label=below:$\bt_1$] {}; 
\node at (1,0) [circle,draw=black,fill=black, inner sep=0pt,minimum size=8,label=below:$\bt_{2}$] {}; 
\node at (-1,0) [place,label=below:$\bt_0$] {}; 
\node at (-2,0) [place,label=below:$\bt_{-1}$] {}; 
\node at (-3,0) [place,label=below:$\bt_{-2}$] {}; 
\end{tikzpicture}
\end{center}

\subsection{The $F_4$ case}
\subsubsection{$FI \equiv F_{4,4}$ (split form)}
\noindent
{\bf Tits-Satake diagram of $F_{4,4}^{+++}$}
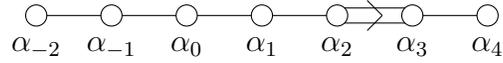
\begin{center}
\begin{tikzpicture} [place/.style={circle,draw=black,fill=white, inner sep=0pt,minimum size=8}] 
\draw(-1,0)--(0,0);
\draw(0,0)--(1,0);
\draw (2,0) -- (3,0); 
\draw (1,.1) -- (2,.1); \draw (1,-.1) -- (2,-.1); 
\draw (1.4,0.2) -- (1.6,0); \draw (1.4,-0.2) -- (1.6,0); 
\draw(-1,0)--(-2,0); \draw(-2,0)--(-3,0);
\node at (0,0) [place,label=below:$\bt_1$] {};
\node at (1,0) [place,label=below:$\bt_2$] {}; 
\node at (2,0) [place,label=below:$\bt_{3}$] {}; 
\node at (3,0) [place,label=below:$\bt_{4}$] {}; 
\node at (-1,0) [place,label=below:$\bt_0$] {}; 
\node at (-2,0) [place,label=below:$\bt_{-1}$] {}; 
\node at (-3,0) [place,label=below:$\bt_{-2}$] {}; 
\end{tikzpicture}
\end{center}

\subsubsection{$FII \equiv F_{4, -20}$}
\noindent
{\bf Tits-Satake diagram of $F_{4,-20}^{+++}$}
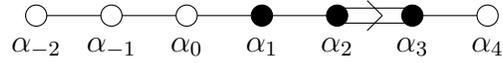
\begin{center}
\begin{tikzpicture} [place/.style={circle,draw=black,fill=white, inner sep=0pt,minimum size=8}] 
\draw(-1,0)--(0,0);
\draw(0,0)--(1,0);
\draw (2,0) -- (3,0); 
\draw (1,.1) -- (2,.1); \draw (1,-.1) -- (2,-.1); 
\draw (1.4,0.2) -- (1.6,0); \draw (1.4,-0.2) -- (1.6,0); 
\draw(-1,0)--(-2,0); \draw(-2,0)--(-3,0);
\node at (0,0) [circle,draw=black,fill=black, inner sep=0pt,minimum size=8,label=below:$\bt_1$] {};
\node at (1,0) [circle,draw=black,fill=black, inner sep=0pt,minimum size=8,label=below:$\bt_2$] {}; 
\node at (2,0) [circle,draw=black,fill=black, inner sep=0pt,minimum size=8,label=below:$\bt_{3}$] {}; 
\node at (3,0) [place,label=below:$\bt_{4}$] {}; 
\node at (-1,0) [place,label=below:$\bt_0$] {}; 
\node at (-2,0) [place,label=below:$\bt_{-1}$] {}; 
\node at (-3,0) [place,label=below:$\bt_{-2}$] {}; 
\end{tikzpicture}
\end{center}

\vspace{.3cm}

\noindent
{\bf Action on $\mu$ and $\bt_0$}

 \begin{equation}
\theta({\mu}) = - \bt_2 - 2 \bt_3 - 2 \bt_4, \; \; \; \; \theta({\bt_0}) = - \bt_0  - 2 \bt_1 - 2 \bt_2 - 2 \bt_3
\end{equation}

\vspace{.3cm}

\noindent
{\bf Restricted root system of $F_{4,-20}^{+++}$}

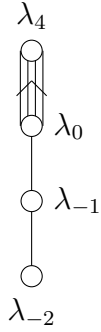
\begin{center}
\begin{tikzpicture} 
[place/.style={circle,draw=black, fill=white, inner sep=0pt,minimum size=8}] 
\draw(1.5,0)--(1.5,1); \draw(1.5,1)--(1.5,2); 
\draw(1.35,2)--(1.35,3); \draw(1.45,2)--(1.45,3); \draw(1.55,2)--(1.55,3); \draw(1.65,2)--(1.65,3);
\draw (1.3,2.4)--(1.5,2.6); \draw(1.7,2.4)--(1.5,2.6);
\node at (1.5,0) [place,label=below:$\lambda_{-2}$] {};
\node at (1.5,1) [place,label=right:$\lambda_{-1}$] {};
\node at (1.5,2) [place,label=right:$\lambda_{0}$] {};
\node at (1.5,3) [place,label=above:$\lambda_{4}$] {};
\end{tikzpicture}
\end{center}
The restricted root system is the same as for $su(1,n)^{+++}$ and of twisted $A_{2}^{(2)++}$-type. However, the multiplicities are different. The restricted root $\lambda_4$ has multiplicity $8$ and $2 \lambda_4$ is a root with multiplicity $7$.  The root $\lambda_0$ has also multiplicity 7.  Accordingly, contrary to what happens  for  $su(1,n)^{+++}$, it is not part of the gravity line and the system cannot be oxidized to $4$ dimensions.

\subsubsection{Compact real form $F_{4, -52}$}

\noindent
{\bf Tits-Satake diagram of $F_{4,-52}^{+++}$}
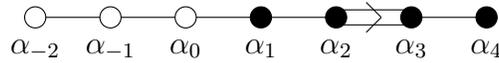
\begin{center}
\begin{tikzpicture} [place/.style={circle,draw=black,fill=white, inner sep=0pt,minimum size=8}] 
\draw(-1,0)--(0,0);
\draw(0,0)--(1,0);
\draw (2,0) -- (3,0); 
\draw (1,.1) -- (2,.1); \draw (1,-.1) -- (2,-.1); 
\draw (1.4,0.2) -- (1.6,0); \draw (1.4,-0.2) -- (1.6,0); 
\draw(-1,0)--(-2,0); \draw(-2,0)--(-3,0);
\node at (0,0) [circle,draw=black,fill=black, inner sep=0pt,minimum size=8,label=below:$\bt_1$] {};
\node at (1,0) [circle,draw=black,fill=black, inner sep=0pt,minimum size=8,label=below:$\bt_2$] {}; 
\node at (2,0) [circle,draw=black,fill=black, inner sep=0pt,minimum size=8,label=below:$\bt_{3}$] {}; 
\node at (3,0) [circle,draw=black,fill=black, inner sep=0pt,minimum size=8,label=below:$\bt_{4}$] {}; 
\node at (-1,0) [place,label=below:$\bt_0$] {}; 
\node at (-2,0) [place,label=below:$\bt_{-1}$] {}; 
\node at (-3,0) [place,label=below:$\bt_{-2}$] {}; 
\end{tikzpicture}
\end{center}

\subsection{The $E_6$ case}
\subsubsection{$EI \equiv E_{6,6}$ (split form)}
\noindent
{\bf Tits-Satake diagram of $E_{6,6}^{+++}$}
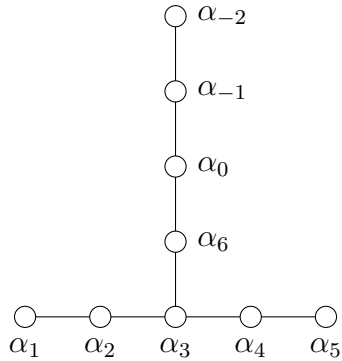
\begin{center}
\begin{tikzpicture} [place/.style={circle,draw=black,fill=white, inner sep=0pt,minimum size=8}] 
\draw(-1,0)--(0,0);
\draw(0,0)--(1,0);
\draw (2,0) -- (3,0); 
\draw (1,0) -- (2,0); \draw(1,0)--(1,1);
\draw(1,1)--(1,2);\draw(1,2)--(1,3);\draw(1,3)--(1,4);
\node at (1,0) [place,label=below:$\bt_3$] {};
\node at (2,0) [place,label=below:$\bt_{4}$] {}; 
\node at (3,0) [place,label=below:$\bt_{5}$] {}; 
\node at (0,0) [place,label=below:$\bt_2$] {}; 
\node at (-1,0) [place,label=below:$\bt_{1}$] {}; 
\node at (1,1) [place,label=right:$\bt_{6}$] {}; 
\node at (1,2) [place,label=right:$\bt_{0}$] {}; 
\node at (1,3) [place,label=right:$\bt_{-1}$] {}; 
\node at (1,4) [place,label=right:$\bt_{-2}$] {}; 
\end{tikzpicture}
\end{center}

\subsubsection{$EII$}
\noindent
{\bf Tits-Satake diagram of $E_{6,2}^{+++}$}
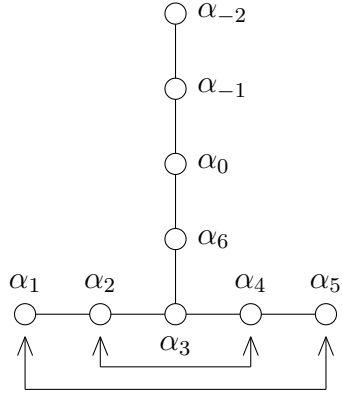
\begin{center}
\begin{tikzpicture} [place/.style={circle,draw=black,fill=white, inner sep=0pt,minimum size=8}] 
\draw(-1,0)--(0,0);
\draw(0,0)--(1,0);
\draw (2,0) -- (3,0); 
\draw (1,0) -- (2,0); \draw(1,0)--(1,1);
\draw(1,1)--(1,2);\draw(1,2)--(1,3);\draw(1,3)--(1,4);
\draw(0,-.7)--(2,-.7); \draw(0,-.7)--(0,-.3); \draw(2,-.7)--(2,-.3);
\draw(-1,-1)--(3,-1); \draw(-1,-1)--(-1,-.3); \draw(3,-1)--(3,-.3);
\draw (1.9,-.55) -- (2,-.3); \draw (2.1,-.55) -- (2,-.3);
\draw (-0.1,-.55) -- (0,-.3); \draw (0.1,-.55) -- (0,-.3);
\draw (2.9,-.55) -- (3,-.3); \draw (3.1,-.55) -- (3,-.3);
\draw (-1.1,-.55) -- (-1,-.3); \draw (-.9,-.55) -- (-1,-.3);
\node at (1,0) [place,label=below:$\bt_3$] {};
\node at (2,0) [place,label=above:$\bt_{4}$] {}; 
\node at (3,0) [place,label=above:$\bt_{5}$] {}; 
\node at (0,0) [place,label=above:$\bt_2$] {}; 
\node at (-1,0) [place,label=above:$\bt_{1}$] {}; 
\node at (1,1) [place,label=right:$\bt_{6}$] {}; 
\node at (1,2) [place,label=right:$\bt_{0}$] {}; 
\node at (1,3) [place,label=right:$\bt_{-1}$] {}; 
\node at (1,4) [place,label=right:$\bt_{-2}$] {}; 
\end{tikzpicture}
\end{center}

\vspace{.3cm}

\noindent
{\bf Action on $\mu$ and $\bt_0$}

\begin{equation}
\theta({\mu}) = - \mu, \; \; \; \; \theta({\bt_0}) = - \bt_0
\end{equation}

\vspace{.3cm}

\noindent
{\bf Restricted root system of $E_{6,2}^{+++}$}

The restricted root system of $E_{6,2}^{+++}$ is of $F_4$-type,
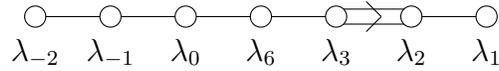
\begin{center}
\begin{tikzpicture} [place/.style={circle,draw=black,fill=white, inner sep=0pt,minimum size=8}] 
\draw(-1,0)--(0,0);
\draw(0,0)--(1,0);
\draw (2,0) -- (3,0); 
\draw (1,.1) -- (2,.1); \draw (1,-.1) -- (2,-.1); 
\draw (1.4,0.2) -- (1.6,0); \draw (1.4,-0.2) -- (1.6,0); 
\draw(-1,0)--(-2,0); \draw(-2,0)--(-3,0);
\node at (0,0) [place,label=below:$\lb_6$] {};
\node at (1,0) [place,label=below:$\lb_3$] {}; 
\node at (2,0) [place,label=below:$\lb_{2}$] {}; 
\node at (3,0) [place,label=below:$\lb_{1}$] {}; 
\node at (-1,0) [place,label=below:$\lb_0$] {}; 
\node at (-2,0) [place,label=below:$\lb_{-1}$] {}; 
\node at (-3,0) [place,label=below:$\lb_{-2}$] {}; 
\end{tikzpicture}
\end{center}
The roots $\lb_2$ and $\lb_1$ have multiplicity $2$, the others are non-degenerate.

\subsubsection{$EIII$}
\noindent
{\bf Tits-Satake diagram of $E_{6,-14}^{+++}$}
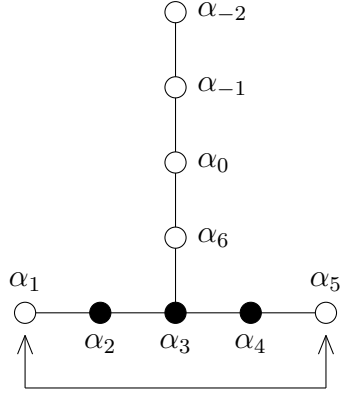
\begin{center}
\begin{tikzpicture} [place/.style={circle,draw=black,fill=white, inner sep=0pt,minimum size=8}] 
\draw(-1,0)--(0,0);
\draw(0,0)--(1,0);
\draw (2,0) -- (3,0); 
\draw (1,0) -- (2,0); \draw(1,0)--(1,1);
\draw(1,1)--(1,2);\draw(1,2)--(1,3);\draw(1,3)--(1,4);
\draw(-1,-1)--(3,-1); \draw(-1,-1)--(-1,-.3); \draw(3,-1)--(3,-.3);
\draw (2.9,-.55) -- (3,-.3); \draw (3.1,-.55) -- (3,-.3);
\draw (-1.1,-.55) -- (-1,-.3); \draw (-.9,-.55) -- (-1,-.3);
\node at (1,0) [circle,draw=black,fill=black, inner sep=0pt,minimum size=8,label=below:$\bt_3$] {};
\node at (2,0) [circle,draw=black,fill=black, inner sep=0pt,minimum size=8,label=below:$\bt_{4}$] {}; 
\node at (3,0) [place,label=above:$\bt_{5}$] {}; 
\node at (0,0) [circle,draw=black,fill=black, inner sep=0pt,minimum size=8,label=below:$\bt_2$] {}; 
\node at (-1,0) [place,label=above:$\bt_{1}$] {}; 
\node at (1,1) [place,label=right:$\bt_{6}$] {}; 
\node at (1,2) [place,label=right:$\bt_{0}$] {}; 
\node at (1,3) [place,label=right:$\bt_{-1}$] {}; 
\node at (1,4) [place,label=right:$\bt_{-2}$] {}; 
\end{tikzpicture}
\end{center}
\vspace{.3cm}

\noindent
{\bf Action on $\mu$ and $\bt_0$}

\begin{equation}
\theta({\mu}) = - \mu, \; \; \; \; \theta({\bt_0}) = - \bt_0
\end{equation}

\vspace{.3cm}

\noindent
{\bf Restricted root system of $E_{6,-14}^{+++}$}

The restricted root system of $E_{6,-14}$  is of $BC_2$-type, leading to the twisted extension $A_{4}^{(2)++}$.
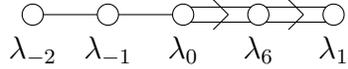
\begin{center}
\begin{tikzpicture} [place/.style={circle,draw=black,fill=white, inner sep=0pt,minimum size=8}] 
\draw(-1,.1)--(0,.1);  \draw(-1,-.1)--(0,-.1);  
\draw (0,.1) -- (1,.1); \draw (0,-.1) -- (1,-.1); 
\draw (-0.6,0.2) -- (-0.4,0); \draw (-0.6,-0.2) -- (-0.4,0); 
\draw (0.4,0.2) -- (0.6,0); \draw (0.4,-0.2) -- (0.6,0); 
\draw(-1,0)--(-2,0);\draw(-2,0)--(-3,0);
\node at (0,0) [place,label=below:$\lambda_6$] {}; 
\node at (1,0) [place,label=below:$\lambda_{1}$] {};     
\node at (-1,0) [place,label=below:$\lambda_0$] {}; 
\node at (-2,0) [place,label=below:$\lambda_{-1}$] {}; 
\node at (-3,0) [place,label=below:$\lambda_{-2}$] {}; 
\end{tikzpicture}
\end{center}
The root $\lambda_6$ is degenerate $6$ times.  The root $\lambda_1$ is degenerate $8$ times.  $2 \lambda_1$ is a root with multiplicity $1$.  The affine root has also multiplicity $1$ and is part of the gravity line.

\subsubsection{$EIV$}
\noindent
{\bf Tits-Satake diagram of $E_{6,-26}^{+++}$}
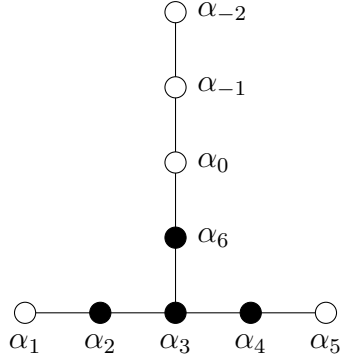
\begin{center}
\begin{tikzpicture} [place/.style={circle,draw=black,fill=white, inner sep=0pt,minimum size=8}] 
\draw(-1,0)--(0,0);
\draw(0,0)--(1,0);
\draw (2,0) -- (3,0); 
\draw (1,0) -- (2,0); \draw(1,0)--(1,1);
\draw(1,1)--(1,2);\draw(1,2)--(1,3);\draw(1,3)--(1,4);
\node at (1,0) [circle,draw=black,fill=black, inner sep=0pt,minimum size=8,label=below:$\bt_3$] {};
\node at (2,0) [circle,draw=black,fill=black, inner sep=0pt,minimum size=8,label=below:$\bt_{4}$] {}; 
\node at (3,0) [place,label=below:$\bt_{5}$] {}; 
\node at (0,0) [circle,draw=black,fill=black, inner sep=0pt,minimum size=8,label=below:$\bt_2$] {}; 
\node at (-1,0) [place,label=below:$\bt_{1}$] {}; 
\node at (1,1) [circle,draw=black,fill=black, inner sep=0pt,minimum size=8,label=right:$\bt_{6}$] {}; 
\node at (1,2) [place,label=right:$\bt_{0}$] {}; 
\node at (1,3) [place,label=right:$\bt_{-1}$] {}; 
\node at (1,4) [place,label=right:$\bt_{-2}$] {}; 
\end{tikzpicture}
\end{center}
\vspace{.3cm}

\noindent
{\bf Action on $\mu$ and $\bt_0$}

\begin{equation}
\theta({\mu}) =  -\bt_1 - \bt_2 - \bt_3 - \bt_4 - \bt_5, \; \; \; \; \theta({\bt_0}) = - \bt_0 - 2 \bt_6 - 2 \bt_3 - \bt_2 - \bt_4
\end{equation}
One recognizes $-\theta(\bt_0)$ as the highest root of the $D_5$-subalgebra associated with $\{\bt_0, \bt_6, \bt_3, \bt_2, \bt_4\}$.
\vspace{.3cm}

\noindent
{\bf Restricted root system of $E_{6,-26}^{+++}$}

The restricted root system of $E_{6,-26}^{+++}$ is identical with that of $\su^*(6)^{+++}$ (but the multiplicities are different) and given by
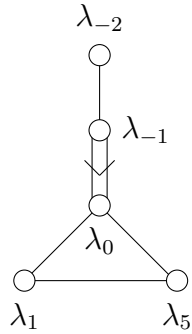
\begin{center}
\begin{tikzpicture} 
[place/.style={circle,draw=black, fill=white, inner sep=0pt,minimum size=8}] 
\draw (.5,0) -- (2.5,0); 
\draw (0.5,0) -- (1.5,1);\draw (2.5,0) -- (1.5,1);\draw (1.4,1)--(1.4,2); \draw (1.6,1)--(1.6,2); \draw (1.5,2) -- (1.5,3);
\draw (1.3,1.6)--(1.5,1.4); \draw(1.7,1.6)--(1.5,1.4);
\node at (.5,0) [place,label=below:$\lambda_1$] {};
\node at (2.53,0) [place,label=below:$\lambda_{5}$] {};
\node at (1.5,1) [place,label=below:$\lambda_0$] {};
\node at (1.5,2) [place,label=right:$\lambda_{-1}$] {};
\node at (1.5,3) [place,label=above:$\lambda_{-2}$] {};
\end{tikzpicture}
\end{center}
The multiplicity of $\lb_0$, $\lb_1$ and $\lb_5$ is $8$. The gravity line contains only $\lb_{-2}$ and $\lb_{-1}$.

\subsubsection{Compact real form $E_{6,-78}^{+++}$}
\noindent
{\bf Tits-Satake diagram of $E_{6,-78}^{+++}$}
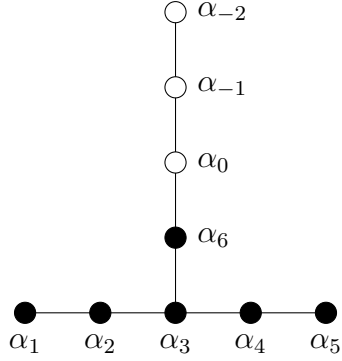
\begin{center}
\begin{tikzpicture} [place/.style={circle,draw=black,fill=white, inner sep=0pt,minimum size=8}] 
\draw(-1,0)--(0,0);
\draw(0,0)--(1,0);
\draw (2,0) -- (3,0); 
\draw (1,0) -- (2,0); \draw(1,0)--(1,1);
\draw(1,1)--(1,2);\draw(1,2)--(1,3);\draw(1,3)--(1,4);
\node at (1,0) [circle,draw=black,fill=black, inner sep=0pt,minimum size=8,label=below:$\bt_3$] {};
\node at (2,0) [circle,draw=black,fill=black, inner sep=0pt,minimum size=8,label=below:$\bt_{4}$] {}; 
\node at (3,0) [circle,draw=black,fill=black, inner sep=0pt,minimum size=8,label=below:$\bt_{5}$] {}; 
\node at (0,0) [circle,draw=black,fill=black, inner sep=0pt,minimum size=8,label=below:$\bt_2$] {}; 
\node at (-1,0) [circle,draw=black,fill=black, inner sep=0pt,minimum size=8,label=below:$\bt_{1}$] {}; 
\node at (1,1) [circle,draw=black,fill=black, inner sep=0pt,minimum size=8,label=right:$\bt_{6}$] {}; 
\node at (1,2) [place,label=right:$\bt_{0}$] {}; 
\node at (1,3) [place,label=right:$\bt_{-1}$] {}; 
\node at (1,4) [place,label=right:$\bt_{-2}$] {}; 
\end{tikzpicture}
\end{center}

\subsection{The $E_7$ case}
\subsubsection{$EV \equiv E_{7,7}$ (split form)}
\noindent
{\bf Tits-Satake diagram of $E_{7,7}^{+++}$}
\begin{center}
\begin{tikzpicture} [place/.style={circle,draw=black,fill=white, inner sep=0pt,minimum size=8}] 
\draw(-1,0)--(0,0);
\draw(0,0)--(1,0);
\draw (2,0) -- (3,0); 
\draw (1,0) -- (2,0); \draw(1,0)--(1,1);
\draw (3,0) -- (4,0);
\draw(-2,0)--(-1,0);
\draw(-2,0)--(-3,0);
\draw(-3,0)--(-4,0);
\node at (1,0) [place,label=below:$\bt_3$] {};
\node at (2,0) [place,label=below:$\bt_{4}$] {}; 
\node at (3,0) [place,label=below:$\bt_{5}$] {}; 
\node at (4,0) [place,label=below:$\bt_{6}$] {};
\node at (0,0) [place,label=below:$\bt_2$] {}; 
\node at (-1,0) [place,label=below:$\bt_{1}$] {}; 
\node at (1,1) [place,label=right:$\bt_{7}$] {}; 
\node at (-2,0) [place,label=below:$\bt_{0}$] {};
\node at (-3,0) [place,label=below:$\bt_{-1}$] {};
\node at (-4,0) [place,label=below:$\bt_{-2}$] {};
\end{tikzpicture}
\end{center}
\begin{equation}
\theta({\bt_0}) = - \bt_0
\end{equation}

\subsubsection{$EVI \equiv E_{7,-5}$}
\noindent
{\bf Tits-Satake diagram of $E_{7,-5}^{+++}$}
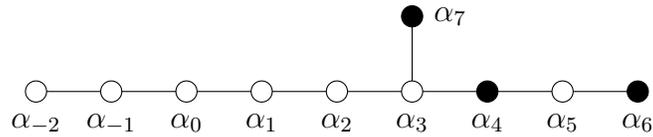
\begin{center}
\begin{tikzpicture} [place/.style={circle,draw=black,fill=white, inner sep=0pt,minimum size=8}] 
\draw(-1,0)--(0,0);
\draw(0,0)--(1,0);
\draw (2,0) -- (3,0); 
\draw (1,0) -- (2,0); \draw(1,0)--(1,1);
\draw (3,0) -- (4,0);
\draw(-2,0)--(-1,0);
\draw(-2,0)--(-3,0);
\draw(-3,0)--(-4,0);
\node at (1,0) [place,label=below:$\bt_3$] {};
\node at (2,0) [circle,draw=black,fill=black, inner sep=0pt,minimum size=8,label=below:$\bt_{4}$] {}; 
\node at (3,0) [place,label=below:$\bt_{5}$] {}; 
\node at (4,0) [circle,draw=black,fill=black, inner sep=0pt,minimum size=8,label=below:$\bt_{6}$] {};
\node at (0,0) [place,label=below:$\bt_2$] {}; 
\node at (-1,0) [place,label=below:$\bt_{1}$] {}; 
\node at (1,1) [circle,draw=black,fill=black, inner sep=0pt,minimum size=8,label=right:$\bt_{7}$] {}; 
\node at (-2,0) [place,label=below:$\bt_{0}$] {};
\node at (-3,0) [place,label=below:$\bt_{-1}$] {};
\node at (-4,0) [place,label=below:$\bt_{-2}$] {};
\end{tikzpicture}
\end{center}
\noindent
{\bf Action on $\mu$ and $\bt_0$}

\begin{equation}
\theta({\mu}) = - \mu, \; \; \; \; \theta({\bt_0}) = - \bt_0
\end{equation}

\vspace{.3cm}

\noindent
{\bf Restricted root system of $E_{7,-5}^{+++}$}

The restricted root system of $E_{7,-5}^{+++}$ is of $F_4$-type,
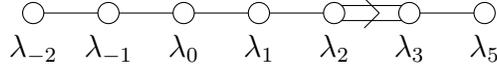
\begin{center}
\begin{tikzpicture} [place/.style={circle,draw=black,fill=white, inner sep=0pt,minimum size=8}] 
\draw(-1,0)--(0,0);
\draw(0,0)--(1,0);
\draw (2,0) -- (3,0); 
\draw (1,.1) -- (2,.1); \draw (1,-.1) -- (2,-.1); 
\draw (1.4,0.2) -- (1.6,0); \draw (1.4,-0.2) -- (1.6,0); 
\draw(-1,0)--(-2,0); \draw(-2,0)--(-3,0);
\node at (0,0) [place,label=below:$\lb_1$] {};
\node at (1,0) [place,label=below:$\lb_2$] {}; 
\node at (2,0) [place,label=below:$\lb_{3}$] {}; 
\node at (3,0) [place,label=below:$\lb_{5}$] {}; 
\node at (-1,0) [place,label=below:$\lb_0$] {}; 
\node at (-2,0) [place,label=below:$\lb_{-1}$] {}; 
\node at (-3,0) [place,label=below:$\lb_{-2}$] {}; 
\end{tikzpicture}
\end{center}
The roots $\lb_3$ and $\lb_5$ have multiplicity $4$, the others are non-degenerate.

\subsubsection{$EVII \equiv E_{7,-25}$}
\noindent
{\bf Tits-Satake diagram of $E_{7,-25}^{+++}$}
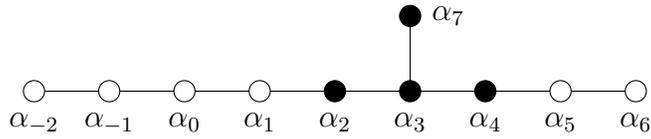
\begin{center}
\begin{tikzpicture} [place/.style={circle,draw=black,fill=white, inner sep=0pt,minimum size=8}] 
\draw(-1,0)--(0,0);
\draw(0,0)--(1,0);
\draw (2,0) -- (3,0); 
\draw (1,0) -- (2,0); \draw(1,0)--(1,1);
\draw (3,0) -- (4,0);
\draw(-2,0)--(-1,0);
\draw(-2,0)--(-3,0);
\draw(-3,0)--(-4,0);
\node at (1,0) [circle,draw=black,fill=black, inner sep=0pt,minimum size=8,label=below:$\bt_3$] {};
\node at (2,0) [circle,draw=black,fill=black, inner sep=0pt,minimum size=8,label=below:$\bt_{4}$] {}; 
\node at (3,0) [place,label=below:$\bt_{5}$] {}; 
\node at (4,0) [place,label=below:$\bt_{6}$] {};
\node at (0,0) [circle,draw=black,fill=black, inner sep=0pt,minimum size=8,label=below:$\bt_2$] {}; 
\node at (-1,0) [place,label=below:$\bt_{1}$] {}; 
\node at (1,1) [circle,draw=black,fill=black, inner sep=0pt,minimum size=8,label=right:$\bt_{7}$] {}; 
\node at (-2,0) [place,label=below:$\bt_{0}$] {};
\node at (-3,0) [place,label=below:$\bt_{-1}$] {};
\node at (-4,0) [place,label=below:$\bt_{-2}$] {};
\end{tikzpicture}
\end{center}
\noindent
{\bf Action on $\mu$ and $\bt_0$}

\begin{equation}
\theta({\mu}) = - \mu, \; \; \; \; \theta({\bt_0}) = - \bt_0
\end{equation}

\vspace{.3cm}

\noindent
{\bf Restricted root system of $E_{7,-25}^{+++}$}

The restricted root system is of $C_3^{+++}$-type,
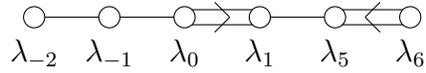
\begin{center}
\begin{tikzpicture} [place/.style={circle,draw=black,fill=white, inner sep=0pt,minimum size=8}] 
\draw(-1,.1)--(0,.1);  \draw(-1,-.1)--(0,-.1);
\draw (0,0) -- (1,0);  
\draw (1,.1) -- (2,.1); \draw (1,-.1) -- (2,-.1); 
\draw (-0.6,0.2) -- (-0.4,0); \draw (-0.6,-0.2) -- (-0.4,0); 
\draw (1.6,0.2) -- (1.4,0); \draw (1.6,-0.2) -- (1.4,0); 
\draw(-1,0)--(-2,0);\draw(-2,0)--(-3,0);
\node at (0,0) [place,label=below:$\lambda_1$] {}; 
\node at (1,0) [place,label=below:$\lambda_{5}$] {}; 
\node at (2,0) [place,label=below:$\lambda_{6}$] {}; 
\node at (-1,0) [place,label=below:$\lambda_0$] {}; 
\node at (-2,0) [place,label=below:$\lambda_{-1}$] {}; 
\node at (-3,0) [place,label=below:$\lambda_{-2}$] {}; 
\end{tikzpicture}
\end{center}
The roots $\lambda_1$ and $\lb_5$ are degenerate $8$ times.  The other roots are non-degenerate. The gravity line contains $\lambda_{-2} $, $\lambda_{-1}$ and $\lambda_{0}$.

\subsubsection{Compact real form $E_{7,-133}$}
\noindent
{\bf Tits-Satake diagram of $E_{7,-133}^{+++}$}
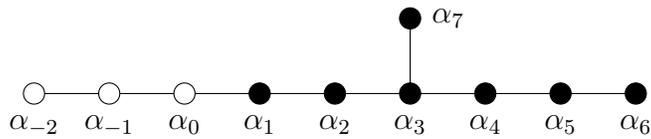
\begin{center}
\begin{tikzpicture} [place/.style={circle,draw=black,fill=white, inner sep=0pt,minimum size=8}] 
\draw(-1,0)--(0,0);
\draw(0,0)--(1,0);
\draw (2,0) -- (3,0); 
\draw (1,0) -- (2,0); \draw(1,0)--(1,1);
\draw (3,0) -- (4,0);
\draw(-2,0)--(-1,0);
\draw(-2,0)--(-3,0);
\draw(-3,0)--(-4,0);
\node at (1,0) [circle,draw=black,fill=black, inner sep=0pt,minimum size=8,label=below:$\bt_3$] {};
\node at (2,0) [circle,draw=black,fill=black, inner sep=0pt,minimum size=8,label=below:$\bt_{4}$] {}; 
\node at (3,0) [circle,draw=black,fill=black, inner sep=0pt,minimum size=8,label=below:$\bt_{5}$] {}; 
\node at (4,0) [circle,draw=black,fill=black, inner sep=0pt,minimum size=8,label=below:$\bt_{6}$] {};
\node at (0,0) [circle,draw=black,fill=black, inner sep=0pt,minimum size=8,label=below:$\bt_2$] {}; 
\node at (-1,0) [circle,draw=black,fill=black, inner sep=0pt,minimum size=8,label=below:$\bt_{1}$] {}; 
\node at (1,1) [circle,draw=black,fill=black, inner sep=0pt,minimum size=8,label=right:$\bt_{7}$] {}; 
\node at (-2,0) [place,label=below:$\bt_{0}$] {};
\node at (-3,0) [place,label=below:$\bt_{-1}$] {};
\node at (-4,0) [place,label=below:$\bt_{-2}$] {};
\end{tikzpicture}
\end{center}

\subsection{The $E_8$ case}
\subsubsection{$EVIII \equiv E_{8,8}$ (split form)}
\noindent
{\bf Tits-Satake diagram of $E_{8,8}^{+++}$}
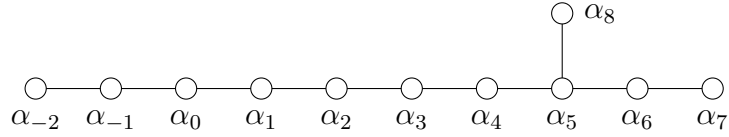
\begin{center}
\begin{tikzpicture} [place/.style={circle,draw=black,fill=white, inner sep=0pt,minimum size=8}] 
\draw(-1,0)--(0,0);
\draw(0,0)--(1,0);
\draw (2,0) -- (3,0); 
\draw (1,0) -- (2,0); \draw(6,0)--(6,1);
\draw (3,0) -- (4,0); \draw (4,0) -- (5,0); \draw (5,0) -- (6,0); \draw (6,0) -- (7,0);
\draw (7,0) -- (8,0);
\node at (1,0) [place,label=below:$\bt_0$] {};
\node at (2,0) [place,label=below:$\bt_{1}$] {}; 
\node at (3,0) [place,label=below:$\bt_{2}$] {}; 
\node at (4,0) [place,label=below:$\bt_{3}$] {}; 
\node at (5,0) [place,label=below:$\bt_{4}$] {};
\node at (6,0) [place,label=below:$\bt_{5}$] {};
\node at (7,0) [place,label=below:$\bt_{6}$] {};
\node at (8,0) [place,label=below:$\bt_{7}$] {};
\node at (0,0) [place,label=below:$\bt_{-1}$] {}; 
\node at (-1,0) [place,label=below:$\bt_{-2}$] {}; 
\node at (6,1) [place,label=right:$\bt_{8}$] {}; 
\end{tikzpicture}
\end{center}

\subsubsection{$EIX \equiv E_{8,-24}$}
\noindent
{\bf Tits-Satake diagram of $E_{8,-24}^{+++}$}
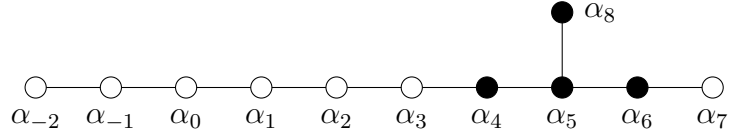
\begin{center}
\begin{tikzpicture} [place/.style={circle,draw=black,fill=white, inner sep=0pt,minimum size=8}] 
\draw(-1,0)--(0,0);
\draw(0,0)--(1,0);
\draw (2,0) -- (3,0); 
\draw (1,0) -- (2,0); \draw(6,0)--(6,1);
\draw (3,0) -- (4,0); \draw (4,0) -- (5,0); \draw (5,0) -- (6,0); \draw (6,0) -- (7,0);
\draw (7,0) -- (8,0);
\node at (1,0) [place,label=below:$\bt_0$] {};
\node at (2,0) [place,label=below:$\bt_{1}$] {}; 
\node at (3,0) [place,label=below:$\bt_{2}$] {}; 
\node at (4,0) [place,label=below:$\bt_{3}$] {}; 
\node at (5,0) [circle,draw=black,fill=black, inner sep=0pt,minimum size=8,label=below:$\bt_{4}$] {};
\node at (6,0) [circle,draw=black,fill=black, inner sep=0pt,minimum size=8,label=below:$\bt_{5}$] {};
\node at (7,0) [circle,draw=black,fill=black, inner sep=0pt,minimum size=8,label=below:$\bt_{6}$] {};
\node at (8,0) [place,label=below:$\bt_{7}$] {};
\node at (0,0) [place,label=below:$\bt_{-1}$] {}; 
\node at (-1,0) [place,label=below:$\bt_{-2}$] {}; 
\node at (6,1) [circle,draw=black,fill=black, inner sep=0pt,minimum size=8,label=right:$\bt_{8}$] {}; 
\end{tikzpicture}
\end{center} 
\noindent
{\bf Action on $\mu$ and $\bt_0$}

\begin{equation}
\theta({\mu}) = - \mu, \; \; \; \; \theta({\bt_0}) = - \bt_0
\end{equation}

\vspace{.3cm}

\noindent
{\bf Restricted root system of $E_{8,-24}^{+++}$}

The restricted root system of $E_{8,-24}^{+++}$ is of $F_4$-type,
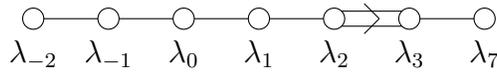
\begin{center}
\begin{tikzpicture} [place/.style={circle,draw=black,fill=white, inner sep=0pt,minimum size=8}] 
\draw(-1,0)--(0,0);
\draw(0,0)--(1,0);
\draw (2,0) -- (3,0); 
\draw (1,.1) -- (2,.1); \draw (1,-.1) -- (2,-.1); 
\draw (1.4,0.2) -- (1.6,0); \draw (1.4,-0.2) -- (1.6,0); 
\draw(-1,0)--(-2,0); \draw(-2,0)--(-3,0);
\node at (0,0) [place,label=below:$\lb_1$] {};
\node at (1,0) [place,label=below:$\lb_2$] {}; 
\node at (2,0) [place,label=below:$\lb_{3}$] {}; 
\node at (3,0) [place,label=below:$\lb_{7}$] {}; 
\node at (-1,0) [place,label=below:$\lb_0$] {}; 
\node at (-2,0) [place,label=below:$\lb_{-1}$] {}; 
\node at (-3,0) [place,label=below:$\lb_{-2}$] {}; 
\end{tikzpicture}
\end{center}
The roots $\lb_3$ and $\lb_7$ have multiplicity $8$, the others are non-degenerate.

\subsubsection{Compact real form $E_{8,-248}^{+++}$}
\noindent
{\bf Tits-Satake diagram of $E_{8,-248}^{+++}$}
\begin{center}
\begin{tikzpicture} [place/.style={circle,draw=black,fill=white, inner sep=0pt,minimum size=8}] 
\draw(-1,0)--(0,0);
\draw(0,0)--(1,0);
\draw (2,0) -- (3,0); 
\draw (1,0) -- (2,0); \draw(6,0)--(6,1);
\draw (3,0) -- (4,0); \draw (4,0) -- (5,0); \draw (5,0) -- (6,0); \draw (6,0) -- (7,0);
\draw (7,0) -- (8,0);
\node at (1,0) [place,label=below:$\bt_0$] {};
\node at (2,0) [circle,draw=black,fill=black, inner sep=0pt,minimum size=8,label=below:$\bt_{1}$] {}; 
\node at (3,0) [circle,draw=black,fill=black, inner sep=0pt,minimum size=8,label=below:$\bt_{2}$] {}; 
\node at (4,0) [circle,draw=black,fill=black, inner sep=0pt,minimum size=8,label=below:$\bt_{3}$] {}; 
\node at (5,0) [circle,draw=black,fill=black, inner sep=0pt,minimum size=8,label=below:$\bt_{4}$] {};
\node at (6,0) [circle,draw=black,fill=black, inner sep=0pt,minimum size=8,label=below:$\bt_{5}$] {};
\node at (7,0) [circle,draw=black,fill=black, inner sep=0pt,minimum size=8,label=below:$\bt_{6}$] {};
\node at (8,0) [circle,draw=black,fill=black, inner sep=0pt,minimum size=8,label=below:$\bt_{7}$] {};
\node at (0,0) [place,label=below:$\bt_{-1}$] {}; 
\node at (-1,0) [place,label=below:$\bt_{-2}$] {}; 
\node at (6,1) [circle,draw=black,fill=black, inner sep=0pt,minimum size=8,label=right:$\bt_{8}$] {}; 
\end{tikzpicture}
\end{center}

\end{document}